\documentclass[onecolumn, secnumarabic, amssymb, noshowpacs,aps,prb]{revtex4-2}
\usepackage[english]{babel}
\usepackage[dvips]{graphicx}
\usepackage{natbib}
\usepackage{dcolumn}
\usepackage{amsmath}
\DeclareMathOperator{\Tr}{Tr}
\usepackage[colorinlistoftodos, color=green!40, prependcaption]{todonotes}
\usepackage{xcolor}
\usepackage[style=plain]{floatrow}
\usepackage{graphicx}
\begin{document}
\title{The Transfer Matrix Method and The Theory of Finite Periodic Systems. From Heterostructures to Superlattices}

\author{Pedro Pereyra}
    \email[Correspondence email address: ]{pereyrapedro@gmail.com, ppereyra@azc.uam.mx}
    \affiliation{ Depto Ciencias B\'asicas, \'Area de F\'isica Te\'orica y Materia Condensada, Universidad Aut\'onoma Metropolitana, Azcapotzalco, Ciudad de M\'exico, C.P. 02200, Mexico.}

\date{\today} 

\begin{abstract}
Long-period systems and superlattices, with additional periodicity, have new effects on the energy spectrum and wave functions. Most approaches adjust theories for infinite systems, which is acceptable for large but not small number of unit cells $n$. In the past 30 years, a theory based entirely on transfer matrices was developed, where the finiteness of $n$ is an essential condition. The theory of finite periodic systems (TFPS) is also valid for any number of propagating modes, and arbitrary potential profiles (or refractive indices). We review this theory, the transfer matrix definition, symmetry properties, group representations, and relations with the scattering amplitudes. We summarize the derivation of multichannel matrix polynomials  (which reduce to Chebyshev polynomials in the one-propagating mode limit), the analytical formulas for resonant states, energy eigenvalues, eigenfunctions, parity symmetries, and discrete dispersion relations, for superlattices with different confinement characteristics. After showing the inconsistencies and limitations of hybrid approaches that combine the transfer-matrix method with Floquet's theorem, we review some applications of the TFPS to multichannel negative resistance, ballistic transistors, channel coupling, spintronics, superluminal, and optical antimatter effects. We review two high-resolution experiments using superlattices: tunneling time in photonic band-gap and optical response of blue-emitting diodes, and show extremely accurate theoretical predictions.

\vspace{0.2in}


{\it Keywords}: Theory of Finite Periodic Systems, Electronic Transport, Transmission Coefficients, Tunneling Time, Eigenvalues and Eigenfunctions of Superlattices, Optical Response of Superlattices, Transfer Matrix Method
\end{abstract}
\maketitle


\section{Introduction}

The continuous development of semiconductor devices and their simultaneous reduction to nanometric dimensions, has increased the need for precise and efficient calculations to solve the Maxwell or Schr\"odinger equation for a variety of systems, among them are the multiple quantum wells and superlattices whose complexity is intermediate between simple systems that are almost textbook examples and intricate three-dimensional molecular heterostructures that require heavy numerical calculations. Multiple quantum wells (MQWs) and superlattices (SLs) have become important and appealing structures for applications in optical and electronic devices. Different theoretical approaches have emerged to describe optical and semiconductor structures. Some, like the envelope function theory,\cite{Bastard1981,WhiteSham1981,Bastard1982,Altarelli1982} evolved by adjusting known theories (rigorously  valid for infinite periodic systems) that work well for large systems, others like the transfer matrix method, closely related to the scattering theory\cite{Borland1961} that was useful in other branches of physics, such as nuclear physics, electromagnetism  and elementary particles physics, was adapted and evolved into a rigorous formalism suitable for superlattices. The understanding of the effects that the `additional periodicity' of superlattices has on the energy spectrum and wave functions has also evolved, from the appearance of minigaps in the bands with continuous dispersion relations to the calculation of truly discrete subbands with  discrete dispersion relations and well defined surface state energies. Similarly, the apparent need to assume Bloch-type functions evolved into the real possibility of determining the true eigenfunctions of finite periodic systems.

The development of growing techniques of semiconductor structures such as the metalorganic chemical vapor deposition (MOCVD) and molecular beam  epitaxi (MBE)\cite{Cho1971} reached in the 70s the ability to produce heterolayered structures  (first suggested by H. Kroemer\cite{Kroemer1957}), quantum wells (QWs), multiple quantum wells and superlattices, \cite{Keldysh1962, EsakiTsu1970,Dupuis1978}.  It also opened up an intense race in search of better optical devices\cite{Kroemer2,Alferov1967,Ruprecht1967,Alferov1970,PanishHayashi1970} and a new era of artificial and endless growing family of heterostructures with new properties and a new spectrum of application possibilities which settled and expanded the research universe of physics. Finally new semiconductor superlattices has become much more appealing than metallic alloys superlattices that already have a longer history.\cite{Seemann1929,Bradley1932,Peierls1936,Wilson1938,Hultgren1938,Wang1945, Slater1949,
Nicholas1953,Flinn1960,Marcinkowski1963,Levdik1965} The growth techniques of semiconductor structures have long ago reached the level of atomic-layer precision.

The emerging field of semiconductor superlattices, originated in the seminal research papers published in the 1970s, led to the study of a variety of basic properties, including, among others, the electron tunneling,\cite{EsakiTsu1970,Movaghar1988} semiconductor lasers,\cite{Alferov1970,Xu1983,Ludowise1983,Sakaki1984,Wu1984,Hayashi1984,
Kirilov1984,Cai1994,Alferov1988,Ji1987,Narukawa1997} injection current thresholds,\cite{Alferov1967,Tsang1979,Rezek} the temperature effect in growing processes,\cite{Dupuis1977} enhancement of charge carries mobility,\cite{Holonyak1978,Dingle1978} recombination and stimulated emission processes,\cite{Kazarinov1972,Vojak1979,Kolbas1979,Smith1983} exciton dimensionality,\cite{Peyghambarian1984,Chemla1984,So1991,Citrin1995,Narukawa1999} impurity effects and photoluminescence,\cite{Petroff1981} and photoreflectance.\cite{Glembocki1985,Pan1988} In recent years, the field of metallic superlattices has also seen a renewed momentum with the emergence of photonic crystals, with an overwhelming amount of theoretical and experimental work published.\cite{Yablonovitch1987,Pendry1994,Baba1999,Gadner2004,BottenI2000,
BottenII2000,Joanopoulos1995,Kawakami2003,Jia-Yasumoto2005, Pereyra2020,Pereyra2021,Camley1984,Vigneron1985,Xue1985,Wallis1987,
Nazarov1994,Quinn1995,Bria2004} A common feature of these papers is that they end up dealing with infinite or semi-infinite superlattices by introducing the approximate Bloch periodicity condition, whose first drawback is the derivation of {\it continuous} subbands, ie. Kronig-Penney\cite{Kronig1931} like bands  with dispersion relations that give, at best, the widths of the allowed and forbidden subbands. Almost simultaneously, other fields of interest for the theoretical and experimental physics of periodic systems have grown. Among  them are the tunneling time through optical superlattices, triggered by direct measurements of tunneling times of photons and optical pulses;\cite{Steinberg1993,Spielmann1994}  the blue laser diodes based on GaN superlattices in the active region,\cite{NakamuraBook,Nakamura2,Nawakami1997,Narukawa1999} with interesting features in the optical response due, particularly, to emissions from surface states. In the 90s and first years of this century there has been also much research activity in the field of spintronics,\cite{Datta1990,Baibich1988, VanEsch1997,Ohno1998, Johnston2001, Kohda2001, Dietl2000, Oiwa2002, Zutic2004,Matos2007} as well as the transport properties in magnetic superlattices.\cite{Itoh1993,Schad1999,Schuller1999,Asano1993,
Hillebrands1988,Leiner2003,Guo1998,Zeng1999,Cardoso2001,Cardoso2008,Levy1991, Polushkin2006,Huo2012,Zeng2002, Ji1997,Barnas1992,Mardaani2006, Lu2012,Ye2009,Zhu2009, Cardoso2001} Although quantum dots has become also another hot field, the interest on periodic arrays of quantum dots is scarce.   Lately, graphene and 2D systems\cite{Barbier2010,Killi2011} become important fields and most of the theoretical approaches rely on the Bloch theorem.

The field of optical and electronic periodic structures, both experimental and theoretical, is so vast that it is impossible to cover everything. As mentioned in the abstract, we will focus on the theory of finite periodic systems based entirely on transfer matrices and their properties, valid for any number of propagation modes, any number of unit cells, and arbitrary potential or refractive indices profiles. We explicitly exclude theoretical approaches\cite{Kohn1959,Kronig1931, YuCardona1996, Kochelap1999, Bastard1993} that in one way or another are based on the Bloch and Floquet theorem\cite{Bloch1928} which imply the assumption of infinite or semi-infinite systems,\cite{Levin,Tamm1,Ritov,Yeh,Cottam,Nkoma,Haupt,Camley1984,Vigneron1985,Xue1985,
Wallis1987,Wendler1987,Mochan1988,Mills1989, Trutschel1989,Sheng1992,
Nazarov1994,Pendry1994,Quinn1995,BottenI2000,BottenII2000,Lyndin,Bria2004,Inan1999} where relevant physical variables, such as the transmission or reflection coefficients, cannot be conceived without being inconsistent. Since the exclusion of the theoretical approaches for periodic systems, that use transfer matrices and are based on Kramers' argument to determine their dispersion relations,\cite{Kramers1935} implies neglecting most of the theoretical papers in this branch, we will include a section that justifies this decision and shows why these approaches are not consistent with the finite periodic systems theory. For this purpose, we will also outline the group structure of the transfer matrices to make clear that it contains a compact and a non-compact subgroup. We will show that the transfer matrices that are compatible with  Kramers' eigenvalue argument and fulfill the Floquet theorem, belong to the compact subgroup, are diagonal, imply local transmission coefficients equal to 1, no reflection, and no attenuation of the wave functions.

Often the band energies and Bloch functions were mistaken for the energy eigenvalues and eigenfunctions of {\it finite} periodic systems\cite{Ivchenko1997}, and a rigorous treatment of the frequency problem in the physical theory of crystals, was considered {\it impracticable}\cite{Ledermann}, and approximation methods, like the Born-von Karman cyclic boundary conditions were widely applied. Fairly accurate experiments\cite{Chang,EMendez,Luo,Kroemer2} and fanciful applications using superlattices in mesoscopic and nanoscopic domains stimulated the development of theoretical approaches to account for the fine structure inside energy bands. The full quantization of electrons and photons became a focal characteristic, relevant in a number of attractive applications as the foreseen ``zero-threshold lasers", where the electron-hole transitions couple with a single spontaneous emission mode\cite{Nakamura,Ponce}.

The layered characteristic of MQWs and SLs structures makes the quasi-one-dimensional scattering approach suitable for studying transport properties in these systems.
The transfer matrices method was widely used in the 40s and 50s of the last century to study wave propagation and electronic structure in 1D alloys,\cite{Jones1948,Abeles1948,James1949,Luttinger1951,Heavens1954,Born1959,Borland1961} and later to study resonant tunnelling and transmission coefficients in heterostructures and superlattices.\cite{Chang,EMendez,Luo,Kroemer2,Nakamura1996,Narukawa1999,
Ponce,Agacy,Cvetic1981, Claro1982, Vezzetti,Ricco,PerezAlvarez,
Kalotas,Griffiths1992,Sprung,Rozman,Pereyra1998}

The application of transfer matrices for one-dimensional local periodic systems was attractive and obvious, especially due to the simple relation that the transfer matrix has for the scattering amplitudes and the multiplicative property of the transfer matrices. The use of these matrices has been so appealing that one of the most important and well-known results, the $n$-cell transfer matrix, that was first reported by R. Clark Jones in 1941,\cite{Jones1941} and later by Florin Abelès, in 1950 when studying the propagation of electromagnetic waves through layered media,\cite{Abeles1948,Born1959} was rediscovered many times.\cite{Jones1973, Cvetic1981,Claro1982,Vezzetti,PerezAlvarez,Kalotas, Griffiths1992, Sprung,Rozman,Macia1996,Pereyra1998,Pereyra2002} Given the transmission coefficients, the calculation of the Landauer conductance,\cite{Landauer1957,Landauer1970} became a common and  important goal in the analysis of periodic structures. In 1988,  Ram-Mohan et al.,\cite{Agarwall1988} by assuming that when going from layer to layer, the fast-varying periodic parts of the Bloch functions do not differ, developed an algorithm to calculate band structures based on the transfer matrix method together with the envelope function approximation.  Griffiths and Steinke\cite{Griffiths2001} used the transfer matrix approach to study the theory of waves propagation in different kind of 1D locally periodic media. Sprung {\it et al.} \cite{Sprung2003} studied the relation between  bound states and surface states in finite periodic systems. In the last years the theory of finite periodic systems was successfully applied to calculate optical transitions in the active region of (blue) laser devices,\cite{KunoldPereyra, Pereyra2018} to study phonon modes in wurtzite,\cite{Zhang2011} periodic structures, coupled resonators and surface acoustic waves for mode
localization sensors,\cite{Hanley2016} to adjust the coherent transport in finite periodic superlattices,\cite{Gornik} transport through  ultra-thin topological insulator films,\cite{Li2014} to model quantum well solar cells,\cite{Panlesku2010} to study the expectation values for Bloch
functions in finite domains,\cite{Pacher2007} bound states in the continuum,\cite{Cattapan} wave packets on finite lattices and through semiconductor and optical-media superlattices,\cite{Peisakovic2008,Simanjuntak2003,Simanjuntak2007,Simanjuntak2013}, to calculate the magneto-conductance of cylindrical
wires in longitudinal magnetic fields,\cite{Wei2008} persistent currents in small quantum rings,\cite{Szelag2006} spin transport through magnetic superlattices,\cite{Barnas1992,Cardoso2001, Ye2009,Pereyra2009,Lu2012} to explain the spin injection through Esaki barriers in ferromagnetic/nonmagnetic structures,\cite{Dietl2000,Ohno1999,Rashba2000,Ciorga2009,Sato2010,Pereyra2014} to study properties of metamaterial superlattices and the antimatter effect,\cite{RobledoRomero2009,Pereyra2011} to improve the theoretical approach to study electromagnetic
waves through fiber Bragg gratings,\cite{Pereyra2017b} to show why the effective mass approximation works well in nanoscopic structures,\cite{Pereyra2019} and many other physical properties and systems.

Unfortunately the use of the transfer matrix method has been a bit patchy, and a coherent summary of the transfer matrix capability, beyond the pure calculation of $n$-cells transmission coefficients, is lacking. The main purpose of the theory of finite periodic systems has been to use the transfer matrix properties and the physical meaning of this mathematical tool not only for the calculation of the resonant energies and wave functions, but also for determining  fundamental quantities, such as the energy eigenvalues and the corresponding eigenfunctions for bounded superlattices, plus congruous discrete dispersion relations.

We will start in section 2 with an introductory review of Bagwell's\cite{Bagwell1990} quasi-one-dimensional approach of electrons
described by a Schr\"odinger equation with locally periodic potential,  which solutions, describing $N$-propagating modes along the growing direction $z$, are determined in terms of the transfer matrix $W(z_2,z_1)$ that connects the wave functions an their derivatives at any two points.\cite{Pereyra1998,AnzaldoPereyra2007} We will then introduce the transfer matrix $M(z_2,z_1)$, which connects wave functions at $z_1$ and $z_2$, and the similarity transformation that relates $M$ with $W$. We will also introduce the transfer matrix in the WKB approximation for systems whose refractive  index and potential functions are not piecewise constant.\cite{Pereyrabook,Grossel1994} Among the important properties that we will first review are the transfer matrix representations determined by physical and symmetry requirements, such as time reversal invariance, spin-inversion symmetry and flux conservation, and then we will review the group structure of the transfer matrices,\cite{Bargmann,Pereyra1995} and the relation of transfer matrices with the scattering matrix and the scattering amplitudes.\cite{Borland1961} To establish this relation, perhaps the most appealing of the transfer matrix $M$, it will be convenient to define the transfer matrix in the basis  of incoming-outgoing functions. We will also show the relation of the scattering amplitudes with the transfer matrix $W$. In section 3 we review the main objectives of the theory of finite periodic systems: 1) the derivation of the transfer matrix $M_{Nn}$, for a system with $n$ unit cells, provided that the transfer matrix $M$ of a unit cell is known. In this derivation, it is assumed that the number of propagation modes $N$ is arbitrary and that the profile of the potentials or refractive indices is also arbitrary. This leads to determining a generalized recurrence relation for non-commutative polynomials, which solutions, the $N\times N$ matrix polynomials $p_{Nn}$, define the $N\times N$ matrix blocks of the transfer matrix $M_{Nn}$.\cite{Pereyra1998,Pereyra2002} As mentioned before, in the scalar one propagating mode limit, the polynomials $p_{Nn}$ become the well-known Chebyshev polynomials of the second kind $U_n$, and the $2N\times 2N$ transfer matrix $M_{Nn}$ becomes the $2\times 2$ transfer matrix $M_{n}$ of Jones and Abelès; 2) given the $n$-unit cells transfer matrix $M_{Nn}$, the second objective has been the calculation of fundamental physical quantities. Although the best-known relation is with the transmission and reflection amplitudes, \cite {Borland1961} other basic quantities that are naturally sought when solving the Schr\"odinger equation, are the eigenvalues and eigenfunctions, because they are very useful when studying other properties of confined superlattices and as important as the transmission coefficients in open SLs. In fact, although resonant levels and resonant properties in open systems were identified many years ago in the resonant behavior of transmission coefficients, and resonant behavior was observed in optical spectra, only in recent years has it been possible to determine analytic and general expressions for the evaluation of eigenvalues and eigenfunctions in confined superlattices, the parity symmetries of eigenfunctions, new transition selection rules, and closed expressions for the tunneling time. In this review, we will leave out the detailed derivation of non-commutative polynomials. Similarly, we will only briefly refer to the branch of periodic systems theory that uses the transfer matrix method combined with Floquet's theorem, which was first invoked by H. Jones.\cite{Jones1969} In the last sections we will discuss a few examples where the transfer matrix method and the multichhannel TFPS were used as an alternative or as the natural and appropriate description. In section \ref{DBandWannierSplitting}
we will address, first, the application of the transfer matrix method to study the resonant transport properties (transmission coefficients and conductance) in the negative resistance domain of a biased double barrier\cite{Bowen1997} (DB), which transverse dimension implies a number of propagating modes and the need of a multichannel approach.  We will then review the ballistic\cite{Pacher2002} and multichannel transport through superlattices.\cite{Pereyra2002} The theory of finite periodic systems has been also applied with success to study the transmission of electrons and electromagnetic wave packets through semiconductor and optical periodic structures\cite{Simanjuntak2003,Simanjuntak2007,Simanjuntak2013}, of electromagnetic waves through photonic crystals,\cite{Pereyra2019,Pereyra2020,Pereyra2021} and through left-handed SLs.\cite{RobledoRomero2009,Pereyra2011} Interesting results were also  found when studying the spin injection and the transport and manipulation of spin waves in magnetic SLs.\cite{Cardoso2001,Cardoso2005,Cardoso2008} We will present only brief summaries and main results here.  At the end, in section \ref{TunnelingAndBlueLaser}, we will review the application of TFPS to two different types of problems in which high-resolution experimental results were reported. First, we consider the tunneling time of photons and optical pulses in the photonic band gap of superlattices, as an example of application to transport problems, and then the calculation of the optical response of superlattices in the active region of a laser diode as an example of explicit calculation of energy eigenvalues, eigenfunctions, and transition matrix elements for electrons and holes in the conduction and valence bands.

\section{On the propagating modes and transfer matrices in quasi-1D systems}\label{mMultichannelApproach}

Most of the formulas written here are equally valid for electromagnetic systems and quantum systems. For the sake of simplicity and lack of space we will mainly discuss in terms of electronic systems, and the few  examples given in the next sections are, basically, in the one propagating mode limit.

\begin{figure}
    \centering
    \includegraphics[width=11cm]{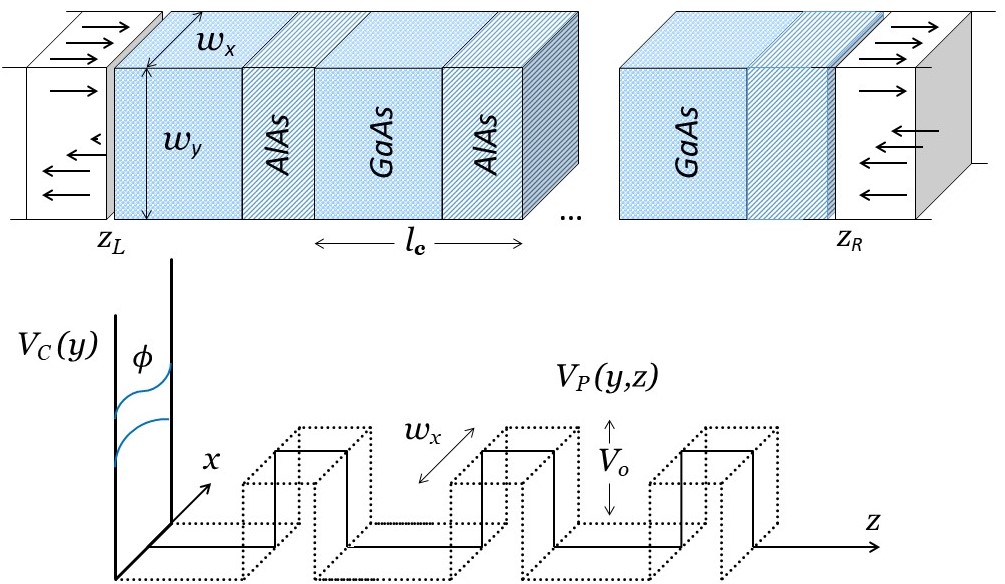}
    \caption{A quasi-1D periodic system $(GaAs/AlGaAs)^{n}$ of period $l_c$ located between  $z_L$ and $z_R$. In the upper panel, arrows indicate right-moving and left-moving waves. Lower panel describes the potential profile along the system with the transverse confining potential $V_C(y)$ and a periodic potential $V_P(y,z)$. $V_o$ is barrier height and $w_x$ is the width of layers along $x$. Figure reproduced with permission.\textsuperscript{[\onlinecite{Pereyra2002}]} 2002, Physical Review B. }
    \label{quasi1DSL}
\end{figure}

When we deal with  transport through a system of length $l=z_{R}-z_{L}$, and transverse cross section $w_{x}w_{y}$ connected to perfect leads of equal cross
section (see Figure \ref{quasi1DSL}),  the potential energy of charge carriers
can be modelled by a confining hard wall
potential $V_{C}(x,y)$ plus a potential $V_{P}(x,y,z)$,  which we will later require to be periodic, at least along the growing direction  $z$. For simplicity, we will consider $V_P(x,y,z)$ as a
stepwise function of $z$, extending from $z_L$ to $z_R$ with
discontinuities at $z=\ell_{r}$ (with $r=0,...,m_{\ell}$), and infinite outside $\{0\leq x\leq
w_{x},\,\,0\leq y\leq w_{y}\}$. The coordinates $\ell_r$
represent the end points of layer's and may coincide with the end points of the unit cells. For the periodic systems with $n$ unit cells along the $z$ axis, the end points of the unit cells will be at $z=\textsl{z}_j$ (with $j=0,...,n$, $\textsl{z}_{0}={z}_{L}$, and
$\textsl{z}_{n}=z_{R}$). When a unit cell contains two discontinuity points, $m_{\ell}=2n$. To solve the Schr\"odinger equation
\begin{equation}\label{FSch}
-\frac{\hbar ^{2}}{2m}\nabla^2\Psi({\bf r})+\left(V_{C}(x,y) +V_P({\bf r})\right) \Psi({\bf r})=E \Psi({\bf r})
\end{equation}
we follow Ref. [\onlinecite{Bagwell1990}] and solve first the
Schr\"odinger equation in the leads
\begin{equation}
-\frac{\hbar ^{2}}{2m}\left(\frac{\partial ^{2}}{\partial x^{2}}+\frac{\partial
^{2}}{\partial y^{2}}\right)\phi _{n_{x}n_{y}}+V_{C}\left( x,y\right) \phi
_{n_{x}n_{y}}=E_{n_{x}n_{y}}\phi _{n_{x}n_{y}},
\end{equation}
which give us the set of orthogonal and normalized functions $\phi _{n_{x}n_{y}}(x,y)$. The quantum numbers $n_x$, $n_y$ and the spin projections define the channel numbers $i=\left\{ n_{x},n_{y},s\right\} =1,2,....,s{\cal N}$. For a given Fermi energy $E$,  the (open) channels or propagating modes, in the leads, are those which threshold energies fulfill the relation
\begin{equation}
E_{Ti}=\frac{\hbar ^{2}\pi ^{2}}{2m}\left( \frac{n_{x}^{2}}{w_{x}^{2}}+\frac{%
n_{y}^{2}}{w_{y}^{2}}\right) \leq E,
\end{equation}
with longitudinal wave numbers
\begin{equation}
k_{i z}^{2}=\frac{2m}{\hbar^{2}}E-\pi^{2}\left (\frac{n_{x}^{2}}{w_{x}^{2}}+\frac{n_{y}^{2}}{w_{y}^{2}}  \right )\geq 0,
\end{equation}
and threshold wave numbers determined by the transverse wave number $k_{Ti}=\sqrt{2mE_{Ti}}/\hbar$.

Following P.F. Bagwell,\cite{Bagwell1990} we can use the set of  functions  $\{\phi _{i}(x,y)\}$ to express the  wave function $\Psi({\bf r})$  as
\begin{equation}
\Psi^r\left( x,y,z\right) =\sum_{i}\phi _{i}\left( x,y\right) \varphi^r
_{i}\left( z\right) \hspace{0.2in}{\rm for}\hspace{0.2in} \ell_{r}<z<\ell_{r+1},
\end{equation}
and substitute into the Schr\"{o}dinger equation  (\ref{FSch}), multiply by $\phi^*_{i}( x,y)$ and integrate to obtain the system of coupled equations
\begin{equation}\label{SistDifEq}
\frac{d^{2}}{dz^{2}}\varphi^r_{i}\left( z\right) +\left(k^{2}-k_{Ti}^{2}\right) \varphi^r_{i}\left( z\right) =\sum_{j}K^r_{ij}(z)\
\varphi^r_{j}\left( z\right) \qquad \qquad i=1,2,...
\end{equation}
where $k =\sqrt{2mE}/\hbar $, $k_{Ti}=\sqrt{2mE_{Ti}}/\hbar $
and $K^r_{ij}(z)$ the coupling matrix elements
\begin{equation}
K^r_{ij}(z)=\frac{2m}{\hbar ^{2}}\int \phi_{i}^{r\ast }\left( x,y\right)
V_{P}\left( x,y,z\right) \phi^r_{j}\left( x,y\right) dx dy \hspace{0.2in}{\rm for}\hspace{0.2in} \ell_{r}<z<\ell_{r+1}.
\end{equation}

The set of coupled equations (\ref{SistDifEq}) is infinite, therefore
impossible to solve in general. Thus, it is natural to cut it at a
finite number $N=s{\cal N}$, which we call the number of channels that,
depending on the Fermi energy, may include only open channels or
open plus some closed channels. For a discussion on the alternatives for choosing the transverse solutions in the leads or inside the system, see Ref. [\onlinecite{AnzaldoPereyra2007}]. Among all the possibilities, we have either coupled or uncoupled channels.

When the potential function $V_{P}\left( x,y,z\right)$  is a stepwise function of $z$, with discontinuities at $z=\ell_r$, and does not couple channels, the matrix $K^r(z)$ is diagonal, and the propagating modes are solutions of
\begin{equation}
\frac{d^{2}}{dz^{2}}\varphi^r_{i}\left( z\right) +\left(k^{2}-k_{Ti}^{2}\right) \varphi^r_{i}\left( z\right) =K^r_{ii}(z)\
\varphi^r_{i}\left( z\right), \qquad \qquad i=1,2,....,N \hspace{0.2in}{\rm and}\hspace{0.2in} \ell_{r}<z<\ell_{r+1}.
\end{equation}
The solutions of these set of equations differ in the threshold wave number $k_{Ti}$. For energies below $E_{Ti}$ the channels are closed and the solutions are evanescent. When channels are open, the solutions are oscillating functions ($e^{\pm ik_{zi}z}$) when $k^{2}-k_{Ti}^{2}-K^r_{ii}>0$, otherwise, the solutions are exponential functions ($e^{\pm q_{zi}z}$). We shall represent the right and left moving $i$-th
propagating mode (with spin $\sigma $)  as  ${\stackrel{\rightarrow }{\varphi}}_{ir}\left( z\right) $ and $\stackrel{\leftarrow }{\varphi}_{ir}\left( z\right)$, respectively. The total wave functions at any point $\ell_r<z<\ell_{r+1}$ in the scattering region, can be written as
\begin{equation}\label{wavefunctionphi}
\varphi (z)=\sum_{i=1}^{{\cal N}}(a_{ir }\stackrel{\rightarrow }{\varphi }_{ir }(z)+b_{ir }\stackrel{\leftarrow }{\varphi }_{ir}(z))=(a_r,b_r)\left(
\begin{array}{c}
\stackrel{\rightarrow }{\phi}_r\!(z)
\stackrel{\leftarrow }{\phi}_r\!(z)
\end{array}
\right)\hspace{0.2in}{\rm with}\hspace{0.2in} \ell_{r}<z<\ell_{r+1}.
\end{equation}
where $a_r$ and $b_r$ are $N$-dimensional coefficients and $\stackrel{\rightarrow }{\phi}_r\!\!(z) $ and $\stackrel{\leftarrow }{\phi}_r\!\!(z)$, $N$ dimensional state vectors.
To determine the wave function at any point $z$ within the system, we use the transfer matrix method, which ensures the rigorous fulfillment of the continuity requirements at each discontinuity point $\ell_r$ of the periodic system. The transfer matrix method is a useful tool, particularly simple in the case of decoupled channels. In this case and for piecewise potentials, we will present the two types of transfer matrices that will be used more in this review, the transfer matrices $M$ and $W$. After introducing these matrices, we will continue with the case of coupled channels.

\subsection{The transfer matrices $M$ and $W$}

To solve the dynamical equations for superlattices, using the transfer matrix method, we need to remember the transfer matrix definitions and some of their properties. There are at least two transfer matrices, generally denoted as $M$ and $W$. They both connect state vectors at any two points $z_1$ and $z_2$, contain the continuity conditions everywhere between $z_1$ and $z_2$, and are related to each other by a similarity transformation. Given the wave function $\varphi(z)$ of equation (\ref{wavefunctionphi}), we can write it as a vector, either as
\begin{equation}
\phi(z)=\left(
\begin{array}{c}
\stackrel{\rightarrow }{\phi_a}\!(z) \\
\stackrel{\leftarrow }{\phi_b}\!(z)
\end{array}
\right)\hspace{0.2in} \textsl{\rm or, as }\hspace{0.2in} f(z)=\left(
\begin{array}{c}
\stackrel{\rightarrow }{\phi_a}\!(z) + \stackrel{\leftarrow }{\phi_b}\!(z)\\
\stackrel{\rightarrow }{\phi_a'}\!(z) + \stackrel{\leftarrow }{\phi'_b}\!(z)
\end{array}
\right),
\end{equation}
where $\stackrel{\rightarrow }{\phi_a}\!(z)$ and  $\stackrel{\rightarrow }{\phi_b}\!(z)$ are $N$ dimensional vectors with elements $a_1\stackrel{\rightarrow }{\varphi_1}\!(z),...,a_N\stackrel{\rightarrow }{\varphi_N}\!(z)$ and $b_1\stackrel{\rightarrow }{\varphi_1}\!(z),...,b_N\stackrel{\rightarrow }{\varphi_N}\!(z)$, respectively, and $\stackrel{\rightarrow }{\phi'_a}\!(z)$ and $\stackrel{\rightarrow }{\phi'_b}\!(z)$ are their derivatives with respect to $z$, respectively. If  $z_1$ and $z_2$ are any two points in the system,  the transfer matrices $M(z_1,z_2)$ and $W(z_1,z_2)$ that connect the state vectors at these points are defined in general by
\begin{equation}\label{DefMW}
  \phi(z_2)=M(z_1,z_2)\phi(z_1)\hspace{0.2in}{\rm and }\hspace{0.2in}f(z_2)=W(z_1,z_2)f(z_1).
\end{equation}
In the next section we determine specific transfer matrices that fulfil continuity conditions.  It is common to write the transfer matrices in block
notation as
\begin{equation}
M(z_2,z_1)=\left(
\begin{array}{cc}
\alpha & \beta \\
\gamma & \delta
\end{array}
\right)\hspace{0.2in}{\rm and }\hspace{0.2in}W(z_2,z_1) =\left(
\begin{array}{cc}
\vartheta & \,\,\mu \\
\nu & \,\,\chi
\end{array}
\right),
\end{equation}
where $\alpha $, $\beta, \gamma $, $\delta $, $\vartheta$, $\mu$, $\nu$ and $\chi$ are $N\times N$ complex sub-matrices. There are some
constrictions between the submatrices $\alpha $, $\beta ,\gamma $ and $
\delta $, that depend on the physical properties and symmetries
inherent to the Hamiltonian of the system. The number of free parameters
and symmetries of the transfer matrices depend, in general, on the symmetry constrictions. We will refer later to these symmetries.

\subsection{Examples. Transfer matrices of quantum well and rectangular barrier}

The quantum well (QW) and the rectangular barrier (RB) are important structures and building blocks of larger systems, we outline here the calculation of the transfer matrices $M$ and $W$ for these structures.

\subsubsection{Transfer matrices for a rectangular quantum well}\label{TMRQW}

\begin{figure}[ht]
\begin{center}
\includegraphics[angle=0,width=190pt]{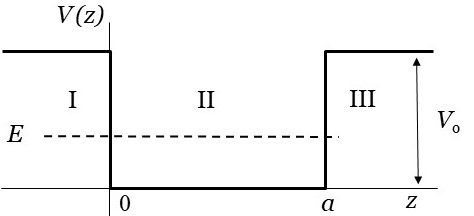}
\caption{Potential well with finite depth $V_o$ and width $a$.}\label{pozoe}
\end{center}
\end{figure}

If we have the quantum well shown in figure \ref{pozoe} and the particle's energy is $E<V_o$,
the solutions of the  Schr\"odinger equation  in regions
{\small I, II} and {\small III}  are, respectively,
\begin{eqnarray}
\varphi_{\rm I}(z)&=&a_{1}e^{qz}+b_{1}e^{-qz},\hspace{0.3in}{\rm for}\hspace{0.2in}  z \leq 0,  \nonumber \\
\varphi_{\rm II}(z)&=&a_{2}e^{ikz}+b_{2}e^{-ikz},\hspace{0.2in}{\rm for}\hspace{0.2in}  0 < z < a, \\
\varphi_{\rm III}(z)&=&a_{3}e^{qz}+b_{3}e^{-qz},\hspace{0.3in}{\rm for}\hspace{0.2in}   a \leq z.
\nonumber\label{funcspozo}
\end{eqnarray}
with $q=\sqrt{2m(V_o-E)/\hbar^2}$ and $k=\sqrt{2mE/\hbar^2}$. Although the wave functions $\varphi_{\rm I}(z)$ and $\varphi_{\rm III}(z)$ diverge when $z \rightarrow -\infty$ and $z \rightarrow \infty$, respectively, we keep temporarily the coefficients $b_1$ and $a_3$. Once the transfer matrices are determined, one can take $b_1=a_3=0$, if necessary. Before we determine the transfer matrices, it is worth writing the state vectors and some relations that will be used below. For the transfer matrix $M$ we need the state vectors
\begin{eqnarray}
\phi_{I}(z)=\left(\!\begin{array}{c}a_{1}e^{qz} \\b_{1}e^{-qz}\end{array}\!\right),\hspace{0.2in}\phi_{II}(z)=\left(\!\begin{array}{c}a_{2}e^{ikz} \\b_{2}e^{-ikz}\end{array}\!\right)\hspace{0.2in}{\rm and}\hspace{0.2in}\phi_{III}(z)=\left(\!\begin{array}{c}a_{3}e^{qz} \\b_{3}e^{-qz}\end{array}\!\right)
\end{eqnarray}
For the transfer matrix $W$ we need the following relation
\begin{eqnarray}\label{funcf2}
f_2(z)=\left(\!\begin{array}{c}\phi_{II}(z) \\\phi'_{II}(z)\end{array}\!\right)=\left(\!\begin{array}{cc}e^{ikz} & e^{-ikz} \\ike^{ikz}&-ike^{-ikz}\end{array}\!\right)\left(\begin{array}{c}a_{2} \\
b_{2}\end{array} \right)=Q_2(z)\left(\begin{array}{c}a_{2} \\
b_{2}\end{array} \right)
\end{eqnarray}

The well known procedure to solve the Schr\"odinger equation of determining the unknown coefficients by successive replacements in all equations  (all the equations that result from the continuity requirements on the  wave functions and  their first order derivatives), is done also in the transfer matrix method but in a more efficient and systematic way. At $z=0$, the continuity conditions imply the following equations
\begin{eqnarray}a_{1}+b_{1}=a_{2}+b_{2},\nonumber \\
q(a_{1}-b_{1})=ik(a_{2}-b_{2}),\label{continuipozo}
\end{eqnarray}
which, in matrix representation, can be written as
\begin{eqnarray}\label{e17}
\left(\begin{array}{c}a_{2} \\b_{2}\end{array} \right)= \frac{1}{2k}
\left(\begin{array}{cc}k-iq & k+iq \\k+iq
&k-iq\end{array}\right)\left(\begin{array}{c} a_{1} \\ b_{1}\end{array} \right)=
M_{l}(0^+,0^-) \left(\begin{array}{c}a_{1} \\
b_{1}\end{array} \right)
\end{eqnarray}
The transition matrix  that we have here, is a transfer matrix that connects the state vectors $\phi_{I}(0^{-})$ and $\phi_{II}(0^{+})$. Here and in the following $z^{\pm}$ means  $z \pm \epsilon$, in the limit of $\epsilon \rightarrow 0$. These relations could be used to determine the coefficients $a_2$ $b_2$, but in the transfer matrix method this is not really the purpose. For the type of quantities of interest, almost all the coefficients are unnecessary, and the TMM leaves us with functions that carry the relevant information of the physical processes. As in $z=0$, the continuity conditions at $z=a$, in matrix representation, take the form
\begin{eqnarray}\left(\!\begin{array}{c}a_{3}e^{qa} \\b_{3}e^{-qa}\end{array}\!\right)&=&
\frac{1}{2q}\left(\begin{array}{cc}q+ik & q-ik \\q-ik & q+ik\end{array} \right)
\left(\!\begin{array}{c}a_{2}e^{ika} \\b_{2}e^{-ika}\end{array} \!\right)=
M_{r}(a^+,a^-)\left(\!\begin{array}{c}a_{2}e^{ika} \\b_{2}e^{-ika}\end{array} \!\right), \label{e18}
\end{eqnarray}
with the transition matrix $M_{r}(a^+,a^-)$ equal (when the QW is symmetric) to the inverse  of the transition matrix $M_{l}(0^+,0^-)$.
To connect the state vector $\phi_{\rm II}(0^+)$ on the left  end of the well with the state vector $\phi_{\rm II}(a^-)$ on the right end, we need another transfer matrix that propagates the state vector in a constant potential region. It is easy to verify that
\begin{eqnarray}
\left(\begin{array}{c}a_{2}e^{ika} \\b_{2}e^{-ika}\end{array}\right)=
\left(\begin{array}{cc}e^{ika} & 0 \\0 & e^{-ika}\end{array} \right)\left(\begin{array}{c}a_{2}
\\b_{2}\end{array}\right)=M_{a}(a^-,0^+)\left(\begin{array}{c}a_{2}
\\b_{2}\end{array}\right).\label{e19}
\end{eqnarray}
With this matrix, we have all the necessary relations to connect the state vector in region { \small III} with the state vector in region {\small I}. Indeed, combining (\ref{e17}), (\ref{e18}) and (\ref{e19}), we obtain
\begin{eqnarray}
\left(\!\!\begin{array}{c}a_{3}e^{qa} \\b_{3}e^{-qa}\end{array} \!\!\right)\!=\!
\frac{1}{4qk}\left(\!
\begin{array}{cc}q+ik & q-ik \\q-ik & q+ik\end{array} \!\right)\!\left(\!\begin{array}{cc}\!e^{ika}\! &\! 0\! \\
\!0\! &\! e^{-ika}\!\end{array} \!\right)\!\left(\!\begin{array}{cc}k-iq & k+iq \\k+iq & k-iq\end{array}
\!\right)\!\left(\!\begin{array}{c}a_{1} \\b_{1}\end{array} \!\right).\hspace{0.4in}\label{e4.94}
\end{eqnarray}
The sequence of transition and transfer matrices define the QW transfer matrix
\begin{eqnarray}
M_{w}(a^+,0^-)=M_{r}(a^+,a^-)M_{p}(a^-,0^+)M_{l}(0^+,0^-).
\end{eqnarray}
After multiplying, and simplifying,  the {\it transfer matrix of the rectangular quantum well} is
\begin{eqnarray}\label{TransfMatrixMa}
M_{w}(a^+,0^-)\!=\!\left(\!\!\begin{array}{cc}\cos ka+\frac{q^2-k^2}{2qk}\sin ka & -\frac{k^2+q^2}{2qk}\sin ka \\
\frac{k^2+q^2}{2qk}\sin ka & \cos ka-\frac{q^2-k^2}{2qk}\sin ka \\\end{array} \!\right)=\left(\!
\begin{array}{cc}\alpha_{a} & \beta_{a} \\-\beta_{a} & \delta_{a} \end{array}\!\right),
\end{eqnarray}

For the calculation of the transfer matrix $W_w$ we can start from the second order differential equation or, given the solutions and the transfer matrix definition (\ref{DefMW}) we can obtain the transfer matrix $W_w(a^+,0^-)$, which satisfies the relation
\begin{eqnarray}\label{DefWw}
f_3(a^+)=W_w(a^+,0^-)f_1(0^-).
\end{eqnarray}
The functions $f_1(z)$ and $f_3(z)$, at the lateral barriers of the quantum well, are
\begin{eqnarray}\label{Relsfqwl}
f_1(z)=\left(\!\begin{array}{c}\phi_{I}(z) \\\phi'_{I}(z)\end{array}\!\right)=\left(\!\begin{array}{cc}e^{qz} & e^{-qz} \\qe^{qz}&-qe^{-qz}\end{array}\!\right)\left(\begin{array}{c}a_{1} \\
b_{1}\end{array} \right)=Q_1(z)\left(\begin{array}{c}a_{1} \\
b_{1}\end{array} \right)
\end{eqnarray}
\begin{eqnarray}\label{Relsfqwr}
f_3(z)=\left(\!\begin{array}{c}\phi_{III}(z) \\\phi'_{III}(z)\end{array}\!\right)=\left(\!\begin{array}{cc}e^{qz} & e^{-qz} \\qe^{qz}&-qe^{-qz}\end{array}\!\right)\left(\begin{array}{c}a_{3} \\
b_{3}\end{array} \right)=Q_3(z)\left(\begin{array}{c}a_{3} \\
b_{3}\end{array} \right)
\end{eqnarray}
Since the continuity conditions at $z=0$ and $z=a$ are
\begin{eqnarray}
f_3(a^+)=f_2(a^-)\hspace{0.2in}{\rm and}\hspace{0.2in} f_2(0^+)=f_1(0^-),
\end{eqnarray}
and the function $f_2(z)$, of equation (\ref{funcf2}), evaluated at these points is
\begin{eqnarray}
f_2(a^-)=Q_2(a^-)\left(\begin{array}{c}a_{2} \\
b_{2}\end{array} \right) \hspace{0.2in}{\rm and}\hspace{0.2in} f_2(0^+)=Q_2(0^+)\left(\begin{array}{c}a_{2} \\
b_{2}\end{array} \right),
\end{eqnarray}
equation (\ref{DefWw}) becomes
\begin{eqnarray}
f_3(a^+)=Q_2(a^-)Q^{-1}_2(0^+)f_1(0^-)=W_w(a^+,0^-)f_1(0^-)
\end{eqnarray}
Thus
\begin{eqnarray}
W_w(a^+,0^-)=\left(\!\begin{array}{cc}\cos ka & k^{-1}\sin  ka \\-k\sin ka& \cos ka\end{array}\!\right)
\end{eqnarray}
Using the relations (\ref{Relsfqwl}) and (\ref{Relsfqwr}), for $f_1(z)$ and $f_3(z)$ , it is easy to show that
\begin{eqnarray}
M_w(a^+,0^-)=\left(\!\begin{array}{cc}1 & 1\\q & -q\end{array}\!\right)^{-1}W_w(a^+,0^-)\left(\!\begin{array}{cc}1 & 1\\q & -q\end{array}\!\right).
\end{eqnarray}

\subsubsection{Transfer matrices of a rectangular potential barrier}

\begin{figure}[ht]
\begin{center}
\includegraphics[angle=0,width=170pt]{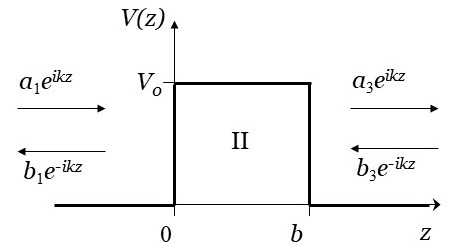}
\caption{The rectangular potential barrier, its potential parameters and the propagating solutions at the left and right sides.}\label{barrier}
\end{center}
\end{figure}
Let us now consider the rectangular potential barrier shown in figure \ref{barrier}.
Again, we will assume that $E < V_{o}$, and the solutions of the Schr\"odinger equations in each of the three regions are:
\begin{eqnarray}
\varphi_{\rm I}(z) = a_{1}e^{ikz} + b_{1}e^{-ikz}, \hspace{0.2in}{\rm{for}} \hspace{0.2in} z \leq 0,
\end{eqnarray}
\begin{eqnarray}
\varphi_{\rm II}(z) = a_{2}e^{qz} + b_{2}e^{-qz},\hspace{0.2in}{\rm{for}}\hspace{0.2in} 0< z < b,
\end{eqnarray}
\begin{eqnarray}
\varphi_{\rm III}(z) = a_{3}e^{ikz} + b_{3}e^{-ikz},\hspace{0.2in}{\rm{for}}\hspace{0.2in} z \geq b,
\end{eqnarray}
with $k = \sqrt{2m E/\hbar^{2}}$ and  $q = \sqrt{2m(V_{o} - E)/\hbar^{2}}$. The fulfillment of the continuity conditions, at $z=0$ and $z=b$, leads
to  establish the relation
\begin{eqnarray}
\phi_{\rm III}(b^+) = M_{r}(b^+,b^-)M_p(b^-,0^+)M_{l}(0^+,0^-)\phi_{\rm I}(0^-).\label{e4.60}
\end{eqnarray}
The transfer matrix barrier that connects state vectors at the left and right hand sides of the rectangular barrier is
\begin{eqnarray}
M_{b}(b^{+},0^{-}) = \frac{1}{4qk}\left(\begin{array}{cc} k - iq  &k + iq\cr k + iq  &k -
iq\end{array}\right)\left(\begin{array}{cc} e^{qb}  & 0 \cr  0  & e^{-qb}\end{array}\right)\left(\begin{array}{cc}
q + ik  &q - ik\cr  q - ik &q+ ik\end{array}\right)\!,\hspace{0.3in}\label{barriermatrix}
\end{eqnarray}
which after multiplying  and simplifying becomes
\begin{eqnarray}
M_{b}(b^{+},0^{-}) \!=\! \left(\begin{array}{cc} \cosh
qb + i\frac{k^{2} - q^{2}}{2qk}\sinh qb  & -i\frac{k^{2} + q^{2}}{2qk}\sinh qb\cr i\frac{k^{2} + q^{2}}{2qk}\sinh
qb  &\cosh qb - i\frac{k^{2} - q^{2}}{2qk}\sinh qb\end{array}\right)= \left(\begin{array}{cc} \alpha_{b} &\beta_{b}\cr  \beta^{\ast}_{b}
& \alpha^{\ast}_{b}\end{array}\right).
\end{eqnarray}
In the same way as for the quantum well, we can determine the transfer matrix $W_b(b^+,0^-)$ defined by
\begin{eqnarray}\label{DefWb}
f_3(b^+)=W_b(b^+,0^-)f_1(0^-).
\end{eqnarray}
With the continuity conditions at $z=0$ and $z=b$
\begin{eqnarray}
f_3(b^+)=f_2(b^-)\hspace{0.2in}{\rm and}\hspace{0.2in} f_2(0^+)=f_1(0^-).
\end{eqnarray}
We need now the function
\begin{eqnarray}\label{Relsfpr}
f_2(z)=\left(\!\begin{array}{c}\phi_{II}(z) \\\phi'_{II}(z)\end{array}\!\right)=\left(\!\begin{array}{cc}e^{qz} & e^{-qz} \\qe^{qz}&-qe^{-qz}\end{array}\!\right)\left(\begin{array}{c}a_{2} \\
b_{2}\end{array} \right)=Q_2(z)\left(\begin{array}{c}a_{2} \\
b_{2}\end{array} \right),
\end{eqnarray}
that allows us to write the relation
\begin{eqnarray}
f_2(b^-)=\left(\!\begin{array}{cc}e^{qb} & e^{-qb} \\qe^{qb}&-qe^{-qb}\end{array}\!\right)\frac{1}{2q}\left(\!\begin{array}{cc}q & 1 \\q&-1\end{array}\!\right)f_2(0^+)=Q_2(b^-)Q^{-1}_2(0^+)f_2(0^+).
\end{eqnarray}
Therefore
\begin{eqnarray}
f_3(b^+)=Q_2(b^-)Q^{-1}_2(0^+)f_1(0^-)=W_b(b^+,0^-)f_1(0^-),
\end{eqnarray}
with
\begin{eqnarray}
W_b(b^+,0^-)=\left(\!\begin{array}{cc}\cosh qb & q^{-1}\sinh  qb \\q\sinh qb& \cosh qb\end{array}\!\right),
\end{eqnarray}
and the relation between this matrix and the transfer matrix $M_b(b^+,0^-)$ is given by
\begin{eqnarray}\label{SimTransfBarrier}
M_b(b^+,0^-)=\left(\!\begin{array}{cc}1 & 1\\ik & -ik\end{array}\!\right)^{-1}W_b(b^+,0^-)\left(\!\begin{array}{cc}1 & 1\\ik & -ik\end{array}\!\right).
\end{eqnarray}

\subsection{Coupled channels and the transfer matrix W}

When the propagating modes are coupled,  we use the reduction of order method
of the theory of differential equations. For this purpose we need the state vector $f$, defined before,
with elements $f_{j}=a_j \varphi _{j}+b_j \varphi _{j}$ and $f_{j+N}=\phi _{j}^{\prime }$ for
$j=1,2,...N$. Using these functions, the system of coupled equations can be written as
\begin{equation}\label{feq}
f^{\prime }(z)=U_r f(z)\hspace{0.2in}{\rm for}\hspace{0.2in} z_{r}<z<z_{r+1}
\end{equation}
with
\begin{equation}
U_r=\left(
\begin{array}{cc}
0 & I_N \\
 K^r(z)- k^2 I_N+k_{T}^{2} & 0
\end{array}
\right)
\end{equation}
a $2N\times 2N$ matrix and $k_{T}=diag(k_{T1},k_{T2},...,k_{TN})$.
Since $K^r$ is symmetric and real, $U_r$ corresponds to an
infinitesimal symplectic transformation. For details see Ref. [\onlinecite{AnzaldoPereyra2007}]. It is simple to verify that the first order differential equation (\ref{feq}) has the solution
\begin{equation}
f(z)=exp[(z-z_{r})U_{r}] f(z_{r})=W(z,z_{r})f(z_{r}),\hspace{0.2in}{\rm for}\hspace{0.2in} z_{r}<z<z_{r+1}.
\end{equation}
If we define the symmetric matrix
\begin{equation}
u_{r}^{2}=\frac{2m}{\hbar ^{2}}(V^{r}-E I_N)+k_{T}^{2}
\hspace{0.2in}
{\rm such \;\, that}\hspace{0.2in}
U_r=\left(
\begin{array}{cc}
0 & I_N \\
u_{r}^2& 0
\end{array}
\right),
\end{equation}
and expand $W$ in power series, we obtain, for the transfer matrix $W(z,z_{r})$, the following representation
\begin{equation}
\label{hyper}
W^{(r)}(z)=\left(
\begin{array}{cc}
\cosh zu_{r} & u_{r}^{-1}\sinh zu_{r} \\
u_{r}\sinh zu_{r} & \cosh zu_{r}
\end{array}
\right),
\end{equation}
which is well known in the 1D-one channel
approaches, with $u_{r}$ scalar functions. In Ref. [\onlinecite{AnzaldoPereyra2007}], the matrix functions $\cosh (zu_{r})$
and $u_{r}^{\pm 1}\sinh (zu_{r})$ are written as polynomials of
degree $N$-$1$ in the matrix variable $u_{r}$. In block
notation we write the transfer matrix $W$ as
\begin{equation}
W =\left(
\begin{array}{cc}
\vartheta & \,\,\mu \\
\nu & \,\,\chi
\end{array}
\right),
\end{equation}
with $\vartheta$, $\mu$, $\nu$ and $\chi$, $N\times N$ sub-matrices. For some purposes, it is convenient to deal with the transfer matrix $M$. Based on the transfer matrix $W$ and transfer matrix $M$ definitions, it is easy to show that
\begin{equation}
M=\left(
\begin{array}{cc}
\kappa^{-1/2} & \kappa^{-1/2} \\
i\kappa^{1/2} & -i\kappa^{1/2}
\end{array}
\right) ^{-1}W\left(
\begin{array}{cc}
\kappa^{-1/2} & \kappa^{-1/2} \\
i\kappa^{1/2} & -i\kappa^{1/2}
\end{array}
\label{MandW} \right).
\end{equation}
where $\kappa =diag(k_{1},k_{2},...,k_{N})$.

Before we review the relation with the scattering amplitudes, we will consider the transfer matrices for potential functions which are not piecewise constant. This means transfer matrices in the WKB approximation, an approximation that works well for  an important class of potentials.

\subsection{Transfer matrices in the WKB approximation}

    \begin{figure}[ht]
        \CommonHeightRow{%
            \begin{floatrow}[2]%
                \ffigbox[\FBwidth]
                {\includegraphics[width=8.5cm]{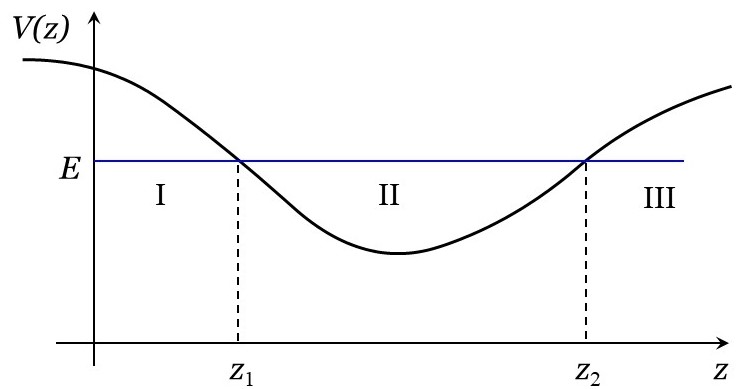}}
                {\caption{An arbitrary potential well.}}
                \ffigbox[\FBwidth]
                {\includegraphics[width=8.5cm]{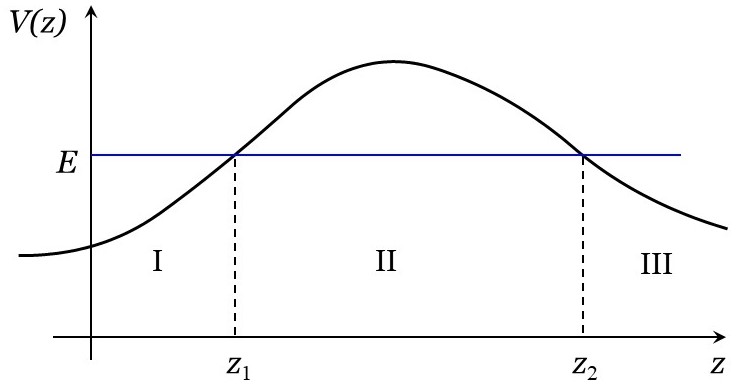}}
                {\caption{An arbitrary potential barrier.}}
            \end{floatrow}}%
    \end{figure}

Thanks to the atomic-layer precision reached in the growing techniques of heterostructures, the abrupt and ideal transition in the potential profiles can be justified for many real systems, however, in most of the actual systems the change in the gap energies is gradual and the potential profile at the interfaces is better modeled by continuous functions. In these cases, the transfer matrices in the WKB approximation are suited and convenient. The explicit derivation of these matrices are given in Ref. [\onlinecite{Pereyrabook}]. As can be seen there, the derivation, similar to that of a quantum well or a barrier, is based on the matrix representation of the continuity conditions, and careful cancellation of zeros of equal order. The main result for the transfer matrix of an arbitrary quantum well, as the one shown in figure 4, is
\begin{eqnarray}\label{transfmatpozo}
M_w(z_2^+,z_1^-)=\left(
\begin{array}{cc}
  \cos \xi & \sin \xi \\
  -\sin \xi & \cos \xi
\end{array}
\right),
\end{eqnarray}
where $z_1^{-}=$ $z_1$$- \epsilon$ and $z_2^{+}=$$z_2$ $+ \epsilon$, with $\epsilon$ infinitesimal, correspond to the classical return points $z_1$ and $z_2$, and
\begin{eqnarray}
   \xi(z_2^+,z_1^-)=\int_{z_{1}^-}^{z_{2}^+}k(z)dz.
\end{eqnarray}
The transfer matrix for an arbitrary potential, as the one shown in figure 5, is
\begin{eqnarray}\label{transfmatbarrier}
 M_b(z_2^+,z_1^-)=\left(
\begin{array}{cc}
  \cosh \zeta & -i\sinh \zeta \\
  i\sinh \zeta & \cosh \zeta
\end{array}
\right)
\end{eqnarray}
where
\begin{eqnarray}
   \zeta(z_{2}^+,z_{1}^-)=\int_{z_{1}^-}^{z_{2}^+}q(z)dz.
\end{eqnarray}

\section{Transfer matrix symmetries, group structure and the scattering amplitudes}\label{SecSimetScat}

The conservation of flux or current is an important principle, and the most common symmetries underlying the interactions are the time reversal invariance (TRI) and spin-rotation invariance (SRI). These symmetries
may not be present, but the requirement of flux conservation (FC)  must always hold. This requirement
implies that the transfer matrices should always fulfill the pseudo-unitarity condition\cite{MPK}
\begin{equation}
M\ \Sigma _z\,M^{\dagger }=\Sigma _z\ \hspace{0.2in} {\rm with}\hspace{0.2in} \Sigma
_z=\left(
\begin{array}{cc}
I_N & 0 \\
0 & -I_N
\end{array}
\right) .
\end{equation}
Here $I_N$ is the unit  matrix of dimension $N\times N$. In the absence of TRI, the Hamiltonians for both spin-dependent and
spin-independent interactions can be diagonalized by a unitary
transformation, and the system belongs to the {\it unitary universality
class}. The transfer matrices for this kind of systems are the most general
ones and will be represented as
\begin{equation}
M_u=\left(
\begin{array}{cc}
\alpha & \beta \\
\gamma & \delta
\end{array}
\right)
\end{equation}
with $\alpha \alpha ^{\dagger }-\beta \beta ^{\dagger }=I_N$, $\delta \delta
^{\dagger }-\gamma \gamma ^{\dagger }=I_N$ and $\alpha \gamma -\beta \delta
=0$, to satisfy the FC requirement, the matrix $M_u$ must fulfill the constraint $M_u\ \Sigma _z\,M_u^{\dagger }=\Sigma _z$. Here the
superscript $\dagger$ stands for the transpose conjugate, and $\Sigma _z$ is the Pauli matrix $\sigma_z$ of dimension $2N\times 2N$. When the interactions are time reversal invariant, the Hamiltonians for both spin-dependent and
spin-independent interactions can be diagonalized by an orthogonal
transformation, and the system belongs to the {\it orthogonal universality
class}. The transfer matrices for  spin-independent systems of the orthogonal universality class should fulfill the condition (see Refs. [\onlinecite{MPK,Pereyra1995}])
\begin{equation}
M=\Sigma_x M^* \Sigma_x
\end{equation}
where $\Sigma_x$ is the Pauli matrix $\sigma_x$ of dimension $2N \times 2N$. In this case, the transfer matrices
can be represented as
\begin{equation}
M_o=\left(
\begin{array}{cc}
\alpha & \beta \\
\beta^* & \alpha^*
\end{array}
\right)
\end{equation}
with $\alpha \alpha ^{\dagger }-\beta \beta ^{\dagger }=I_N$ and $\alpha \beta
^{T }-\beta^T \alpha =0$. Here the
superscript $T$ stands for the transpose of the matrix. The transfer matrices of the orthogonal class that fulfill both TRI and FC belong to the symplectic group $Sp$(2$N,\mathcal{C}$) and satisfy the requirement
\begin{equation}
M_o \mathcal{F}\,M_o^T=\mathcal{F}\hspace{0.3in}{\rm with}\hspace{0.3in}\mathcal{F}=\Sigma_z\Sigma_x=\left(\begin{array}{cc}
0& I_N  \\
-I_N&0
\end{array}
\right).
\end{equation}
Besides the physical symmetries and the requirements on the transfer matrices, the group structure and the possible representations, in terms of linearly independent parameters, are very important properties and we will now refer to this topic briefly. It has been shown in Ref. [\onlinecite{Pereyra1995}] that every transfer matrix $M_o$ can be written as the product of two matrices $M_{oc}$ and $M_{on}$ that belong to a compact and a noncompact subgroup, respectively, i.e.
\begin{equation}
M_o=M_{oc}M_{on}=\left(
\begin{array}{cc}
u & 0 \\
0 & u^*
\end{array}
\right)\left(
\begin{array}{cc}
\sqrt{I_N+\xi\xi^{\dagger}} & \xi \\
\xi^{\dagger} & \sqrt{I_N+\xi^{\dagger}\xi}
\end{array}
\right)
\end{equation}
where $u$ and $\xi$ are unitary and symmetric matrices, respectively. Similarly, if we are dealing with systems of the symplectic class with spin dependent interactions, the invariance under spin inversion and time reversal, for spin 1/2 particles, imply that the transfer matrices $M_s$ fulfill the requirement (see Ref. [\onlinecite{Pereyra1995}])
\begin{equation}
M_s^*=K^TM_sK\hspace{0.3in}{\rm with}\hspace{0.3in}K=\left(\begin{array}{cc}
0& \Sigma_y  \\
\Sigma_y&0
\end{array}
\right)\hspace{0.3in}{\rm and}\hspace{0.3in} \Sigma_y=\left(\begin{array}{cc}
0& -i I  \\
i I &0
\end{array}
\right).
\end{equation}
These matrices belong to the  pseudo-orthogonal $spO$(4$N,\mathcal{C}$) group, and decompose also as the product of a compact $M_{sc}$ and a noncompact matrix $M_{sn}$, i.e.
\begin{equation}
M_s=M_{sc}M_{sn}=\left(
\begin{array}{cc}
w & 0 \\
0 & \Sigma_y^Tw^*\Sigma_y
\end{array}
\right)\left(
\begin{array}{cc}
\sqrt{I_N+\eta\eta^{\dagger}} & \eta \\
\eta^{\dagger} & \sqrt{I_N+\eta^{\dagger}\eta}
\end{array}
\right)
\end{equation}
where $w$ is a unitary matrix and $\eta \Sigma_y=v(\sinh \lambda )\Sigma_yv^T$ is an anti-symmetric product, with $v$ a unitary matrix. When the interactions are not time reversal and spin rotation invariants, the systems belongs to the unitary class. The transfer matrices fulfill only the FC requirement, and belong to the pseudo-unitary $spU$(2$N,\mathcal{C}$) group, and every transfer matrix of this group can also be decomposed as the product of a compact $M_{uc}$ and a noncompact matrix $M_{un}$, i.e.
\begin{equation}
M_u=M_{uc}M_{un}=\left(
\begin{array}{cc}
w_1 & 0 \\
0 & w_2
\end{array}
\right)\left(
\begin{array}{cc}
\sqrt{I_N+\zeta\zeta^{\dagger}} & \zeta \\
\zeta^{\dagger} & \sqrt{I_N+\zeta^{\dagger}\zeta}
\end{array}
\right)
\end{equation}
where $w_1$ and $w_2$ are unitary and $\zeta =v_1 (\sinh \lambda)v_2$ an arbitrary square matrix, with $v_1$ and $v_2$ unitary. Notice that $\det M_o=\det M_s=\det M_u=1$.

\subsection{Transfer matrix and the scattering amplitudes}
\begin{figure}[ht]
\begin{center}
\includegraphics[angle=0,width=300pt]{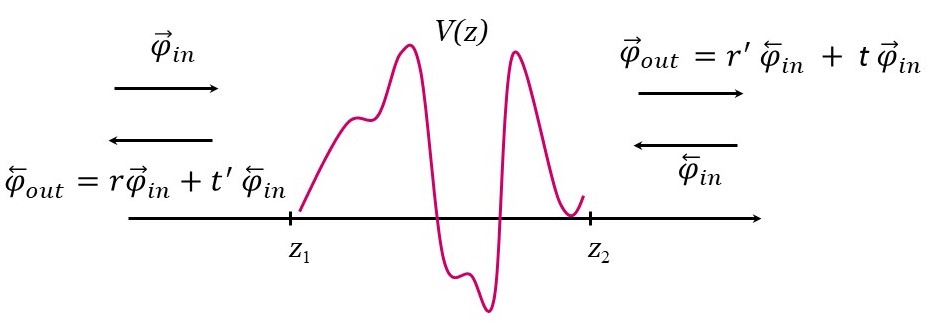}
\caption{Incoming and outgoing wave functions and the scattering amplitudes for incoming from the left and right hand sides. Figure reproduced with permission.\textsuperscript{[\onlinecite{Pereyra2002}]} 2002, Physical Review B.}\label{ScattAmps}
\end{center}
\end{figure}
To describe transport properties based on transfer matrices, it is worth recalling the well known relations between the transfer
matrices $M$ and $W$, and the scattering matrix $S$. This relation, first derived to our knowledge by Borland\cite{Borland1961,MPK} to show that atomic potentials could be replaced by $\delta$-functions surrounded by two regions of zero potential, is one of the
best known within the theoretical and experimental researches that use  transfer  matrices frequently. For a scattering process as the one
sketched in figure \ref{ScattAmps}, with incident amplitudes $\stackrel{\rightarrow }{\varphi }_{in}$ and $\stackrel{\leftarrow }{\varphi }_{in}$, from the left and the right hand side, respectively, the outgoing amplitudes are  $\stackrel{\leftarrow }{\varphi }_{out}=r \stackrel{\rightarrow }{\varphi }_{in}+t^{\prime }\stackrel{\leftarrow }{\varphi }_{in}$ and $\stackrel{\rightarrow }{\varphi }_{out}=r^{\prime }\stackrel{\leftarrow }{\varphi }_{in}+t \stackrel{\rightarrow }{\varphi }_{in}$, where $r,$ $t,$ $r^{\prime }$ and $t^{\prime }$ are the reflection and transmission amplitudes
for particles coming from the left and right, respectively. These relations define
the scattering matrix $S$, that connects the incoming amplitudes
with the outgoing ones,
and depends on the scattering amplitudes as follows
\begin{equation}
S=\left(
\begin{array}{cc}
r & t^{\prime } \\
t & r^{\prime }
\end{array}
\right) .
\end{equation}

The scattering matrix is an amply studied mathematical and physical quantity. It is well-known that flux conservation implies that the $S$ matrix is unitary and $S^{\dagger}S=I$, while time reversal invariance implies that $S=S^T$. When time reversal symmetry is present, one has to distinguish
spin-dependent from spin-independent systems. The TRI
requirement for spin-independent systems implies that $t^{\prime }=t^{T}$ while
for spin-dependent and TRI systems, the transmission amplitude should satisfy the condition $t^{\prime }=\Sigma_yt^{T}\Sigma_y^T.$  These global relations (valid independently of the size of the system, the number of unit cells, the number of propagating modes and the potential
profiles), are important and appealing properties of the transfer matrix method and
provide the possibility to establish a bridge between
mathematically well defined objects, as the transfer matrices, and physical quantities.

The relation between the $S$ and $M$ matrix, can easily be obtained based on their  definitions\footnote{It is worth emphasizing that in order to establish the relation between $S$ and $M$, we have to use the same basis of wave functions.}
\begin{equation}
\left(
\begin{array}{c}
\stackrel{\leftarrow }{\varphi }_{out} \\
\stackrel{\rightarrow }{\varphi }_{out}
\end{array}
\right)=S\left(
\begin{array}{c}
\stackrel{\rightarrow }{\varphi }_{in} \\
\stackrel{\leftarrow }{\varphi }_{in}
\end{array}
\right) \hspace{0.3in}{\rm and}\hspace{0.3in}\left(
\begin{array}{c}
\stackrel{\rightarrow }{\varphi }_{out} \\
\stackrel{\leftarrow }{\varphi }_{in}
\end{array}
\right)=M\left(
\begin{array}{c}
\stackrel{\rightarrow }{\varphi }_{in} \\
\stackrel{\leftarrow }{\varphi }_{out}
\end{array}
\right).
\end{equation}

When  the transfer matrix is of the unitary universality class $M_{u}$  (in the symplectic and TRI case, one has  $\gamma =\beta ^{\ast }$ and $\delta
=\alpha ^{\ast }$), one obtains the following equations
\begin{equation}
\begin{array}{c}
t-\alpha -\beta r=0, \\
r^{\prime }-\beta t^{\prime }=0, \\
\gamma +\delta r=0, \\
1-\delta t^{\prime }=0,
\end{array}
\end{equation}
whose solutions, using the relations $r^{\dagger}r+t^{\dagger}t=1$ and $r^{\dagger}t'+t^{\dagger}r'=0$, are \cite{Pereyra2002}
\begin{equation}
\alpha=(t^{\dagger})^{-1}, \hspace{0.2in}\beta=r'(t')^{-1},\hspace{0.2in}\gamma=-(t')^{-1}r,\hspace{0.2in}{\rm and}\hspace{0.2in}\delta=(t')^{-1}
\end{equation}
and
\begin{equation}
t=(\alpha^{\dagger})^{-1}, \hspace{0.2in}r=-(\delta)^{-1}\gamma,\hspace{0.2in}t'=(\delta)^{-1},\hspace{0.2in}{\rm and}\hspace{0.2in}r'=\beta(\delta)^{-1} .
\end{equation}
Thus, the transfer matrix of the unitary universality class can be written
as
\begin{equation}\label{MandSA}
M_{u}=\left(
\begin{array}{cc}
\left( t^{\dagger }\right) ^{-1} & r^{\prime }\left( t^{\prime }\right) ^{-1}
\\
-\left( t^{\prime }\right) ^{-1}r & \left( t^{\prime }\right) ^{-1}
\end{array}
\right)
\end{equation}
while a transfer matrix in the orthogonal universality class, takes the form
\begin{equation}
M_{o}=\left(
\begin{array}{cc}
\left( t^{\dagger }\right) ^{-1} & r^{\ast }\left( t^{T}\right) ^{-1} \\
-\left( t^{T}\right) ^{-1}r & \left( t^{T}\right) ^{-1}
\end{array}
\right) .
\end{equation}
It can also be shown that the scattering amplitudes in terms of
the transfer matrix $W$ blocks are given by the following
relations
\begin{equation}\label{tandW}
t=2\kappa^{1/2}(\vartheta ^{T}+\kappa \chi ^{T}\kappa ^{-1}-i(\kappa \mu ^{T}-\nu ^{T}\kappa
^{-1}))^{-1}\kappa^{-1/2}=(t^{\prime })^{T},
\end{equation}
\begin{equation}\label{randW}
r=\frac{1}{2}t^{\prime }\kappa^{1/2}(\vartheta -\kappa ^{-1}\chi
\kappa +i(\mu \kappa +\kappa ^{-1}\nu ))\kappa^{-1/2},
\end{equation}
and
\begin{equation}\label{rpandW}
r^{\prime }=-\frac{1}{2}\kappa^{1/2}(\vartheta -\kappa ^{-1}\chi
\kappa -i(\mu \kappa +\kappa ^{-1}\nu ))\kappa^{-1/2}t^{\prime}.
\end{equation}

Another important attribute of the transfer matrices that makes them
appropriate quantities to describe systems of finite but in principle
arbitrary size is the multiplicative property. Indeed, if $M(z_2,z_1)$ connects state vectors at $z_1$ and $z_2$, and $M(z_3,z_2)$ connects state vectors at $z_2$ and $z_3$, the transfer matrix that connects state vectors at  $z_1$ and $z_3$ is given by the product
\begin{equation}
M(z_3,z_1)=M(z_3,z_2)M(z_2,z_1).
\end{equation}
This property and the possibility of relating the
matrix with the scattering amplitudes, have been broadly used; they
constitute the principal components of the transfer matrix approach to the
quantum description of finite periodic systems.

It should be noted that the dispersion and transfer matrices contain all the physics of the dispersion processes. This is why a theory based on these quantities is capable of describing physical systems whose geometries allow us not only to define transfer matrices but also to determine, {\it analytically}, new
results for larger systems. This is the goal of the next section. We will establish a general method and derive general formulas that can be applied directly to determine the physical quantities of specific finite periodic systems. Although most systems belong to the orthogonal universality class, we will assume in the derivations reviewed here that all transfer matrices belong to the unitary universality class. All of our results can be easily adjusted for the other universality classes. For example, for the orthogonal universality class we have just to consider $\gamma=\beta^*$ and $\delta=\alpha^*$.

\section{The theory of finite periodic systems, $N$ propagating modes, $n$-unit cells}

The theory of finite periodic systems aims to describe and to determine the physical properties of a layered periodic system  based on the transfer matrix method. \footnote{Most of the content in this section was published in Refs. [\onlinecite{Pereyra1998,Pereyra2002,AnzaldoPereyra2007}]}
The multiplicative property of transfer matrices make them suitable
quantities to describe layered systems. As mentioned before, if we put together two
identical cells of length $L/n$ and the transfer matrix of each unit-cell is $M$, the  transfer matrix $M_{2}$ of the
resulting system, of length $2L/n$, is $M_{2}=MM=M^{\,2}$. It is well established  that knowing the unit-cell transfer matrix, we have all the information about the wave functions in the unit cell. In the same way, knowing the transfer matrix $M_{2}$, we have the whole information of the wave functions of the two unit-cell system, and the possibility of determining other quantities such as eigenvalues or scattering amplitudes, which formal relations with the transfer matrix remain unchanged. Applying the multiplicative property over and over, we can express the
global ($n$-cell) transfer matrix as
\begin{equation}
M_{n}=M^{n}=\left(
\begin{array}{cc}
\alpha & \beta \\
\gamma & \delta
\end{array}
\right) ^{n}\equiv \left(
\begin{array}{cc}
\alpha _{n} & \beta _{n} \\
\gamma _{n} & \delta _{n}
\end{array}
\right).
\end{equation}
The relation with the scattering amplitudes is
\begin{equation}
\left(
\begin{array}{cc}
\alpha _{n} & \beta _{n} \\
\gamma _{n} & \delta _{n}
\end{array}
\right) =\left(
\begin{array}{cc}
\left( t_{n}^{\dagger }\right) ^{-1} & r_{n}^{\prime }\left( t_{n}^{\prime
}\right) ^{-1} \\
-\left( t_{n}^{\prime }\right) ^{-1}r_{n} & \left( t_{n}^{\prime }\right)
^{-1}
\end{array}
\right) .
\end{equation}
An important leap in the transfer matrix method is, precisely, the
possibility of analytically determining the matrices $\alpha_{n},$ $\beta_{n}$, etc., and hence, to deduce analytical expressions for
global $n$-cell physical quantities. It is clear that for the purpose of numerical evaluations it may be
sufficient to diagonalize $M$ as $U\Lambda U^{\dagger }$ and to write the $n$%
-cell transfer matrix as $U\Lambda ^{n}U^{\dagger }$. However, by doing this
one losses a great deal of the power of the transfer matrix method and spoils the possibility of deriving new expressions for
fundamental physical quantities. It is worth mentioning that transfer matrices are frequently used to study structures with few layers as well as in non-periodic structures, such as Fibonacci systems \cite{Macia1996} and self-similar fractal structures. \cite{Chuprikov2000} Our interest here is rather in periodic structures.

Let us now consider some transfer-matrix properties and derive fundamental
relations in this approach. In the following we will be concerned
with $M_u$, but for an easy notation the subindex $u$ will be omitted.

Since
\begin{equation}
M_n=M M_{n-1}
\end{equation}
it is clear that
\begin{equation}\label{Prealp}
\alpha _n=\alpha \ \alpha _{n-1}+\beta \ \gamma _{n-1}
\end{equation}
\begin{equation}\label{Prebet}
\beta _n=\alpha \ \beta _{n-1}+\beta \ \delta _{n-1}
\end{equation}
\begin{equation}\label{Pregam}
\gamma _n=\gamma \ \alpha _{n-1}+\delta \ \gamma _{n-1}
\end{equation}
\begin{equation}\label{Predel}
\delta _n=\gamma \ \beta _{n-1}+\delta \ \delta _{n-1}
\end{equation}
with $\alpha _0=\delta _0=I_{s{\cal N}}$ and $\beta _0=\gamma _0=0.$
Starting from these relations one can easily obtain the {\it matrix
recurrence relation (MRR)}
\begin{equation}\label{PreRRbet}
\beta _n=(\alpha +\beta \delta \beta ^{-1})\ \beta _{n-1}+(\beta \gamma
-\beta \delta \beta ^{-1}\alpha )\ \beta _{n-2}\ \ ,
\end{equation}
and a similar one for $\alpha _n$. We also obtain
\begin{equation}\label{PreRRgam}
\gamma _n=(\delta +\gamma a\gamma ^{-1})\ \gamma _{n-1}+(\gamma \beta
-\gamma a\gamma ^{-1}\delta )\ \gamma _{n-2}\ \ ,
\end{equation}
and a similar one for $\delta _n$. All these relations are three-term
recurrence relations with matrix coefficients of dimension $N\times N$. If
we define the matrix-functions
\begin{equation}
p_{N,n-1}^{(1)}=\beta ^{-1}\beta _n \hspace{0.2in}\rightarrow \hspace{0.2in} \beta_n=\beta p_{N,n-1}^{(1)}
\end{equation}
and
\begin{equation}
p_{N,n-1}^{(2)}=\gamma ^{-1}\gamma _n \hspace{0.2in}\rightarrow \hspace{0.2in} \gamma_n=\gamma p_{N,n-1}^{(2)}
\end{equation}
we can write  equations (\ref{Prebet}) and (\ref{Pregam}) as
\begin{equation}\label{deltan}
\delta_n=p_{N,n}^{(1)}-\beta^{-1}\alpha \beta p_{N,n-1}^{(1)}
\end{equation}
\begin{equation}\label{alphan}
\alpha_n=p_{N,n}^{(2)}-\gamma^{-1}\delta \gamma p_{N,n-1}^{(2)}
\end{equation}
and equations (\ref{PreRRbet}) and (\ref{PreRRgam}), dropping the index $N$ to simplify notation,  become the non-commutative polynomials recurrence relation $(NCPRR)$
\begin{equation}\label{NCPRReq}
p_n^{(i)}+\zeta _i\ p_{n-1}^{(i)}+\eta _i\ p_{n-2}^{(i)}\
=0 \hspace{0.2in}{\rm for}\hspace{0.2in}n\geq 1\hspace{0.2in}
i=1,2.
\end{equation}
Here $\zeta _1=-(\beta ^{-1}\alpha \beta +\delta )$, $\eta _1=(\delta \beta
^{-1}\alpha \beta -\gamma \beta )$, $\zeta _2=-(\gamma ^{-1}\delta \gamma
+\alpha )$ and $\eta _2=(\alpha \gamma ^{-1}\delta \gamma -\beta \gamma )$
are the matrix coefficients. The subindex $N$ has been dropped for
simplicity.  It is easy to see that the initial conditions are $p_{-1}^{(i)}=0$ and $p_0^{(i)}=I_N$. An important achievement of this theory, extremely important to obtain analytical expressions for the physical quantities, has been the solution of equation (\ref{NCPRReq}). The explicit derivation of the matrix polynomials $p_{n}^{(1)}$ and  $p_{n}^{(2)}$, can be seen in Refs [\onlinecite{Pereyra1998,Pereyra2002,AnzaldoPereyra2007}]. The polynomials in terms of the unit-cell transfer matrix eigenvalues $\lambda_i$ are
\begin{equation}
p_{N,m}=\sum\limits_{k=0}^{m}\sum\limits_{l=0}^{k}p_{N,l}g_{k-l}q_{m-k}
\hspace{0.2in}  {\rm for}\hspace{0.2in} m<2N,
\end{equation}
and
\begin{equation}
p_{N,m}=\sum\limits_{k=0}^{2N-1}\sum\limits_{l=0}^{k}p_{N,l}g_{k-l}q_{m-k}
\hspace{0.2in} {\rm for}\hspace{0.2in} m\geq 2N.
\end{equation}
where
\begin{equation}
g_{m}=(-)^{m}\sum\limits_{l_{1}<l_{2}<...<l_{m}}^{2N}\lambda _{l_{1}}\lambda
_{l_{2}}...\lambda _{l_{m}},\qquad \qquad g_{0}=1.
\end{equation}
and
\begin{equation}
q_{n}=\sum_{i=1}^{2N}\frac{\lambda _{i}^{2N+n-1}}{\prod\limits_{j\neq
i}^{2N}\left(\lambda _{i}-\lambda _{j}\right) }I_{N}.
\end{equation}
Using these results, we can now write the most general $n$-cell transfer matrix $M_{n}$ as
\begin{equation}
M_{n}= \left(
\begin{array}{cc}
 p_{N,n}^{(2)}-\gamma^{-1}\delta \gamma p_{N,n-1}^{(2)} & \beta p_{N,n-1}^{(1)} \\
\gamma p_{N,n-1}^{(2)} &  p_{N,n}^{(1)}-\beta^{-1}\alpha \beta p_{N,n-1}^{(1)}
\end{array}
\right)
\end{equation}
By solving the matrix recurrence relation the TFPS extends the capabilities of
describing the transport properties of  multichannel systems.
From the mathematical point of view, the generalized recurrence relations
have special implications which go beyond the purpose of this paper. The matrix representations of the
generalized orthogonal polynomials and the noncommutative algebras, are similar to
those discussed by I. Gelfand \cite{Gelfand}.

We will see below that the NCPRR becomes, in the limit $N$=1, the recurrence relation of the well-known Chebyshev polynomials of the second kind.

\subsection{The scattering amplitudes, transport coefficients, Landauer conductance}\label{ScatAmpConductance}

Given the non-commutative polynomials and using equations (\ref{deltan}) and (\ref{alphan}), together with the relation (\ref{MandSA}) we can write the global multichannel
transmission and reflection amplitudes as
\begin{equation}
t_{n}^{\dagger}=\left( p_{n}-p_{n-1}\,(\gamma ^{-1}\delta \gamma )\right) ^{-1}
\end{equation}
\begin{equation}
t_{n}^{\prime }=\left( p_{n}-(\beta ^{-1}\alpha \beta )\ p_{n-1}\right) ^{-1}
\end{equation}
\begin{equation}
r_{n}=-\left( p_{n}-(\beta ^{-1}\alpha \beta )\ p_{n-1}\right) ^{-1}\gamma \
p_{n-1}
\end{equation}
\begin{equation}
r_{n}^{\prime }=\beta \ p_{n-1}\left( p_{n}-(\beta ^{-1}\alpha \beta
)p_{n-1}\right) ^{-1}.
\end{equation}
These interesting results show that the $n$-cell scattering amplitudes can
be expressed entirely in terms of single-cell transfer-matrix blocks (or
single-cell transmission and reflection amplitudes $r,t,r^{\prime }$ and $%
t^{\prime }$) and the polynomials $p_{n}$. For time reversal invariant and
spin-independent systems, $t_{n}$ is just the transpose of $t_{n}^{\prime }$%
, and $\gamma =\beta ^{\ast }$, $\delta =\alpha ^{\ast }$. For
spin-dependent systems $t^{\prime }=k^{T}t^{T}k$ and $\gamma =k^{T}\beta
^{\ast }k$, $\delta =k^{T}\alpha ^{\ast }k$. The previous relations are
simple and of general validity at the same time.

Especially simple, in its functional appearance, are the global Landauer
multichannel resistance amplitudes $R_{N,n}^{^{\prime }}=r_{N,n}^{\prime
}\left( t_{N,n}\right) ^{-1}$ and $R_{N,n}=-\left( t_{N,n}^{\prime }\right)
^{-1}r_{N,n}$. These quantities, in terms of the polynomials $p_{N,n},$ are
just
\begin{equation}
R_{N,n}^{\prime }=\ R_{N,1}^{\prime }p_{N,n-1}\ \quad {\rm and\quad }%
R_{N,n}=R_{N,1}p_{N,n-1}.
\end{equation}
Here, the most important properties, tunneling and interference phenomena, appear nicely factorized.

A quantity often used in the transport theory is the Landauer multichannel
conductance matrix
\begin{equation}
G_N=\frac{e^2}{h}t_N\frac{1}{r_N^{\dagger }r_N}t_N^{\dagger }
\end{equation}
which for
the $n$ cell system becomes
\begin{equation}\label{ConductanceNn}
G_{N,n}=\frac 1{p_{N,n-1}}\ G_{N,1}\ \left( \frac 1{p_{N,n-1}}\right)
^{\dagger }.
\end{equation}

So far, we have given a number of non-trivial but extremely appealing
relations and results. {\it The }$n$-{\it cell} {\it Landauer resistance amplitude is
just the product of the one-cell Landauer resistance amplitude }$R${\it \
and the polynomial }$p_{n-1}$. The polynomial $p_n$ has the information on
the number of layers $n$, the number of channels $N$ and, more importantly,
on the complex but precise phase interference phenomena that happens along the n-period structures.

\section{The TFPS in the one-propagating mode limit, and $n$-unit cells}

So far we have presented a general approach for quasi-1D, multichannel periodic systems, Since the more common systems are well modeled as one propagating mode systems, we will consider in this section the one-propagating mode limit, and given  the scalar polynomials $p_{n}$, we will deduce general expression for the most common superlattice configurations. For open systems, which are the most known systems, we will review the  resonant energies, eigenfunctions and dispersion relations. For bounded superlattices we will obtain formulas for the evaluation of energy eigenvalues, eigenfunctions and discrete dispersion relations.

\subsection{The polynomial recurrence relation and the transfer matrix for $n$-unit cells in the one-channel limit}

In the one-propagating mode limit, the functions $\zeta $ and $\eta $ defined before become
$\alpha +\delta =TrM$ and $\delta \alpha -\gamma \beta =\det M$, respectively. Thus, for the {\it one-dimensional } systems the matrix NCPRR becomes the scalar commutative  recurrence relation
\begin{equation}\label{NCPRR1D}
p_n^{(i)}+(\alpha +\delta)\ p_{n-1}^{(i)}+ \det M \ p_{n-2}^{(i)}\
=0 \hspace{0.2in}{\rm for}\hspace{0.2in}n\geq 1\hspace{0.2in}
i=1,2.
\end{equation}
which, for the orthogonal universality class, where $\alpha =\delta^*$, reduces to
\begin{equation}\label{NCPRR1Do}
p_n+ 2\alpha_R \; p_{n-1}+  p_{n-2} =0 \hspace{0.2in}{\rm for}\hspace{0.2in}n\geq 1\hspace{0.2in}
\end{equation}
with $\alpha_R=\Re e \alpha$ (i.e., the real part of $\alpha$), and initial conditions  $p_{-1}=0$ and $p_{0}=1$. This is precisely the recurrence relation of the Chebyshev polynomials of the second kind $U_n$ evaluated at $\alpha_R$.

In the one-propagating mode limit, the transfer matrix $M_{n}$ becomes
\begin{equation}\label{TM1Dncells}
M_{n}= \left(
\begin{array}{cc}
 U_{n}-\delta  U_{n-1} & \beta U_{n-1} \\
\gamma U_{n-1} &  U_{n}-\alpha  U_{n-1}
\end{array}
\right)
\end{equation}
which for the orthogonal universality class, I.E., for the time reversal invariant systems becomes the well-known  Jones-Abelès' transfer matrix
\begin{equation}
M_{n}= \left(
\begin{array}{cc}
 U_{n}-\alpha^*  U_{n-1} & \beta U_{n-1} \\
\beta^* U_{n-1} &  U_{n}-\alpha  U_{n-1}
\end{array}
\right).
\end{equation}

This matrix, reported initially for electromagnetic fields through  layered media, has been repeatedly rediscovered\cite{Jones1973, Cvetic1981, Claro1982, Vezzetti, PerezAlvarez,  Kalotas, Griffiths1992, Sprung, Rozman, Macia1996, Pereyra1998} and frequently used to calculate transmission coefficients  through semiconductor superlattices, metallic superlattices, photonic crystals and many other types of periodic systems, even though the theoretical approaches generally also introduce the Floquet theorem which is rigorously valid for $n=\infty$.

Although the one propagating mode approach is the simplest version in the TFPS, it has been frequently applied to calculate transmission coefficients for different types of systems.  When the channel coupling is weak, these matrices may be also useful for a first order approximation. Other systems as the magnetic superlattices are at least two-mode systems.

\subsection{Scattering amplitudes and transport properties in the one-channel limit}

In the particular but very
much used 1-D one channel case, the transmission amplitude
\begin{equation}
t_{n}=\frac{t^{\dagger }}{p_{n}t^{\dagger }-p_{n-1}}
\end{equation}
takes the form
\begin{equation}\label{1DScattAmplit}
t_{n}=\frac{t^{\ast }}{t^{\ast }U_{n}-\ U_{n-1}}.
\end{equation}
This is an extremely simple function of the Chebyshev polynomials of the
second kind, $U_{n}(\alpha _{R})$ and $U_{n-1}(\alpha _{R})$ (evaluated at
the real part of $\alpha )$, and of the single cell transmission amplitude $%
t $. Using the identity $U_{n}U_{n-2}=U_{n-1}^{2}-1,$ or alternatively $|\alpha_n|^2$=1+$|\beta_n|^2$, it is easy to show
that the transmission coefficient $T_{n}=\left| t_{n}\right| ^{2}$can be
written as \cite{Pereyra2005}
\begin{equation}\label{TransmiTn}
T_{n}=\frac{T}{T+U_{n-1}^{2}(1-T)}
\end{equation}
with an evident resonant behavior. Here $T=\left| t\right| ^{2}$ is the
single-cell transmission coefficient. The transmission resonances occur
precisely when the polynomial $U_{n-1}$ becomes zero. Therefore the $\nu $-th resonant energy $E^r_{\mu ,\nu }$ is the solution of
\begin{equation}\label{ResonantDispRel}
(\alpha _{R})_{\nu }=\cos \frac{\nu \pi }{n}
\end{equation}
with $\nu =1,2,3...,n-1.$ The index $\mu $ labels the bands and $\nu $ labels the intraband states. This equation is a dispersion relation for the resonant states, a discrete dispersion relation, at variance with the continuous dispersion relations that result in the hybrid approaches that combine transfer matrices and Bloch functions. We will refer to these approaches below.  It is worth mentioning here that   60 years ago some theoretical approaches studying ordered and mainly disordered one-dimensional systems, obtained similar equations for eigenfrequencies of simple linear chains.\cite{Hori,Matsuda}

In the one-channel case, the $n$-cell Landauer conductance is just
\begin{equation}
G_n=\frac 1{\left( U_{n-1}\right) ^2}\ G.
\end{equation}
The zeros of the polynomial determine both the points of divergence of $G_n$
and the zeros of the resistance $R_n$. They also determine the resonant
energies $E_{\mu ,\nu }$ where the
global-transmission-coefficient $T_n$ is resonant.

\begin{figure}
    \centering
    \includegraphics[width=210pt]{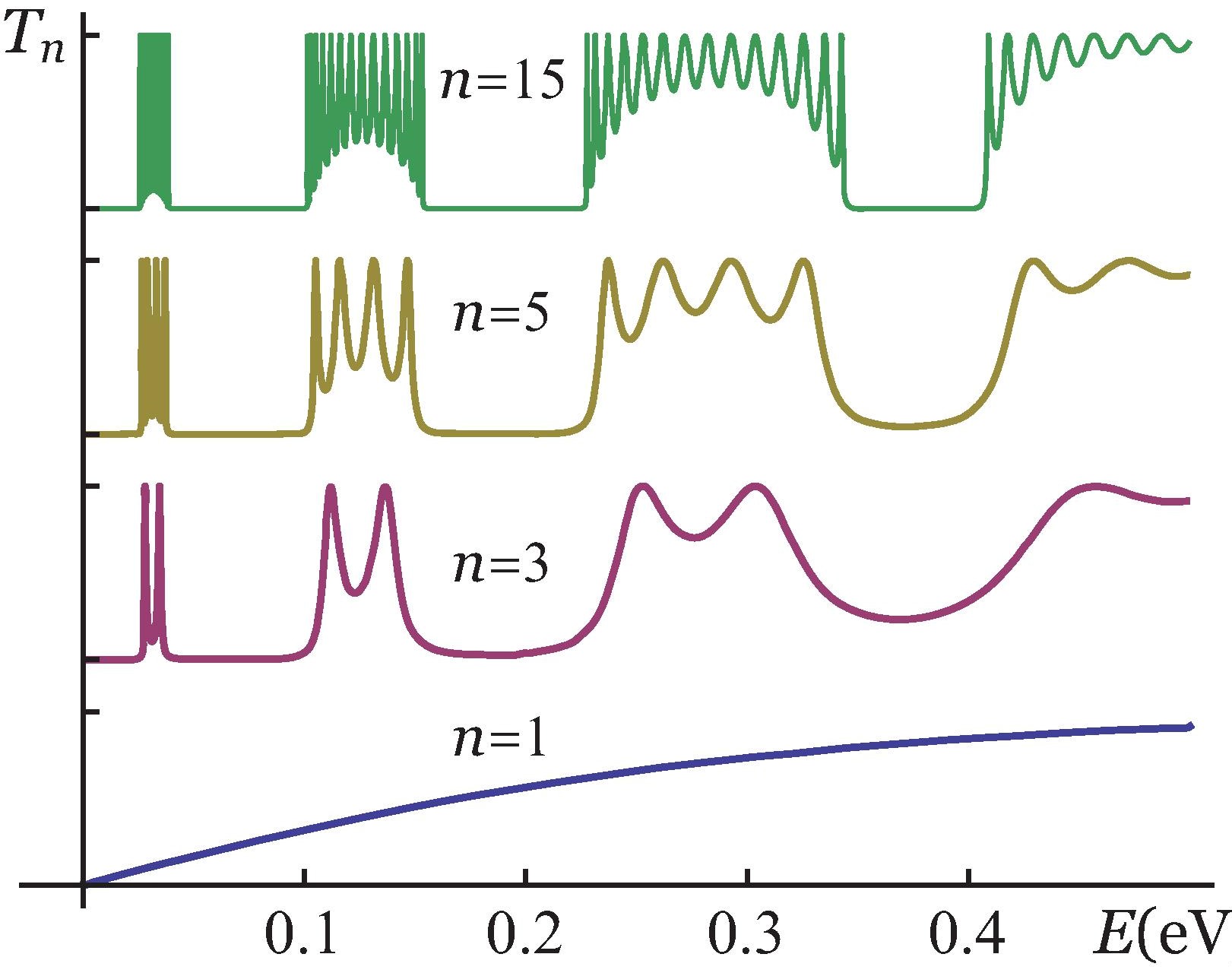}
    \caption{The transmission coefficient $T_n$ of a periodic potential for different number of cells $n$. We plot here for $n=1,3,5$ and $15$,
and rectangular barrier parameters $a=2$nm, $b=10$nm and $V_o=0{\rm{.}}23$
eV. Figure reproduced with permission.\textsuperscript{[\onlinecite{Pereyra2002}]} 2002,  Physical Review B.}\label{transSuperlattice}
\end{figure}

Before we continue with the most recent advances of the TFPS that make possible the calculation of basic physical
quantities, such as the optical transitions, let us consider a Kronig-Penney-like
sequence of square barrier potentials in the conduction band of a $(GaAs/AlGaAs)^{n}$
superlattice. It is easy to calculate, using equation (\ref{TransmiTn}) and fixed unit-cell
parameters,  the transmission coefficients shown in Figure \ref{transSuperlattice}.
The series of graphs of the transmission coefficient $T_{n}$, plotted as a
function of the particle's energy $E$ and of the number of unit cells $n$, show
that by increasing $n$ a band structure builds up gradually. It is evident also that when $n$ is of order 5 the gaps in the band structure begin to look better defined.

\section{The TFPS and the eigenvalues, eigenfunctions and dispersion relations of SLs}\label{EigenvalEigenvec}

\begin{figure}[ht]
\floatbox[{\capbeside\thisfloatsetup{capbesideposition={left,top},capbesidewidth=8cm}}]{figure}[\FBwidth]
{\caption{Open, bounded, and quasi-bounded SLs, and a sketch of the $z$ axis intervals for determining the wave function at any point inside the SLs. In a) the SL is open. In this system one can define scattering amplitudes, as well as the resonant energies and wave functions. The SL in b) is bounded by infinite barriers, while the SL in c) is bounded by finite barriers. In the last two cases, we can obtain the energy eigenvalues, the eigenfunctions and dispersion relations. To determine the wave function at any point inside the SLs, say at point $z$ in the $j+1$ cell (with $j=0,1,2,...,n-1$) we need
the transfer matrix $M(z,z_{o})$, which is the product of $M_j=M(z_j,z_o)$ and $M_p=M(z,z_j)$, as depicted below panel a). Figure reproduced with permission.\textsuperscript{[\onlinecite{Pereyra2005}]} 2005, Annals of Physics.}\label{DistinctSLs}}
{\includegraphics[width=9cm]{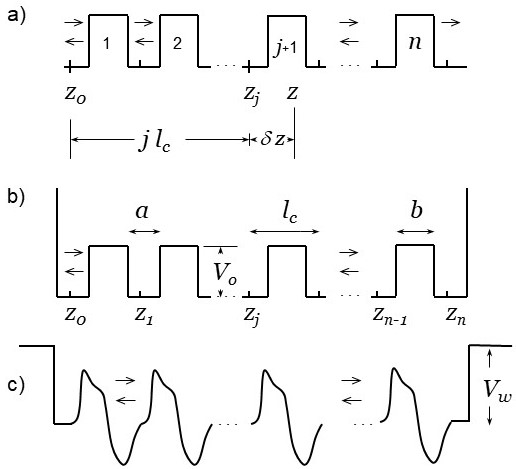}}
\end{figure}

The calculation of eigenvalues and eigenfunctions is one of the most important objectives in solving differential equations of the dynamical systems. The calculation of energy eigenvalues and the corresponding eigenfunctions of bounded quasi-1D periodic systems is also as important as the calculation of transmission coefficients and resonant states in open systems. Many other properties of the periodic systems, such as transition probabilities, and optical response in semiconductor superlattices depend on these quantities. The theory of finite periodic systems (TFPS), originally oriented to calculate scattering amplitudes and resonant energies defined by the zeros of the Chebyshev polynomial $U_{n-1}$ in open superlattices, was expanded and  rigorous and compact analytical expressions for the calculation of energy eigenvalues, $E_{\mu ,\nu }$ and the corresponding eigenfunctions $\Psi _{\mu ,\nu }(z)$ in bounded superlattices were derived. We will present here only the main formulas  for the three possible configurations in which SLs can be found; either as part of an electronic transport system or as a part of a device  where the superlattice is bounded by cladding layers (symmetric or asymmetric) which impose (finite or infinite) lateral barriers, as sketched in figure \ref{DistinctSLs}. All the results obtained in this section are accurate, free of additional assumptions or approximations and they are based on the transfer matrix method as well as on the general formulas derived in the last section.

We will  first consider open superlattices as in Figure \ref{DistinctSLs}a). We will then review the derivation of analytic expressions for eigenvalues and eigenfunction of confined SLs as functions of  the unit-cell transfer matrix elements. In Figures \ref{DistinctSLs} b) and c) we show examples of confined superlattices.  For a detailed derivation of the results presented here, see Ref. [\onlinecite{Pereyra2005}]. We will also see that the eigenfunctions of symmetric SLs posses well defined parity symmetries, and we will derive new selection rules for inter and intra-subband transition probabilities. For detailed analysis of this issue, see Refs. [\onlinecite{Pereyra2017,Pereyra2018}].

\subsection{Resonant energies and resonant functions in open 1D periodic systems}
\begin{figure}
\includegraphics[width=15cm]{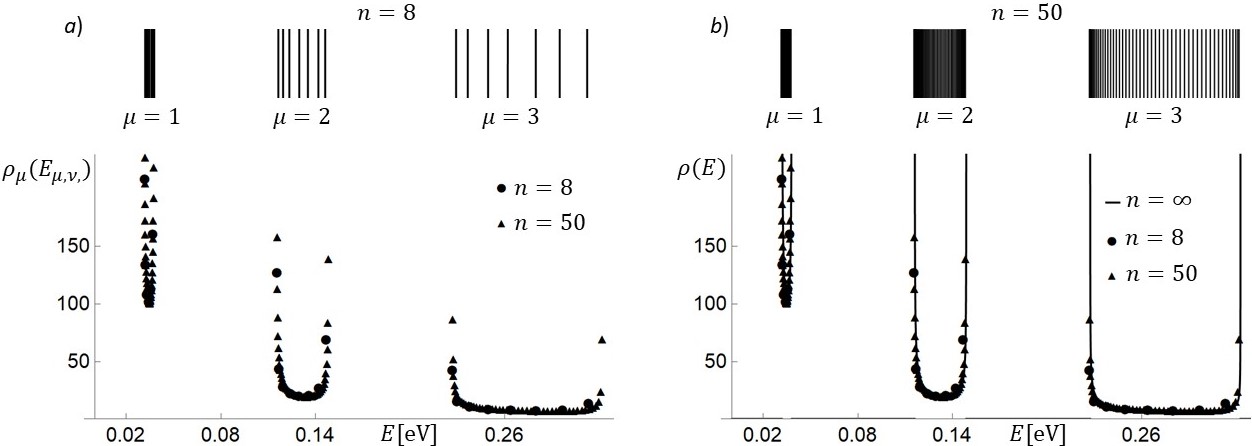}
\caption{$a$) Discret spectra and the level density $\rho_{\mu}(E_{\mu,\nu})$ of resonant states, plotted alongside with the Kronig-Penney level density  $\rho(E)$ valid in the continuous limit ($n=\infty$) $b$). The level densities shown here are for  the first three subbands of $GaAs\left(Al_{0.3}Ga_{0.7}As/GaAs\right)^n$ superlattices, with $a=10 nm$, $b=3 nm$, $V_o=0.23 eV$, and different values of $n$. In the upper panels the resonant energies $E_{\protect\mu,\protect\nu}^r$ obtained  for $n=8$ in $a$) and $n=$50 in $b$) obtained both from (\ref{ResEnerg}) for $\mu=$1, 2 and 3. In the lower panels the level densities $\rho_{\mu}(E_{\mu,\nu})$ for $n=7,70$ (squares and triangles, respectively), and the level density $\rho (E)$ predicted by Kronig-Penney for $n=\infty$.\cite{Kronig1931}  Figure reproduced with permission.\textsuperscript{[\onlinecite{Pereyra2005}]} 2005, Annals of Physics. }\label{ResDispRelRho}
\end{figure}

We will assume that the $n$-cell system is connected to ideal leads. Even though the results that will be obtained here are valid in general, i.e. for any shape of the single cell potential profile, we will, for specific calculations, consider in this section and the coming ones, SLs with piecewise constant potential as shown in figure \ref{quasi1DSL},  known also as the Kronig-Penney model.

In open superlattices, the resonant behavior of the transmission coefficient has been always recognized as a natural feature in these systems, and, at the same time, the continuous energy spectrum and Bloch-functions were assumed as characteristic properties of SLs. In this sections we will present the formulas that allow an exact calculation of the true energies and wave functions for open SLs. Since the characteristics of these results contrast eloquently with the well-established and widely accepted theory, we will present more specific results.

In the previous section, we observed that the resonant transmission occurs when the energy is such that the argument of the Chebyshev polynomial, $\alpha_R$, becomes a zero of the Chebyshev polynomial $U_{n-1}(\alpha_R)$. It is known and was recalled in Ref. [\onlinecite{Pereyra2017}] that, for each value of an integer $\mu=$, 1, 2,..., the number of zeros of the Chebyshev polynomial $U_n$ is $n$,  Thus, the resonant energies  for a periodic system with $n$ unit cells are solutions of
\begin{equation}\label{ResEnerg}
(\alpha_{R})_{\mu,\nu }=\cos \frac{\nu +(\mu-1) n }{n}\pi
\end{equation}
and are characterized by the quantum number $\mu$ that labels the subbands (or cycles in the unit circle) and by the quantum number $\nu$, that labels the intrasubband resonant energies.

Solving this equation we have  the  whole set of resonant energies ${E_{\mu ,\nu }^r}$, with $\nu =1,2,...n-1$ and $\mu=$ 1, 2, 3,... The function $(\alpha_{R})_{\mu,\nu }$ represents the $\nu $-th zero of the $\mu$-th subband. The number of
resonant states per subband equals the number of confining wells in the periodic system.

With the resonant energies that can be easily obtained from this relation, we can determine the density of resonant levels for any number of unit cells $n$ from
\begin{equation}
\rho_{\mu}(E_{\mu,\nu})=\frac{1/n}{E_{\mu,\nu+1}-E_{\mu,\nu}}\hspace{0.3in}{\rm for}\hspace{0.3in}\mu=1,2,3,...\hspace{0.3in}{\rm and}\hspace{0.3in}\nu=1, 2,...,n-1.
\end{equation}
In the continuous limit, the level density becomes
\begin{equation}
\rho(E)=\frac{1}{\pi}\frac{d}{dE}\cos^{-1}\left[\alpha_{R})\right]
\end{equation}
which corresponds to the level density of Kronig and Penney\cite{Kronig1931}.

In figure \ref{ResDispRelRho} a) and b), we show the resonant energies and the level density of the resonant states $\rho_{\mu}(E_{\mu,\nu})$ and $\rho(E)$ for the $GaAs\left(
Al_{0.3}Ga_{0.7}As/GaAs\right)^{n}$ superlattice, modeled as a sequence of sectionally constant potentials, with $GaAs$ layer widths of $10nm $ in the wells,  and $Al_{0.3}Ga_{0.7}As$ layer widths of $3nm$ in the barrier, which  height  is taken as $V_{o}=0.23eV$.  For this system, the explicit form of equation (\ref{ResEnerg}) is
\begin{equation}
\cos k_{\nu }a\cosh q_{\nu }b-\frac{k_{\nu }^{2}-q_{\nu
}^{2}}{2k_{\nu }q_{\nu }}\sin k_{\nu }a\sinh q_{\nu }b=\cos \frac{\nu +(\mu-1) n }{n}\pi \label{e.5}
\end{equation}
where $k_{\nu }^{2}=2m_{v}^{\ast }E_{\mu ,\nu }^r/\hbar ^{2}$ and $q_{\nu }^{2}=2m_{b}^{\ast
}(V_{o}-E_{\mu ,\nu }^r)/\hbar ^{2}$.
In the upper panel of figure \ref{ResDispRelRho}, the energy spectrum the resonant
energies inside the first three conduction subbands are shown, for $n=14$. In the lower panel, the subband level densities, for $n=7$ and $n=70$  (squares and triangles, respectively). As expected, the continuous level density $\rho (E)$ predicted by the Kronig-Penney is reached when the number of cells $n$ $\rightarrow \infty $.

The resonant functions are also in clear contrast with the amply assumed Bloch type functions. Based on the transfer matrix definition, the state vector at any point $z$ of the SL, say inside the $j+1$ cell, is determined  by
and  obtained from
\begin{equation}
\phi (z)=M_{p}M_{j}\left(
\begin{array}{c}
\stackrel{\rightarrow }{\varphi }(z_{o}) \\
\stackrel{\leftarrow}{\varphi }(z_{o})
\end{array}
\right) =M_{j}M_{p}\left(
\begin{array}{c}
\stackrel{\rightarrow }{\varphi }(z_{o}) \\
\stackrel{\leftarrow}{\varphi }(z_{o})
\end{array}
\right). \label{e.8}
\end{equation}
Here $\stackrel{\rightarrow}{\varphi }(z_{o})$ and $\stackrel{\leftarrow}{\varphi }
(z_{o}) $ are the right and left moving wave functions at $z_{o}$,  $M_j=M(z_j,z_o)$ is the transfer matrix for $j$ full cells, and  $M_p$  the transfer matrix $M(z,z_j)$, as shown in the figure \ref{DistinctSLs}.
Assuming that the incidence is only from the left side, we have
\[
\stackrel{\leftarrow}{\varphi }(z_{o})=-\frac{\beta _{n}^{\ast }}{\alpha _{n}^{\ast
}}\stackrel{\rightarrow }{\varphi }(z_{o})=r_n \stackrel{\rightarrow }{\varphi }(z_{o}).
\]
Here $\alpha _{n}$ and $\beta _{n}$ are the transfer matrix elements of the whole $n$-cell system, and
$r_n$ the total reflection amplitude. Thus, the wave function at $z$, for any value of the energy $E$, is given by
\begin{eqnarray}
\Psi (z,E) &=&\stackrel{\rightarrow }{\varphi }(z_{o})[(\alpha _{j}-\beta _{j}\frac{%
\beta _{n}^{\ast }}{\alpha _{n}^{\ast }})(\alpha _{p}+\gamma _{p})
+(\beta _{j}^{\ast }-\alpha _{j}^{\ast }\frac{\beta _{n}^{\ast
}}{\alpha _{n}^{\ast }})(\beta _{p}+\delta _{p})].  \label{e.9}
\end{eqnarray}

By evaluating this function for $E=E^r_{\mu ,\nu }$, we
get the desired $\nu $-th resonant wave function in the $\mu $-th
subband, {\it I.E.},
\begin{equation}
\Psi^r_{\mu ,\nu }(z)=\Psi (z,E^r_{\mu ,\nu }),  \label{e.10}
\end{equation}

\begin{figure}
\includegraphics[width=9cm]{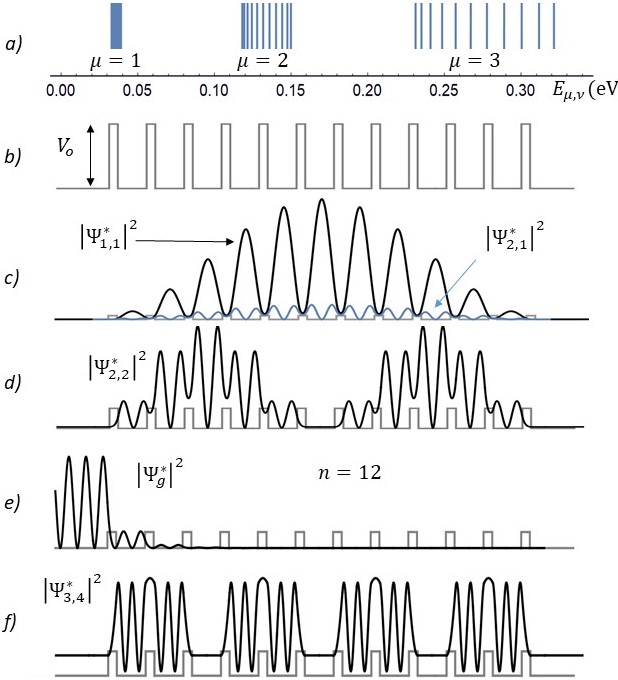}
\caption{Resonant wave functions at different points of the energy spectrum of the open superlattice $GaAs\left(Al_{0.3}Ga_{0.7}As/GaAs\right)^n$. All squared
wave-function amplitudes are plotted using arbitrary units. To get an idea of the relation between the eigenfunction amplitude and the energy, we plot in $c$), using the same scale, the resonant functions $\left| \Psi_{1,1}^r(z)\right|^{2}$ and $\left|
\Psi_{2,1}^r (z)\right| ^{2}$ with different subband indices. In $d)$ and $f)$ we show the resonant functions $\left|\Psi_{2,2}^r(z)\right|^{2}$ and $\left|
\Psi_{3,4}^r (z)\right| ^{2}$, at $E_{2,2}^r=0.11761667862 eV$ and $E_{3,4}^r= 0.24557249944 eV$ , respectively. In $e)$ we have a wave function {\it {in}} a gap, stationary in the left hand side and exponentially decreasing inside the superlattice. Figure reproduced with permission.\textsuperscript{[\onlinecite{Pereyra2005}]} 2005, Annals of Physics. }\label{ResWaveFunc}
\end{figure}

In 1D periodic systems, the resonant wave function $\Psi^{(*)}_{\mu ,\nu }(z)$ is a simple but not trivial combination of Chebyshev polynomials. It is easy to verify that equation (\ref{e.9})
implies
\begin{eqnarray}
\Psi (z_{n},E) =\frac{ 1}{\alpha _{n}^{\ast }}
\stackrel{\rightarrow }{\varphi }(z_{o}) = t_{n} \stackrel{\rightarrow}{\varphi
}(z_{o})=\stackrel{\rightarrow }{\varphi }(z_{n}) \nonumber.
\end{eqnarray}
\begin{eqnarray}
\Psi (z_{o},E) &=&(1-\frac{ \beta _{n}^{\ast }}{\alpha _{n}^{\ast }})
\stackrel{\rightarrow }{\varphi }(z_{o}) = (1+r_{n})
\stackrel{\rightarrow }{\varphi }(z_{o}) \nonumber \\
&=&\stackrel{\rightarrow }{\varphi }(z_{o})+\stackrel{\leftarrow}{\varphi }(z_{o})
\nonumber.
\end{eqnarray}
Here $t_n$ and $r_n$ are the $n$-cell transmission and reflection amplitudes, respectively.

To make even more compelling the difference with the standard approach, we show in Figure \ref{ResWaveFunc} resonant wave functions and a function in the gap. It is clear that, at variance with the Bloch functions, the resonant functions are not periodic. Furthermore, the resonant states are extended wave
functions with particle density different from zero throughout and at
the ends of the system. This will not be the case, of course, for
bounded systems.

\subsection{Eigenvalues and eigenfunctions in bounded 1D periodic systems}

     \begin{figure}
        \CommonHeightRow{%
            \begin{floatrow}[2]%
                \ffigbox[\FBwidth]
                {\includegraphics[width=8.2cm]{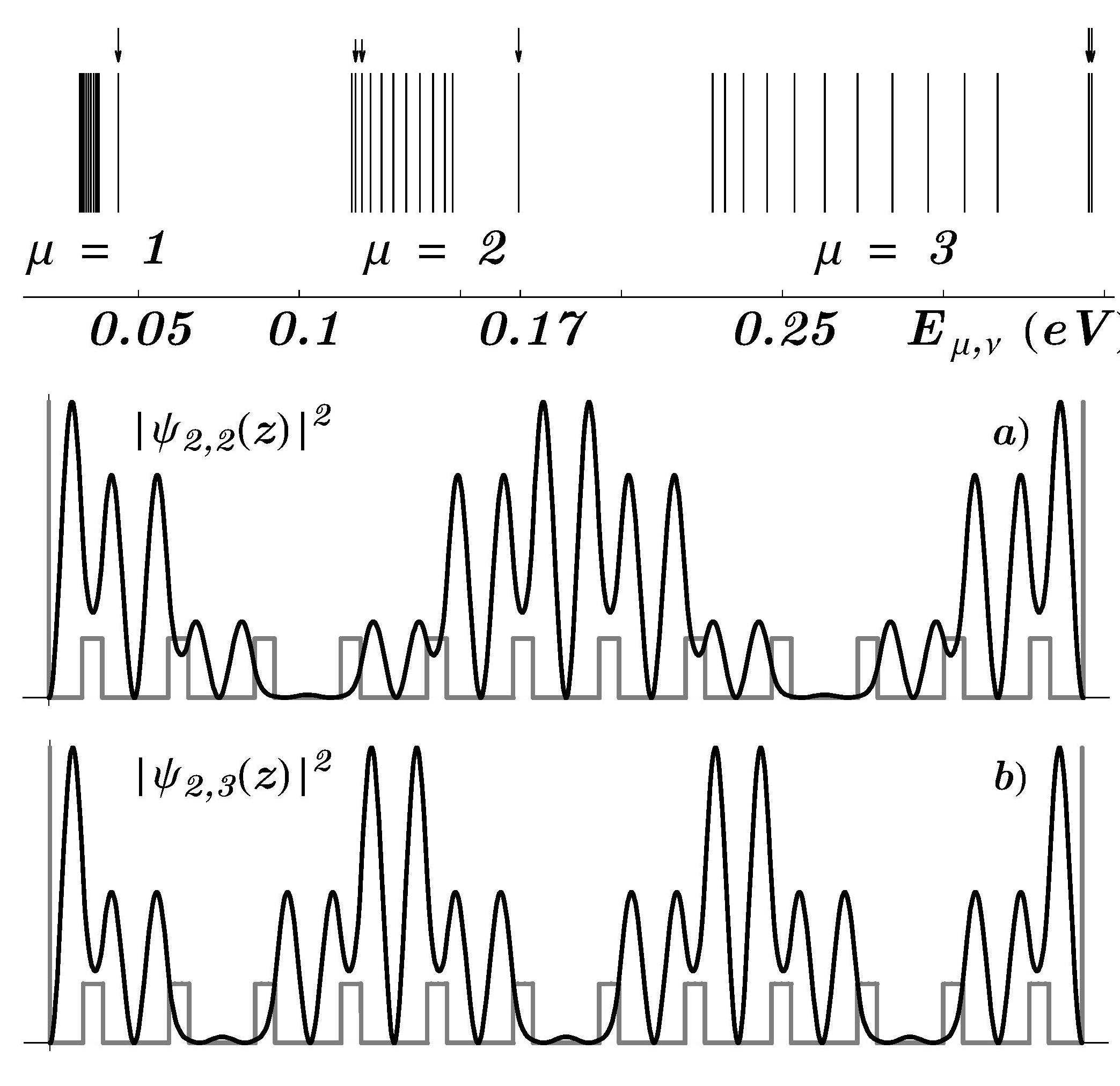}}
                {\caption{ Energy eigenvalues and eigenfunctions $\left|
\Psi_{\mu,\nu}(z)\right| ^{2} $ for a SL of length
$L$=$nl_{c}$ bounded by infinite hard walls. The potential parameters are the same as in figure \ref{ResWaveFunc}. The larger arrows, around $0.1681eV$ and $0.364eV$, indicate
the quasi-degenerate surface energy levels pushed up by the hard
walls. The small arrows indicate the energy eigenvalues $E_{2,2}$
and $E_{2,3}$, whose eigenfunctions are plotted in a) and b). It
is interesting to compare these functions with the corresponding
ones in figure \ref{EigWFBb}, where the system length is $L$=$nl_{c}$+$a$. Figure reproduced with permission.\textsuperscript{[\onlinecite{Pereyra2005}]} 2005, Annals of Physics.}\label{EigWFBa}}
                \ffigbox[\FBwidth]
                {\includegraphics[width=8.2cm]{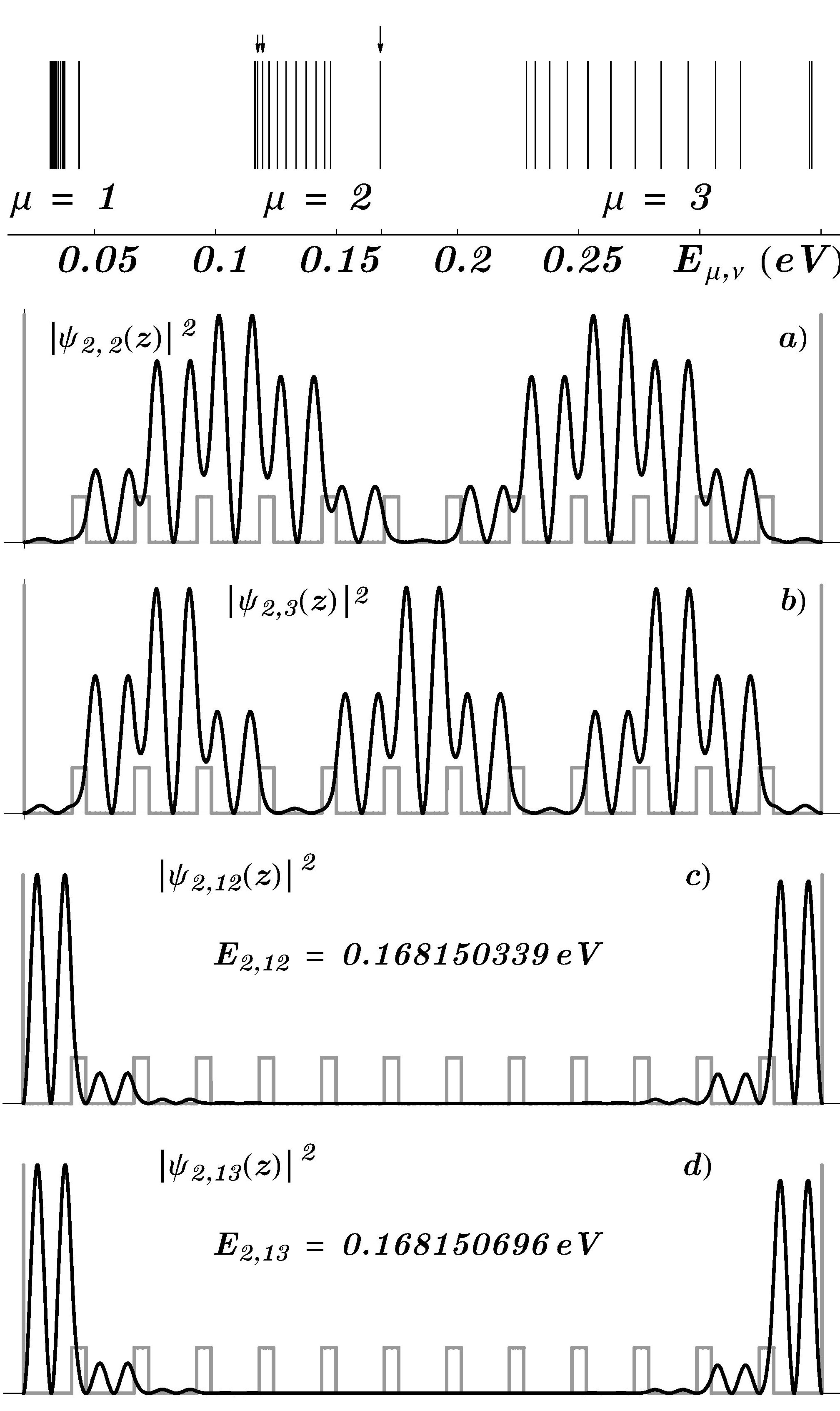}}
                {\caption{Energy eigenvalues and eigenfunctions $\left|
\Psi_{\protect\mu, \protect\nu}(z)\right| ^{2}$ for a system of
length $L$=$nl_{c}$+$a$ bounded by infinite hard walls. The small arrows indicate the energy eigenvalues $E_{2,2}$
and $E_{2,3}$, whose eigenfunctions are plotted in a) and b). The
larger arrows indicate the energy levels pushed up by the {\it
surface repulsion effect}. The corresponding eigenfunctions of
these levels $\left| \Psi_{2,12}
(z)\right| ^{2}$ and $\left| \Psi_{2,13}
(z)\right| ^{2}$, concentrate the particles near the surface. {\it None of these functions is periodic in the Bloch
sense} $\left| \Psi_{\mu,\nu }(z+l_{c})\right| ^{2}=\left|
\Psi_{\mu ,\nu }(z)\right| ^{2}$.  Figure reproduced with permission.\textsuperscript{[\onlinecite{Pereyra2005}]} 2005, Annals of Physics.}\label{EigWFBb}}
            \end{floatrow}}%
    \end{figure}

An important extension in the theory of finite periodic system approach has been accomplished
when  general expressions for the evaluation of
eigenvalues and eigenfunctions were obtained,
independent of the specific single-cell potential parameters, and the number of unit cells. If the superlattice is bounded by infinite height barriers we will consider two cases. Superlattices of length $L$=$nl_c$ and of length $L$=$nl_c$+$a$, with $l_c$ the unit-cell length and $a$ the well width. When the separation of the hard walls is exactly $nl_c$, the boundary conditions on the functions
\begin{equation}
\psi (z_{o})=\sum_{i=1}^{N}\left( \overrightarrow{\varphi }_{i}(z_{o})+%
\overleftarrow{\varphi }_{i}\left( z_{o}\right) \right)
\end{equation}
and
\begin{equation}
\psi (z_{n})=\sum_{i,j=1}^{N}[(\alpha _{n}+\beta _{n}^{\ast })_{i,j}%
\overrightarrow{\varphi }_{j}(z_{o})+(\beta _{n}+\alpha _{n}^{\ast })_{i,j}%
\overleftarrow{\varphi }_{j}\left( z_{o}\right)] ,
\end{equation}
at the ends of the SL, lead to the eigenvalue equation
\begin{equation}
\alpha _{n}-\alpha _{n}^{\ast }+\beta _{n}^{\ast }-\beta _{n}=0.
\label{EigenValB}
\end{equation}
Using the relations $\alpha _{n}=U_{n}-\alpha
^{\ast }U_{n-1}$ and $\beta _{n}=\beta U_{n-1}$, derived before, gives us
\begin{equation}
U_{n-1}(\alpha _{I}-\beta _{I})=0. \label{EigenValBr}
\end{equation}
Here the subscript $I$ refers to the imaginary part. It is clear from this formula that there are $n-1$ of the energy eigenvalues that come from the zeros
of the Chebyshev polynomial $U_{n-1}$, and two other eigenvalues come
from the factor $(\alpha _{I}-\beta _{I})$. This is not a trivial
result; it is remarkable because they correspond to the well-known Tamm
and Shockley\cite{Tamm,Shockley} localized surface states. The hard walls push upwards two of the $n+1$
energy levels of each subband as can be seen in the upper panel of Figure \ref{EigWFBb}.

When the length of the system is $nl_{c}+a$, which can be achieved by adding  two
layers of thickness $a/2$ at the ends of the $n$-cells superlattice, the eigenvalue equation changes slightly into
\begin{equation} \label{EigenValBnp1}
(\alpha _{n}e^{ika}-\alpha _{n}^{\ast }e^{-ika})+\beta _{n}^{\ast
}-\beta _{n}=0,
\end{equation}
assuming that the potential in the additional half layers is constant. In terms of the Chebyshev polynomials the eigenvalue equation is
\begin{equation}
U_{n}\sin ka+(\alpha _{I}\cos ka-\alpha_{R}\sin ka-\beta
_{I})U_{n-1}=0. \label{EigenValBnp1r}
\end{equation}
As for the open systems, the transfer matrix properties and the
boundary conditions lead, for the SL of length $L=nlc$, to the  wave function
\begin{eqnarray}
\Psi ^{b}(z,E) &=&A\left[(\alpha _{p}+\gamma _{p})\left( \alpha _{j}-\beta _{j}%
\frac{\alpha _{n}+\beta _{n}^{\ast }}{\alpha _{n}^{\ast }+\beta
_{n}}\right)
+
(\beta _{p}+\delta _{p})\left( \beta _{j}^{\ast }-\alpha
_{j}\frac{\alpha _{n}+\beta _{n}^{\ast }}{\alpha _{n}^{\ast }+\beta
_{n}}\right)\right].
\end{eqnarray}
Here $A$ is a normalization constant. Evaluating this function at
$E=E_{\mu ,\nu }$, we obtain the corresponding eigenfunction
\begin{eqnarray}
\Psi _{\mu ,\nu }^{b}(z)=\Psi ^{b}(z,E_{\mu ,\nu }).
\end{eqnarray}
This is a rigorous solution of the Schr\"{o}dinger equation for 1D finite periodic systems, bounded by infinite hard walls. In the case of superlattice with length $L=nl_{c}+a$, the wave function gets, if the potential in the additional half layers is constant, an overall factor $e^{ika/2}$, and the term $(\alpha _{n}+\beta_{n}^{\ast })/(\beta _{n}+\alpha_{n}^{\ast })$ is replaced by $(\alpha _{n}+\beta _{n}^{\ast}e^{-ika})/(\beta _{n}+\alpha _{n}^{\ast }e^{-ika})$.

To plot specific eigenvalues and eigenfunctions, we use parameters of the superlattice $
GaAs(Al_{0.3}Ga_{0.7}As/GaAs)^{12}$, bounded by hard walls. In figures \ref{EigWFBa} and \ref{EigWFBb}, we show (in the upper panels) the discrete spectra, of the bounded SLs with lengths $L=nl_c$ and $L=nl_c+a$, respectively. In both figures we plot the
eigenfunctions  $\left| \Psi _{2,2}^{b}(z)\right| ^{2}$, $\left| \Psi
_{2,3}^{b}(z)\right| ^{2}$. The lower panels in figure \ref{EigWFBb} are the surface
functions $\left| \Psi _{2,12}^{b}(z)\right| ^{2}$, $\left| \Psi
_{2,13}^{b}(z)\right| ^{2}$, these functions describe localized particles at the ends of the SL and correspond to the energy levels pushed
upwards by the hard walls. Notice that because of the overall phase, the envelopes of the
eigenfunctions of the SL with length $L=nl_{c}+a$ have a $sine$-like shape, while the envelopes of the eigenfunctions of the SL of length $L=nl_{c}$ are $cosine$-like.

\subsection{Eigenvalues and eigenfunctions for SLs bounded by cladding layers}\label{boundclad}

The superlattices bounded by symmetric or asymmetric cladding layers represent an important class of MQW structures, widely used in optical devices. Assuming that $E<V_o,V_{lb}$, $V_{rb}$, where $V_o$ is the barrier height, $V_{lb}$ and $V_{rb}$ the left and right cladding layer barrier heights, general formulas for the calculation of the energy eigenvalues and their corresponding eigenfunctions have been\cite{Pereyra2005} obtained. For a symmetric SL  of length $L=nl_c$, IE., of exactly $n$-unit cells, the eigenvalue equation is
\begin{equation}\label{EnEigQBSex}
h_{w}U_{n}+f_{w}U_{n-1}=0,
\end{equation}
where
\begin{equation}
h_{w}=1,
\hspace{0.2in}
{\rm and}
\hspace{0.2in}
f_{w}= \alpha _{I}\frac{q_{w}^{2}-k^{2}}{2q_{w}k}-\alpha _{R}-\beta
_{I}\frac{q_{w}^{2}+k^{2}}{2q_{w}k}.
\end{equation}
When $V_{lb}=V_{rb}\equiv V_w $, $q_{w}^{2}=2m(V_{w}-E)/\hbar ^{2}$ and $k$ is the wave vector at
$z_{o}^{+}$, and $z_{n}^{-}$, while $\alpha _{R}$ and $\alpha _{I}$ are the
real and imaginary parts of the single-cell transfer-matrix element
$\alpha $.

\begin{figure}
\begin{center}
\includegraphics[width=9cm]{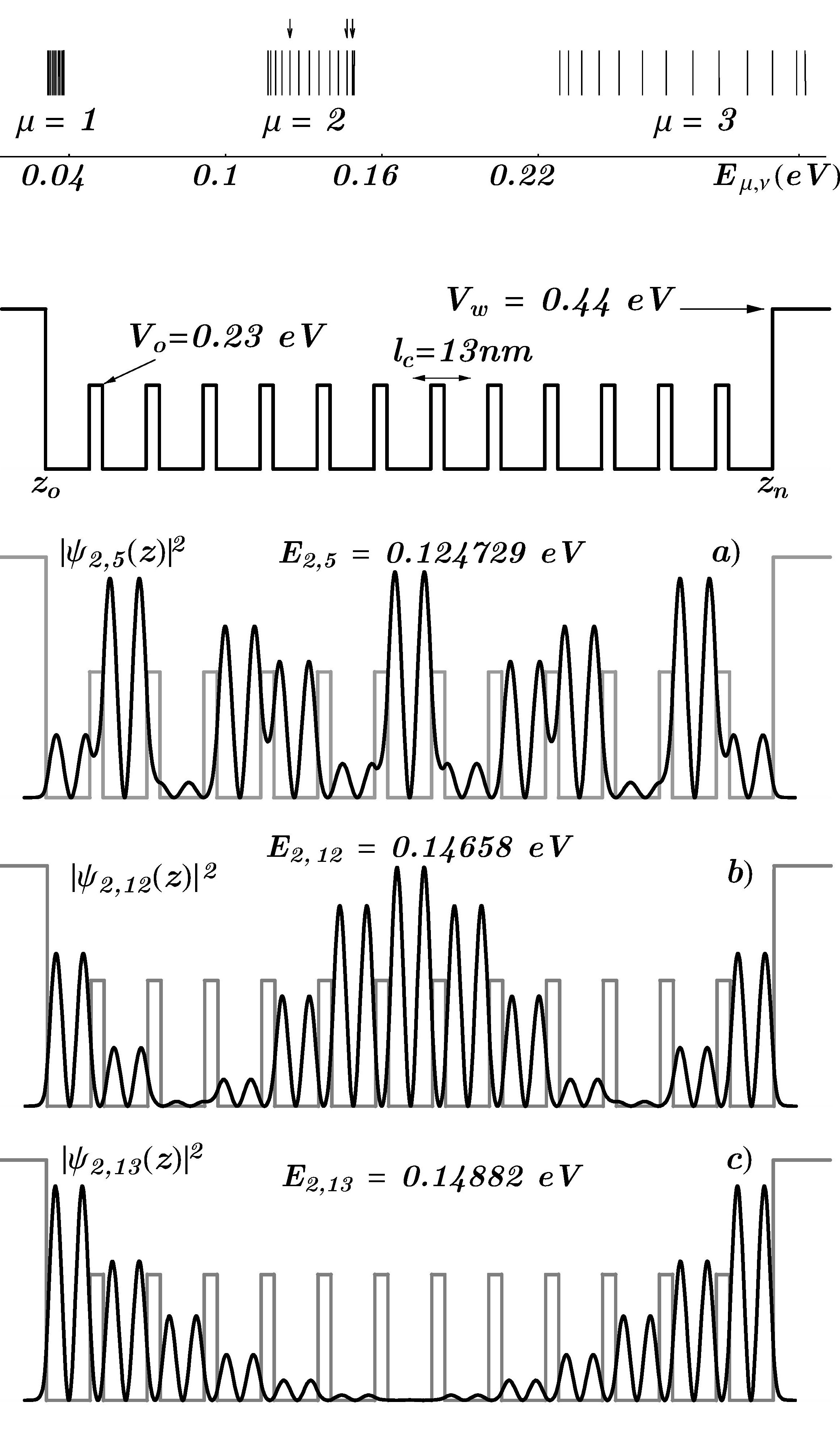}
\end{center}
\caption{Quasi-bounded eigenfunctions $\left|
\Psi_{\protect\mu, \protect\nu}(z)\right| ^{2} $ in the {\it
second} subband of a $GaAs\left(Al_{0.3}Ga_{0.7}As/GaAs\right)^n$
superlattice, with $AlAs$ cladding layers and $n=12$. These
functions are rather similar to the corresponding functions in
figures 5 and 7. For the confining potential height $V_w$ equal to
$0.44eV,$ indicated in the superlattice sketch, the repulsion
effect is weak and the effect slightly visible. Figure reproduced with permission.\textsuperscript{[\onlinecite{Pereyra2005}]} 2005, Annals of Physics.}\label{EigQBS}
\end{figure}

Again, as in the previous section, a slightly more realistic and symmetric structure is a SL of length $L$=$nl_c$+$a$. The eigenvalue equation for this system is again equation (\ref{EnEigQBSex}) but the functions $h_{w}$ and $f_{w}$, change. For the system shown in figure \ref{EigQBS} these functions are
\begin{equation}
h_{w}=\frac{q_{w}^{2}-k^{2}}{2q_{w}k}\sin ka+\cos ka.
\end{equation}
and
\begin{eqnarray}
f_{w} &=&\frac{q_{w}^{2}-k^{2}}{2q_{w}k}(\alpha _{I}\cos ka-\alpha
_{R}\sin ka)
-\alpha _{R}\cos ka-\alpha _{I}\sin ka-\beta _{I}\frac{q_{w}^{2}+k^{2}%
}{2q_{w}k}.
\end{eqnarray}
Using the transfer matrices introduced in previous sections, it is easy
to show that the wave function at any point $z$ in cell $j+1$ is given by
\begin{eqnarray}
\Psi^{qb}(z,E) &=&\frac{A}{g_{n}}\{\left[(\alpha_{p}+\gamma
_{p})\alpha
_{j}+(\beta_{p}+\delta_{p})\beta_{j}^{\ast}\right]\left(1-i\frac{q_{w}
}{k}\right)+\left[(\alpha
_{p}+\gamma_{p})\beta_{j}+(\beta_{p}+\delta _{p})\alpha _{j}^{\ast
}\right]\left(1+i\frac{q_{w}}{k}\right)\} .
\end{eqnarray}
Here $A$ is a normalization constant and
\begin{equation}
g_{n}=\alpha _{ni}\frac{q_{w}^{2}+k^{2}}{2q_{w}k}-\beta
_{ni}\frac{q_{w}^{2}-k^{2}}{2q_{w}k}-\beta _{nr}
\end{equation}
with $\alpha _{ni}=(\alpha_{n}-\alpha_{n}^{\ast})/2$, $\beta _{nr}=(\beta
_{n}+\beta_{n}^{\ast })/2$ and $\beta _{ni}=(\beta _{n}-\beta_{n}^{\ast
})/2$. Again, evaluating the wave function at $E=E_{\mu ,\nu }$, we
obtain the corresponding eigenfunction
\begin{equation}
\Psi _{\mu ,\nu }^{qb}(z)=\Psi ^{qb}(z,E_{\mu ,\nu }).
\end{equation}
With this formula we complete the set of rigorous solutions of the
Schr\"{o}dinger (and analogously Maxwell equations) for 1D finite periodic systems
with different boundary conditions. In figure \ref{EigQBS}, we plot the
eigenfunctions $\Psi _{2,5}^{qb}(z)$, $ \Psi _{2,12}^{qb}(z)$ and
$\Psi _{2,13}^{qb}(z)$ that should be compared with those in figures
\ref{EigWFBa} and \ref{EigWFBb}. The eigenfunction $\Psi _{2,5}^{qb}(z)$
looks rather similar to $\Psi _{2,5}^{b}(z)$ in \ref{EigWFBb}c). In
\ref{EigQBS}b-c) the surface functions start to build. Although
imperceptibly, the wave functions decrease exponentially inside the
potential walls. As in the previous figures, two main characteristics
can be distinguished: i) a remarkable symmetry with respect to the
center of the superlattice, and ii) rapid oscillations modulated by
envelope functions, symmetric with respect to the middle of the
subband. In this figure we plotted three eigenfunctions in the second subband, for the energies indicated in the graphs.

\subsubsection{Superlattices bounded by asymmetric cladding layers}\label{boundasym}

When the cladding layers that bound a SL are asymmetric, the eigenvalue equation remains the same, i.e.,
\begin{equation}
h_{w}U_{n}+f_{w}U_{n-1}=0 \nonumber
\end{equation}
but now the functions $h_w$ and $f_w$ modify a bit and become
\begin{equation}
h_{w}=\frac{q_{lw}q_{rw}-k^{2}}{(q_{lw}+q_{rw})k}\sin ka+\cos ka.
\end{equation}
and
\begin{eqnarray}
f_{w} &=&\frac{q_{lw}q_{rw}-k^{2}}{(q_{lw}+q_{rw})k}(\alpha _{I}\cos ka-\alpha
_{R}\sin ka)
-\alpha _{R}\cos ka-\alpha _{I}\sin ka-\beta _{I}\frac{q_{lw}q_{rw}+k^{2}%
}{(q_{lw}+q_{rw})k}
\end{eqnarray}
with  $q_{lw}^{2}=2m(V_{lb}-E)/\hbar ^{2}$ and $q_{rw}^{2}=2m(V_{rb}-E)/\hbar ^{2}$ the wave numbers in the left and right barriers. As a consequence of this asymmetry the quasi-degeneracy of the surface energy levels is lifted, and the energy levels split. The splitting grows as the asymmetry increases.

The universal formulas reported here, written in terms of Chebyshev polynomials $U_n$ and the single-cell transfer matrix elements, allow US to solve
completely the fine structure in the bands, and can easily be applied to calculate intraband states, photo-transitions \cite{KunoldPereyra,Pereyra2018}, and other properties of finite periodic systems described either by the electromagnetic or the quantum theories. All the
expressions   are
valid for \textit{any profile} of the single-cell potential and arbitrary
number $n$ of  unit cells. In the limit of $n\rightarrow \infty $, these
formulas reproduce the well known results of current theories.

At the time when these resonant energies and the energy eigenvalues were first obtained it was not yet clear that high-resolution optical response measurements revealed the intra-subband energy levels. In Section 8.2, we will present an example that can be fully explained only with the results obtained in this section.

\subsection{Parity symmetries of the SL eigenfunctions and the transition selection rules}

The parity of the resonant functions and particularly of the eigenfunctions is an important property that was analyzed in Ref. [\onlinecite{Pereyra2017}]. We shall now outline the parity symmetries of the three cases considered in the last section. Since the eigenfunctions depend on the Chebyshev polynomials, their symmetry properties are closely related to the Chebyshev polynomials' symmetries. It is worth recalling that all the Chebyshev polynomials, which enter in the physical expressions derived here for 1D superlattices, are evaluated at $\alpha_R$, the real part of $\alpha$.

\subsubsection{The parity symmetries of the resonant wave functions}
\begin{figure}
\begin{center}
\includegraphics[width=13cm]{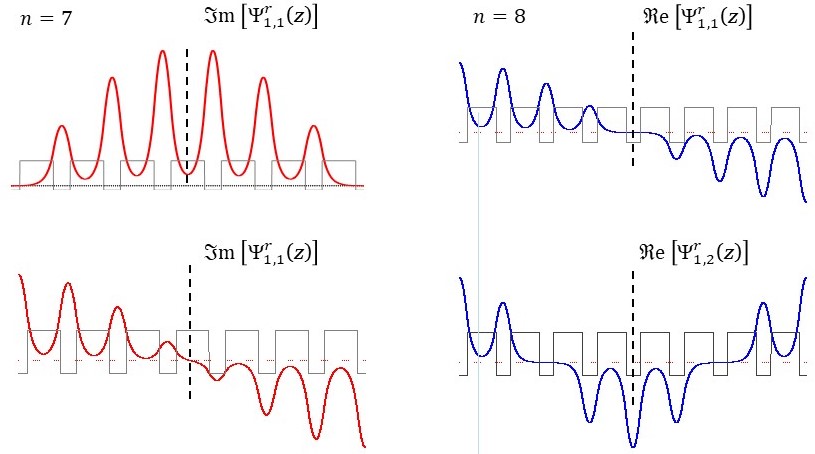}
\end{center}
\caption{Imaginary and real parts of the resonant wave functions of a superlattice with $n$=7 (left column) and $n$=8 (right column). In accord with the parity symmetry relations (\ref{WFSimRelImP}) and (\ref{WFSimRelReP}), in the left column (for $n$=7), $\Im{\rm m}\Psi^r_{1,1}(z)$ is symmetric and $\Im{\rm m}\Psi^r_{1,2}(z)$ is antisymmetric, and in the right column (for $n$=8), $\Re{\rm e}\Psi^r_{1,1}(z)$ is antisymmetric and $\Re{\rm e}\Psi^r_{1,2}(z)$ is symmetric. Figure reproduced with permission.\textsuperscript{[\onlinecite{Pereyra2017}]} 2017, Annals of Physics.}
\end{figure}

While the resonant energies were recognized in open systems associated with the resonant behavior of the transmission coefficients defined in equation (\ref{TransmiTn}), there is no reference to  resonant wave functions of open SLs. In the last section, the resonant wave functions are given by
\begin{eqnarray}
\Psi^r_{\mu,\nu} (z) &=&\stackrel{\rightarrow }{\varphi }(z_{o})[(\alpha_{j}-\beta _{j}\frac{\beta_{n}^{\ast }}{\alpha_{n}^{\ast }})(\alpha_{p}+\gamma _{p})
+(\beta_{j}^{\ast }-\alpha_{j}^{\ast }\frac{\beta_{n}^{\ast
}}{\alpha_{n}^{\ast }})(\beta_{p}+\delta _{p})]\Bigr|_{E^r_{\mu,\nu}}.
\end{eqnarray}
In order to determine the space-inversion symmetries of these functions, it is  useful to  evaluate the resonant functions at two points, symmetric with respect to the middle point  of the SL. Since this function is complex, two parity relations were reported in Ref. [\onlinecite{Pereyra2017}]. One for the real part and one for the imaginary part. Choosing the points $z_n=L/2$ and at $z_o=-L/2$, it IS easily found
that the real part posses the symmetry
\begin{eqnarray}\label{WFSimRelReP}
\Re{\rm e}[\Psi^r_{\mu,\nu}(z_{n-1})]=\frac{1}{U_n}\Re{\rm e}[\Psi^r_{\mu,\nu}(z_1)].
\end{eqnarray}
The imaginary parts of $\Psi_{\mu,\nu}(z_1)$ and $\Psi_{\mu,\nu}(z_{n-1})$ satisfy the relation
\begin{eqnarray}\label{WFSimRelImP}
\Im{\rm m}[\Psi^r_{\mu,\nu}(z_{n-1})]=-\frac{1}{U_n}\Im{\rm m}[\Psi^r_{\mu,\nu}(z_1)]
\end{eqnarray}
These relations together imply the symmetry (here * stands for complex conjugate)
\begin{eqnarray}\label{OpenWFSymms1}
\Psi^r_{\mu,\nu}(L/2)=U_n\Psi^{r*}_{\mu,\nu}(-L/2),
\end{eqnarray}
which depends on the symmetry of the Chebychev polynomial $U_n$ evaluated at the resonant energies. From a simple analysis of the Chebyshev polynomials, it was found in Ref. [\onlinecite{Pereyra2017}] that
\begin{eqnarray}
U_n\Bigl |_{E_{\mu,\nu}}\!=\!\Biggl\{\begin{array}{lc} (-1)^{\nu} & {\rm for}\hspace{0.1in} n \hspace{0.1in}{\rm even}\cr  & \cr (-1)^{\nu+\mu+1} & {\rm for}\hspace{0.1in} n \hspace{0.1in}{\rm odd} . \end{array}\Biggr.
\end{eqnarray}
Therefore
\begin{eqnarray}
\Psi^r_{\mu,\nu}(L/2)\!=\!\Biggl\{\begin{array}{lc} (-1)^{\nu}\Psi^{r*}_{\mu,\nu}(-L/2) & {\rm for}\hspace{0.1in} n \hspace{0.1in}{\rm even}\cr  & \cr (-1)^{\nu+\mu+1}\Psi^*_{\mu,\nu}(-L/2) & {\rm for}\hspace{0.1in} n \hspace{0.1in}{\rm odd} . \end{array}\Biggr.
\end{eqnarray}

\subsubsection{The parity symmetries of eigenfunctions of bounded SLs}
\begin{figure}
\begin{center}
\includegraphics[width=13cm]{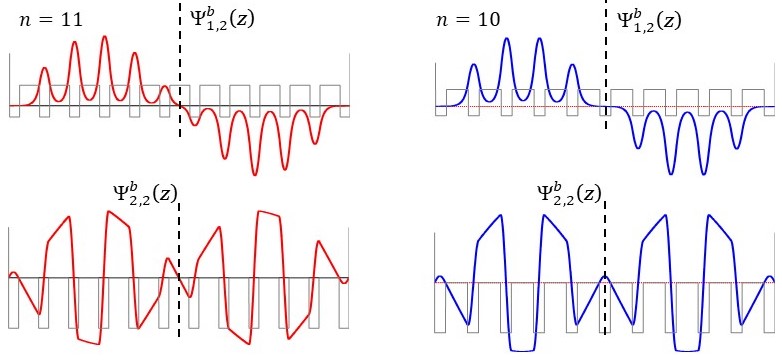}
\caption{Eigenfunctions of superlattices with length $L$=$nl_c$+$a$, bounded by infinite walls. In accord with the parity symmetry relation (\ref{BWFSymnlpa}) the eigenfunction $\Psi^b_{1,2}(z)$ and $\Psi^b_{2,2}(z)$, in the left column (for $n$=11) are both antisymmetric, and the eigenfunctions $\Psi^b_{1,2}(z)$ and $\Psi^b_{2,2}(z)$, in the right column (for $n$=10 are antisymmetric and symmetric, respectively. Figure reproduced with permission.\textsuperscript{[\onlinecite{Pereyra2017}]} 2017, Annals of Physics.}
\end{center}
\end{figure}
In this case we will also distinguish the two cases: bounded SLs of length $L=nl_c$ and bounded SLs of length $L=nl_c+a$.

When the length of the bounded SL is $L=nl_c$, the eigenfunctions are given by
\begin{eqnarray}
\Psi^{b}_{\mu,\nu} &\!=\!&A\left[(\alpha_{p}+\gamma_{p})\left( \alpha_{j}-\beta _{j}%
\frac{\alpha_{n}+\beta_{n}^{\ast }}{\alpha_{n}^{\ast }+\beta_{n}}\right)
 +(\beta _{p}+\delta _{p})\left( \beta_{j}^{\ast }-\alpha_{j}^*\frac{\alpha_{n}+\beta_{n}^{\ast }}{\alpha_{n}^{\ast }+\beta_{n}}\right) \right]\Bigr|_{E_{\mu,\nu}}, \hspace{0.5in}
\end{eqnarray}
where $A$ is a normalization constant. The symmetries are obtained also by  evaluating the eigenfunctions a two points, symmetric with respect to the center of the SL. Choosing the points at $z=z_1=-L/2+l_c$ and at $z=z_{n-1}=L/2-l_c$, the matrix elements are $\alpha_p=\delta_p=1$ and $\beta_p=\gamma_p=0$. At $z_1$,  $\alpha_j=\alpha$ and $\beta_j=\beta$, while at  $z_{n-1}$,  $\alpha_j=\alpha_{n-1}$ and $\beta_j=\beta_{n-1}$. The eigenvalue equation implies also that $(\alpha_{n}+\beta_{n}^{\ast })/\alpha_{n}^{\ast }+\beta_{n}=1$, therefore
\begin{eqnarray}
\Psi_{\mu,\nu}^b(z_1)=A (\alpha-\beta+\beta^*-\alpha^*).
\end{eqnarray}
Similarly, we have
\begin{eqnarray}
\Psi_{\mu,\nu}^b(z_{n-1})=A(\alpha_{n-1}-\beta_{n-1}+\beta^*_{n-1}-\alpha^*_{n-1}).
\end{eqnarray}
Using the relations
\begin{equation}\label{anm1bnm1}
  \alpha_{n-1}=\alpha_n\alpha^*-\beta_n\beta^*\hspace{0.3in}\rm and\hspace{0.3in}\beta_{n-1}=-\alpha_n\beta-\beta_n\alpha
\end{equation}
results IN
\begin{eqnarray}
\Psi_{\mu,\nu}^b(z_{n-1})=-A(\alpha_{n}-\beta^*_{n})(\alpha-\beta+\beta^*-\alpha^*),
\end{eqnarray}
which means
\begin{eqnarray}
\Psi_{\mu,\nu}^b(L/2-l_c)=-(\alpha_n+\beta_n^*)\Psi_{\mu,\nu}^b(-L/2+l_c).
\end{eqnarray}
Since the imaginary part of $\alpha_n+\beta_n^*$, which is proportional to $U_{n-1}$, is zero,  we are left with
\begin{eqnarray}
\Psi_{\mu,\nu}^b(L/2-l_c)=-U_{n}\Psi_{\mu,\nu}^b(-L/2+l_c),
\end{eqnarray}
and the symmetry parities are also related to those of $U_n$ when $U_{n-1}=0$. This means that
\begin{eqnarray}\label{BWFSymExactly}
\Psi_{\mu,\nu}^b(z)\!=\!\Biggl\{\begin{array}{cc} (-1)^{\nu+1}\Psi_{\mu,\nu}^b(-z) & {\rm for}\hspace{0.1in} n \hspace{0.1in}{\rm even}\cr  & \cr (-1)^{\nu+\mu}\Psi_{\mu,\nu}^b(-z) & {\rm for}\hspace{0.1in} n \hspace{0.1in}{\rm odd}.  \end{array}\Biggr.
\end{eqnarray}

When the length of the bounded SL is $L=nl_c+a$, the eigenfunctions are given by
\begin{eqnarray}
\Psi^{b}_{\mu,\nu}(z) &\!\!=\!\!&A e^{ika/2}\left[(\alpha_{p}\!+\!\gamma_{p})\left( \alpha_{j}\!-\!\beta _{j}e^{-ika}\right)
 +(\beta _{p}+\delta _{p})\left( \beta_{j}^{\ast }\!-\!\alpha_{j}^*e^{-ika}\right) \right]\Bigr|_{E_{\mu,\nu}}, \hspace{0.5in}
\end{eqnarray}
where $A$ is a normalization constant and  the eigenvalue equation $\alpha_{n}e^{ika}\!+\!\beta_{n}^{\ast }=\alpha_{n}^{\ast }e^{-ika}\!+\!\beta_{n}$ was used. Also in this case, the symmetries are determined by evaluating the eigenfunctions at two symmetric points, say at $z_o=-L/2$ and at $z_n=L/2$. At these points, $\alpha_p=\delta_p=1$ and $\beta_p=\gamma_p=0$. At $z_o$,  $\alpha_j=1$ and $\beta_j=0$, while, at  $z_n$,  $\alpha_j=\alpha_n$ and $\beta_j=\beta_n$. Thus
\begin{eqnarray}
\Psi^b_{\mu,\nu}(-L/2,)=Ae^{i k a/2}\left (1-e^{-i ka}\right ),
\end{eqnarray}
and
\begin{eqnarray}
\Psi^b_{\mu,\nu}(L/2)=Ae^{i k a/2}\left (\alpha_n +\beta_n^*-(\beta_n+\alpha_n^*)e^{-ika}\right ).\hspace{0.1in}
\end{eqnarray}
AGAIN, using the eigenvalue equation it is possible to write
\begin{eqnarray}
\Psi^b_{\mu,\nu}(L/2)=-Ae^{i k a/2}\left (1-e^{-i ka}\right )\left (\beta_n -\alpha_n^*e^{ika}\right ),\hspace{0.1in}
\end{eqnarray}
with the imaginary part of $\beta_n -\alpha_n^*e^{ika}$ equal to zero. It was shown in Ref. [\onlinecite{Pereyra2017}] that this factor has also the symmetries of $-U_n$ when $U_{n-1}=0$. Therefore
\begin{eqnarray}\label{BWFSymnlpa}
\Psi_{\mu,\nu}^b(z)\!=\!\Biggl\{\begin{array}{cc} (-1)^{\nu+\mu}\Psi_{\mu,\nu}^b(-z) & {\rm for}\hspace{0.1in} n \hspace{0.1in}{\rm even}\cr  & \cr (-1)^{\nu+1}\Psi_{\mu,\nu}^b(-z) & {\rm for}\hspace{0.1in} n \hspace{0.1in}{\rm odd}  \end{array}\Biggr.
\end{eqnarray}

\subsubsection{The parity symmetries of the eigenfunctions of SLs bounded by cladding layers, and selection rules}\label{Quasi-Bound}

\begin{figure}
\begin{center}
\includegraphics[width=13cm]{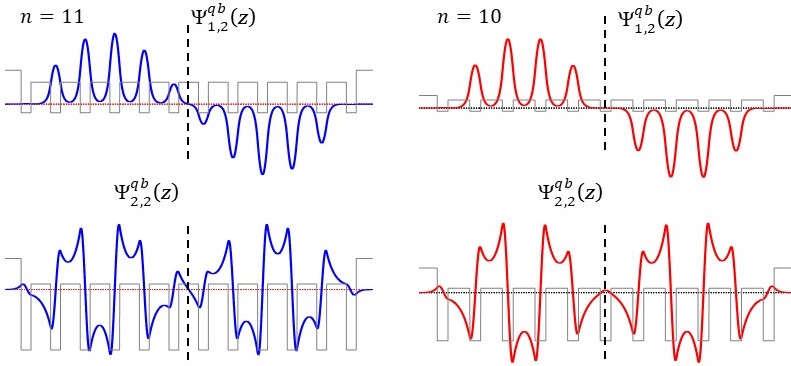}
\end{center}
\caption{Eigenfunctions of superlattices with length $L$=$nl_c$+$a$, bounded by finite walls. According to the parity symmetry relation (\ref{QBWFSym}) the eigenfunction $\Psi^{qb}_{1,2}(z)$ and $\Psi^{qb}_{2,2}(z)$ in the left column (for $n$=11) are both antisymmetric, while the eigenfunctions $\Psi^{qb}_{1,2}(z)$ and $\Psi^{qb}_{2,2}(z)$ IN the right column (for $n$=10) are antisymmetric and symmetric, respectively. Figure reproduced with permission.\textsuperscript{[\onlinecite{Pereyra2017}]} 2017, Annals of Physics.}
\end{figure}

It was shown in Ref. [\onlinecite{Pereyra2017}] that when the SLs are bounded by finite lateral barriers, the eigenfunctions posses the same symmetries as the eigenfunctions of SLs bounded by infinite walls. This means that for a SL of length $L=nl_c+a$ the eigenfunction symmetries are
\begin{eqnarray}\label{QBWFSym}
\Psi_{\mu,\nu}^{qb}(z)\!=\!\Biggl\{\begin{array}{cc} (-1)^{\nu+1}\Psi_{\mu,\nu}^{qb}(-z) & {\rm for}\hspace{0.1in} n \hspace{0.1in}{\rm odd}\cr  & \cr (-1)^{\nu+\mu}\Psi_{\mu,\nu}^{qb}(-z) & {\rm for}\hspace{0.1in} n \hspace{0.1in}{\rm even} . \end{array}\Biggr.
\end{eqnarray}
If we write the parity symmetries of the wave functions, in general, as
\begin{eqnarray}\label{PSym}
\Psi_{\mu,\nu}^{c}(z)\!=\!s^{c}_{\mu,\nu}\Psi_{\mu,\nu}^{c}(-z),
\end{eqnarray}
$s^{c}_{\mu,\nu}$ is a sign factor, which values depend on the type of the  SL (I.E., open ($c=o$), or confined by hard walls ($c=b$), or finite height walls ($c=qb$)), the quantum numbers $\mu$ and $\nu$, and the parity of the number of unit cells $n$. The possible values are summarized in table \ref{signs}.

The wave function parity relations lead to the following selection rules.\cite{Pereyra2018} When the number of unit cells $n$ is  even, the symmetry selection rules are:
\begin{eqnarray}\label{selectrul1}
 \int dz
{\it \Psi}^{q,v}_{\mu',\nu'}(z)\frac{\partial}{\partial z}{\it \Psi}^{q,c}_{\mu,\nu}(z)\left\{ \begin{array}{llrl}
= 0 &\hspace{0.2in}{\rm when} \hspace{0.2in} P[\mu'+\nu']&=& P[\mu+\nu]\cr
\neq 0 &\hspace{0.2in}{\rm when}\hspace{0.2in} P[\mu'+\nu']&=& P[\mu+\nu+1]\end{array}\right. .
\end{eqnarray}
Here $P[\mu+\nu]$ means parity of $\mu+\nu$, $c$ refers to conduction band, $v$ refers to valence band and $q$ stands for quasi-bounded superlattice. When $n$ is odd the symmetry selection rules are
\begin{eqnarray}\label{selectrul2}
 \int dz
{\it \Psi}^{q,v}_{\mu',\nu'}(z)\frac{\partial}{\partial z}{\it \Psi}^{q,c}_{\mu,\nu}(z)\left\{ \begin{array}{llrl}
= 0 &\hspace{0.2in}{\rm when} \hspace{0.2in} P[\nu']&=& P[\nu]\cr
\neq 0 &\hspace{0.2in}{\rm when}\hspace{0.2in} P[\nu']&=& P[\nu+1]\end{array}\right..
\end{eqnarray}

Similar relations are valid for infrared (intra-band) transitions but with
additional restrictions $\mu \geq \mu'$ and, whenever $\mu=\mu'$ we must also have $\nu > \nu'$.

\begin{table}[]
\centering
    \begin{tabular}{lcll}
       $s^{c}_{\mu,\nu}$\hspace{0.15in} & SL length\hspace{0.15in} & $n$ even \hspace{0.15in} &  $n$ odd \\
    \hline  $s^{r}_{\mu,\nu}$  & $n l_c$  & $(-1)^{\nu}$& $(-1)^{\nu+\mu+1}$ \vspace{0.08in} \\ \vspace{0.08in}
    $s^{b}_{\mu,\nu}$  & $n l_c$  & $(-1)^{\nu+1}$& $(-1)^{\nu+\mu}$ \\ \vspace{0.08in}
    $s^{b}_{\mu,\nu}$  & $n l_c+a$  & $(-1)^{\nu+\mu}$& $(-1)^{\nu+1}$ \\ \vspace{0.08in}
    $s^{qb}_{\mu,\nu}$ & $n l_c+a$  & $(-1)^{\nu+\mu}$& $(-1)^{\nu+1}$ \\
     \hline
    \end{tabular}
    \caption{Parity symmetry sign $s^{c}_{\mu,\nu}$ for resonant (r), bounded (b) and quasi-bounded  (qb) SLs, as a function of the parity of the number of unit cells $n$, and the quantum numbers $\mu$ and $\nu$}.
    \label{signs}
\end{table}

\section{On the TMM combined with Floquet Theorem approaches}

The transfer matrix method is one of the most pedestrian methods in the sense that a transfer matrix is built step by step, that is, layer by layer, and therefore the finiteness of the system is an eloquent reality, and becomes an intrinsic quality of the method. On the other hand, the Bloch-Floquet theorem that is known to be rigorously valid for infinite systems, implies necessarily the infiniteness as the intrinsic quality. Any attempt to construct an approach based on concepts valid in different domains falls necessarily into inconsistencies and limited predictions. The number of papers that combine the transfer matrix method with the Bloch theorem to study semiconductor, optical, and other type of superlattices is really overwhelming.

As mentioned above, faced with the problem of describing the physics of layered periodic structures, whose dynamics are determined by differential equations, it is natural to resort to the transfer matrix method, which is becoming gradually a well-known method and particularly useful as a tool to solve quantum equations of simple structures. When applied to larger systems, the Jones-Abelès transfer matrix (that has been multiple times rediscovered)  is also a well-known formula in the literature and highly regarded by its capability of making possible direct calculations of the transmission coefficients. On the other hand, we face the real fact that most physicists have tattooed the mistaken idea that Bloch functions and periodic systems are intimate and inextricable joint concepts that oblige us to invoke the Bloch functions whenever a periodic system appears. Based on the well-established transfer matrix properties and the explicit relations recalled in the first section, it is easy to show the inconsistency of these approaches. Let us first join some important pieces.

\begin{figure}
\begin{center}
\includegraphics[width=16cm]{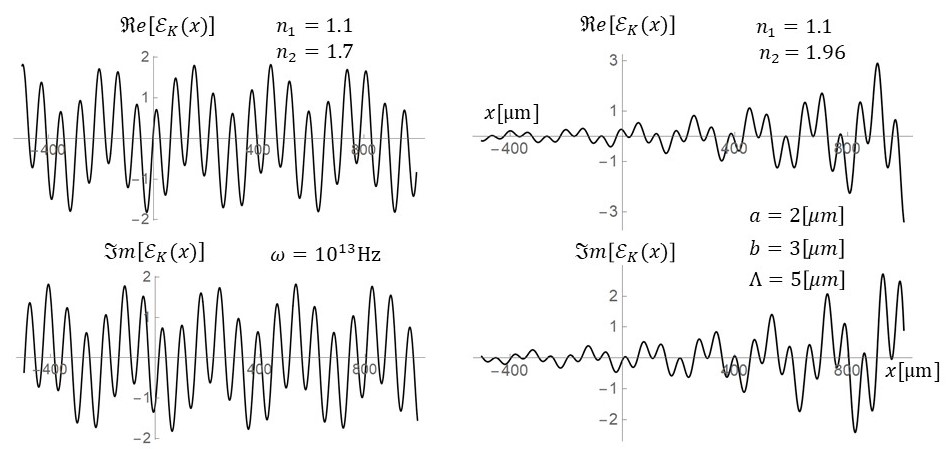}
\caption{Real and Imaginary parts of the presumptive periodic expression inside square brackets of equation (6.2-9) in Ref. [\onlinecite{PYehBook}], evaluated for the TM electromagnetic fields given in the same reference. All parameters are the same, except $n_2$. In the left side panels $n_2$=1.7, while in right hand side panels $n=1.96$. For larger values of $n_2$ the expression grows rapidly and diverges as $x$ grows. It is clear that the expression in the square brackets is not periodic.}\label{PYehPeriodic}
\end{center}
\end{figure}

We outlined that the group of transfer matrices contains the compact subgroup of transfer matrices with the general representation
\begin{equation}
M_{uc}=\left(
\begin{array}{cc}
u_1 & 0 \\
0 & u_2^*
\end{array}
\right)
\end{equation}
with $u_i$ $N$-dimensional unitary matrices. In the orthogonal universality class $u_2=u_1$. In most of the publications dealing with infinite and semi-infinite layered and periodic systems, where the authors end up invoking the Floquet theorem (see for example. Refs. [\onlinecite{Ritov,Yeh,Cottam,Haupt}]) the electromagnetic fields and field vectors in the $n$-th unit cell, are written, respectively, as
\begin{equation}
{\cal{E}}(z,n)=a_ne^{ik_z (z-nl_c)}+b_ne^{ik_z (z-nl_c)} \hspace{0.2in}{\rm and}\hspace{0.2in}\Phi_{\cal{E}}(z,n)=\left(\begin{array}{c}a_n e^{ ik_z(z-nl_c)}\\b_n e^{- ik_z(z-nl_c)}\end{array}\right) .
\end{equation}
It is clear that
\begin{equation}
\Phi_{\cal{E}}(z,n)=\left(\begin{array}{c}a_n e^{ ik_z(z-nl_c)}\\b_n e^{- ik_z(z-nl_c)}\end{array}\right)=\left(\begin{array}{cc}e^{ ik_zl_c} & 0 \\0 & e^{ -ik_zl_c}\end{array}\right)\left(\begin{array}{c}a_{n-1 }e^{ ik_z(z-(n-1)l_c)}\\b_{n-1} e^{- ik_z(z-(n-1)l_c)}\end{array}\right)=M_{oc}\Phi_{\cal{E}}(z,n-1).
\end{equation}
To establish the relation between the transfer matrix and the scattering amplitudes, one has (as emphasized in section \ref{SecSimetScat}) to define the scattering and transfer matrix with the same basis of functions (generally the asymptotic incoming and outgoing radial wave functions in 3D scattering systems, and the plane waves $e^{\pm ikz}$ in the leads, at the ends of the quasi-1D periodic systems). This means that whenever one uses the transfer matrix representation
\begin{equation}\label{Munitc}
M_{u}=\left(
\begin{array}{cc}
\alpha & \beta
\\
\gamma & \delta
\end{array}
\right)=\left(
\begin{array}{cc}
\left( t^{\dagger }\right) ^{-1} & r^{\prime }\left( t^{\prime }\right) ^{-1}
\\
-\left( t^{\prime }\right) ^{-1}r & \left( t^{\prime }\right) ^{-1}
\end{array}
\right)
\end{equation}
or the corresponding one for the orthogonal universality class, the basis of functions is fixed. Dealing with electromagnetic fields or  quantum functions and systems with homogeneous layers with sectionally constant refractive indices or potential energies, the field vectors are written precisely as (cf. Refs. [\onlinecite{Ritov,Cottam,Haupt}])
\begin{equation}
\left(
\begin{array}{c}
a_l e^{ ik_lz}
\\
b_l e^{- ik_lz}
\end{array}
\right),
\end{equation}
and the transfer matrices of unit cell $M$ satisfies the relation
\begin{equation}
\left( \begin{array}{c} a_l e^{ ik(z+l_c)} \\ b_l e^{- ik(z+l_c)} \end{array} \right)=M \left( \begin{array}{c} a_{l-1} e^{ ikz} \\ b_{l-1} e^{- ikz} \end{array} \right).
\end{equation}
Because of the importance that this issue has, and the large number of followers that P. Yeh's book\cite{PYehBook} has, we will briefly analyse the arguments that lead one to write the electromagnetic Bloch wave in equation (6.2-9) of this book. We also show graphically, (see Figure \ref{PYehPeriodic}), that the statement written after his equation  (6.2-9), that the function in square brackets that multiplies $e^{iKx}$ is periodic, is false. In fact, as can be seen in figure \ref{PYehPeriodic}, it is not only the lack of periodicity, it happens that, for example for $n_1$=1, the function inside the square brackets diverges when $z$ grows and $n_2\geq 2$.

The problem posed in this approach, when `the solutions for periodic medium in accord with Floquet theorem' are written as
\begin{equation}\label{PYe1}
  E_K(x,z)={\cal E}_K(x)e^{-i\beta z}e^{iKx}
\end{equation}
where ${\cal E}_K(x)$ is periodic with a period $\Lambda$, and ${\cal E}_K(x+\Lambda)={\cal E}_K(x)$, is to find $K$ and ${\cal E}_K(x)$. We will see below that a correct choice should be to determine the transfer matrix compatible with the Floquet theorem.

Starting from the electromagnetic fields
\begin{eqnarray}
E(x) =\left\{
\begin{array}{llc}
    a_n e^{-ik_1(x-n\Lambda)}+b_n e^{ik_1(x-n\Lambda)}&{\rm for}&\hspace{0.2in}n\Lambda-a<x<n\Lambda\\
  c_n e^{-ik_1(x-n\Lambda+a)}+d_n e^{ik_1(x-n\Lambda+a)}& {\rm for}&\hspace{0.2in}(n-1)\Lambda<x<n\Lambda-a
\end{array} \right.
\end{eqnarray}
and the transfer matrix representation, P. Yeh  ends up writing the relation
\begin{equation}\label{TMYeh}
\left( \begin{array}{c} a_{n-1} \\ b_{n-1} \end{array} \right)=\left(
\begin{array}{cc}
A & B
\\
C & D
\end{array}
\right) \left( \begin{array}{c} a_{n} \\ b_{n} \end{array} \right)
\end{equation}
The transfer matrix here is the inverse of ours, but this in not an issue. An important step in the attempt to obtain Bloch-type electromagnetic fields has been to write the Bloch waves as
\begin{equation}
\left( \begin{array}{c} a_{n-1} \\ b_{n-1} \end{array} \right)_K=e^{-iK\Lambda} \left( \begin{array}{c} a_{n} \\ b_{n} \end{array} \right)_K,
\end{equation}
and to assume that the Bloch vectors, written here with a subindex $K$, are the same as the   electromagnetic fields in (\ref{TMYeh}). This makes one believe that {\it the same} transfer matrix that relates the electromagnetic field vectors also relates the Bloch vectors in the last equation. This misleading assumption leads to write the Kramers eigenvalue equation
\begin{equation}\label{EigenvKramers}
\left(
\begin{array}{cc}
A & B
\\
C & D
\end{array}
\right)\left( \begin{array}{c} a_{n} \\ b_{n} \end{array} \right)_K=e^{iK\Lambda} \left( \begin{array}{c} a_{n} \\ b_{n} \end{array} \right)_K.
\end{equation}
in terms of a matrix that fulfills equation (\ref{TMYeh}) but not necessarily a similar relation when the field vectors are replaced by the eigenvectors.  Solving this equation yields the eigenvalues
\begin{equation}\label{EigvalKramers}
  \lambda_{\pm}=e^{iK_{\pm}\Lambda}=\frac{1}{2}(A+D)\pm \sqrt{\left(\frac{1}{2}(A+D)\right)^2-1}
\end{equation}

\noindent with the corresponding eigenvectors. Both widely accepted and written in terms of the wrong transfer matrix elements. To obtain the periodic factor ${\cal E}_K(x)$, in a kind of backwards step, a second assumption identifying again the eigenvectors with the electromagnetic waves results in equation (6.2-9) for ``the Bloch wave in the $n_1$ layer of the $n$-th unit cell" given by
\begin{equation}\label{Eq629Yeh}
{\cal E}_K(x)e^{-iKx}=\left[\left( a_o e^{-ik_1(x-n\Lambda)}+b_o e^{ik_1(x-n\Lambda)}\right) e^{iK(x-n\Lambda)}\right] e^{-iKx}
\end{equation}
with $a_o=B$ and $b_o=\lambda_+-A$ and  the additional statement that the ``expression inside the square brackets ...is periodic". However, using the same transfer matrix elements given in Yeh's book, we have plotted the expression inside the square brackets and find that it is not periodic. For example, for $n_1=1$ and $n_2=1.2$, we have the behavior shown in Figure a) and for $n_1=1$ and $n_2=2.1$ the divergent behavior shown in Figure b), as $x$ grows. This is a consequence of the unjustified assumptions.

It may be important however to see what kind of transfer matrix is compatible with Bloch-type electromagnetic functions. If instead of  assuming equation  (\ref{EigenvKramers}) as if we were looking for vectors, which are known and fixed, we look for the transfer matrix elements $A_K$, $B_K$... such that
\begin{equation}\label{EigKramers}
\left(
\begin{array}{cc}
A_K & B_K
\\
C_K & D_K
\end{array}
\right)\left( \begin{array}{c} a_{n} e^{ik(x+n\Lambda}\\ b_{n} e^{-ik(x+n\Lambda} \end{array} \right)=e^{iK_{\pm}\Lambda} \left( \begin{array}{c} a_{n}  e^{ik(x+n\Lambda}\\ b_{n}  e^{ik(x+n\Lambda} \end{array} \right)
\end{equation}
where $\lambda_+\lambda_-=1=e^{iK_{+}\Lambda}e^{iK_{-}\Lambda}$, an obvious solution is $K_+=-K_-=K$, $A_K=\lambda_+$, $D_K=\lambda_-$, and $B_K=C_K=0$, thus a Floquet transfer matrix that is compatible with these assumptions is
\begin{equation}\label{FloqueTM}
M_F=\left(
\begin{array}{cc} e^{iK_{+}\Lambda}& 0 \\ 0 & e^{iK_{-}\Lambda} \end{array}
\right),
\end{equation}
which belongs to the compact subgroup of transfer matrices and allows Bloch-type electromagnetic fields such that
\begin{equation}
\left( \begin{array}{c} E^+(z+n\Lambda) \\E^-(z+n\Lambda) \end{array} \right)=M_F^n \left( \begin{array}{c} E^+(z) \\E^-(z) \end{array} \right).
 \end{equation}
In this way, each component of the electromagnetic waves is a Bloch-type wave function, and can be written as
\begin{eqnarray}
  E^+(z+n\Lambda) &=& e^{iK n\Lambda} E^+(z)= e^{iK n\Lambda} e^{ik_zz} \\
  E^-(z+n\Lambda) &=& e^{-iKn\Lambda} E^-(z)= e^{-iKn\Lambda}e^{-ik_zz}.
\end{eqnarray}
The price to pay for an electromagnetic wave which amplitude does not vanish from $-\infty$  to $\infty$ is a transmission amplitude $t_B=e^{iK n\Lambda}$ that only modifies the phase and implies, as should be expected, a local transmission coefficient $T_n$=1.

\section{Multichannel features in double barriers, ballistic resonant tunneling transistors, and some other examples}\label{DBandWannierSplitting}

We will review here, and in the next section,  some examples in which the TFPS has been applied. We will start with the calculation of the conductance for a resonant biased double barrier. We will see that the shoulder that is present in the negative resistance domain of numerous experimental results is easily explained by the multichannel conductance calculation for biased double barrier (DB).\cite{PereyraMendoza} We will then refer to one of the various proposals of ballistic electrons through SLs in resonant tunneling transistors, and we will show also good agreement between the theoretical prediction and the measured transmission coefficients.\cite{Pacher2002} We will review also the ballistic multichannel transmission through a 3D periodic array of $\delta$-scatterer centers,\cite{Pereyra2002} and will see an example of channel coupling and channel interference. In some of the applications of the TFPS the space-time evolution of Gaussian electromagnetic and Gaussian electron wave packets through SLs were studied. We will show few results in the published work, among them, evidences of superluminal transmission of electromagnetic waves packets through optical media SLs,\cite{Simanjuntak2007} optical antimatter effect in electromagnetic wave packets through metamaterial superlattices,\cite{RobledoMorales2008,RobledoRomero2009,RobledoSandoval2009,Pereyra2011}, I-V characteristics in spin injection, spin filter, and spin inversion behavior for spin wave packets through homogeneous magnetic superlattices.\cite{Ibarra2009}

\subsection{Multichannel features in the negative resistance domain of biased double barriers}
\begin{figure}
\begin{center}
\includegraphics[width=15cm]{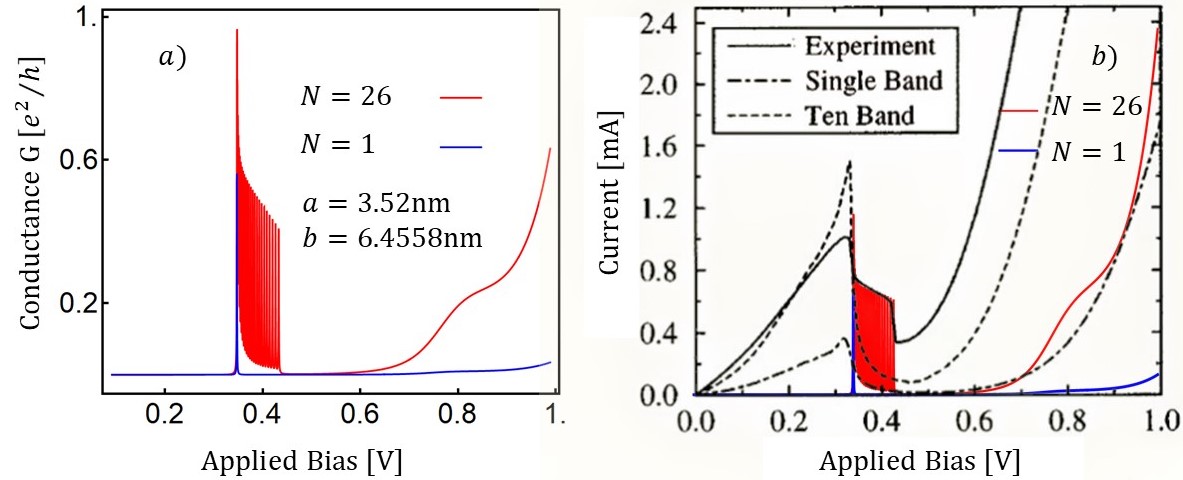}
\caption{Multichannel transmission and conductance of a biased double barrier. In the right, the experimental and theoretical I-V characteristics of the DB reported in Ref. [\onlinecite{Bowen1997}]. In a) and b), the transmission coefficient and current calculated for a single propagating mode (blue) and for the contribution of $N=26$ propagating modes, reproducing faithfully the  shoulder of observed in several experiments. Experimental figure reproduced with permission.\textsuperscript{[\onlinecite{Bowen1997}]} 1997, American Institute of Physics.}\label{MultichannelAiryDB}
\end{center}
\end{figure}

The  biased double barrier is a simple system and the resonant tunneling and negative resistance behavior have been amply and extensively studied, and little or nothing should be added after 50 years of research and applications. However, it turns out that a feature in the negative resistance domain, that appeared in numerous experimental reports,\cite{Sollner1984,Bonnefoi1985,Morkoc1986,Ray1986,Brown1987,Huang1987,
Woodward1987,BoSu1991,Bowen1997,Dai2007,Feigenov2011,Zeng2013} has not been well understood, much less predicted. That feature is a shoulder as the one seen, between 0.32V and 0.43V, in the experimental (black, continuous) curve in figure \ref{MultichannelAiryDB} b) reported by Bowen et al in Ref. [\onlinecite{Bowen1997}]. The dot-dashed line in this figure, is a single band simulation,
while the dashed line
curve is ten bands simulation in a nearest neighbor sp3s* model.\cite{Ko1988}  In Ref. [\onlinecite{Sollner1984}] the shoulder is considered a consequence of a self-detection; in Ref. [\onlinecite{Zeng2013}] it is attributed to serial resistance, and, in Ref. [\onlinecite{Feigenov2011}], to oscillations due to the bias field. The shoulder in the conductance and I-V characteristics of biased double barriers is a good example to test the calculation of these quantities.

In many real systems, the transverse dimensions  imply, as mentioned in section \ref{mMultichannelApproach}, the contribution of several propagating modes.  In Ref. [\onlinecite{PereyraMendoza}], the transfer matrix method and the multichannel (multi-propagating modes) approach, introduced in section \ref{mMultichannelApproach}, was applied to study transmission coefficients and conductance through a double barrier under the influence of longitudinal and transverse electric fields. Some results published there give an insight to explain what is behind  the frequently observed shoulder in the negative resistance domain of DBs. In fact, a simple calculation of the transmission coefficient of a biased DB, with only longitudinal electric field, has the behavior shown in figure \ref{MultichannelAiryDB} a), and  in the current shown in figure \ref{MultichannelAiryDB} b). For this results we consider the Shr\"odinger equation
\begin{equation}
-\frac{\hbar ^{2}}{2m}\nabla^2\Psi({\bf r})+\left(V_{C}(x,y) + V_f(z)\right) \Psi({\bf r})=E \Psi({\bf r}).
\end{equation}
Here
\begin{eqnarray}
V_f(z) = \left\{ \begin{array}{ccc}  -f_bz \hspace{0.2in}\;&  -l_i< z\leq 0    \cr  V_o-f_bz \hspace{0.2in}\;& 0< z < b  \cr -f_bz \hspace{0.2in}\;& b\leq z\leq a+b   \cr  V_o-f_bz \hspace{0.2in}\;& a+b < z < a+2b \cr -f_bz  \hspace{0.2in}\;& a+2b< z < a+2b+l_r \end{array} \right.,
\end{eqnarray}
where $f_b$ is the electric force of a bias $V_b$ potential applied between $z=-l_i$ and $z=a+2b+l_r$, where $a$ and $b$ are the well and barrier widths.   $V_{C}(x,y)$ is a confining hard wall
potential, as in equation (\ref{FSch}). After performing variables separation, and neglecting the evanescent channels contribution, one is left  with the system of coupled equations
\begin{eqnarray}
\frac{d^2}{dz^2}\varphi_i(z)+(k^2-k_{Ti}^2)\varphi_i(z)=\sum_{j=1}^N K_{ij}(z)\varphi_j(z)
\end{eqnarray}
where $N$ is the number of open channels and
 \begin{eqnarray}
K_{ij}(z)=\frac{2 m^* }{\hbar^2}\int_0^{w_yw_x} dx dy \phi^*_i(x,y)V(z)\phi_j(x,y)
\end{eqnarray}
is the coupling channels matrix. The  functions $\phi_i(x,y)$ are the eigenfunctions of the infinite square well in the leads. As explained in section \ref{mMultichannelApproach}, the channel indices $i,j$, characterize the propagating modes and the transverse momenta $k_{Tj}=\pi j/w_y$. Here $w_y$ is the largest transverse width, or the radius when the transverse section is circular. For the DB potential $U(y,z)$, defined above, the coupling matrix is
\begin{eqnarray}
K_{ij}(z)= \frac{2 m^*}{\hbar^2} \bigl(-f_b z +V_o\bigr)\delta_{i,j}
\end{eqnarray}
Notice that in the absence of a transverse field, no channel mixing exists, and one is left with the system of soluble Airy equations
\begin{eqnarray}
\frac{d^2}{dz^2}\varphi_i(z)+(k^2-k_{Ti}^2)\varphi_i(z)= \frac{2 m^*}{\hbar^2} \bigl(-f_b z +V_o \bigr)\varphi_i(z)\hspace{0.2in}i=1, 2, ..., N.
\end{eqnarray}
As was explained in section \ref{TMRQW}, given the solutions and their derivatives, it is possible and some times convenient to define transfer matrices $W(z_2,z_1)$, that connects functions and their derivatives. For the DB, the $N$ channel transfer matrix is a $2N\times 2N$ block-diagonal matrix, with blocks of the form
\begin{eqnarray}
W(z_2,z_1) \!=\! \left(\begin{array}{cc} A_i(z_2,k_{Ti})  & B_i(z_2,k_{Ti})\cr A'_i(z_2,k_{Ti})  & B'_i(z_2,k_{Ti})\end{array}\right)\left(\begin{array}{cc} A_i(z_1,k_{Ti})  & B_i(z_1,k_{Ti})\cr A'_i(z_1,k_{Ti})  & B'_i(z_1,k_{Ti})\end{array}\right)^{-1}
\end{eqnarray}
where $A_i(z_2,k_{Ti})$ and $B_i(z_2,k_{Ti})$ are the Airy functions in channel $i$. After the similarity transformation, see equation (\ref{SimTransfBarrier}), it is easy to calculate the Landauer conductance
\begin{equation}\label{LandCondDB}
  G=\frac{e^2}{h}\Tr tt^{\dagger}
\end{equation}
as well as the current
\begin{equation}\label{CurrentdDB}
  j=e n_e \mu_e f_bL \Tr tt^{\dagger}
\end{equation}
for a given incoming energy $E$. In the last equation $n_e$ is the charge carriers concentration,  $e$ is the electron's charge, $\mu_e$ the mobility, $h$ the Planck's constant, and $L=a+2b+l_i+l_r$ the distance between the point contacts. In figures \ref{MultichannelAiryDB} a) and b) we show the calculated conductance and current for the parameters mentioned in Ref. [\onlinecite{Bowen1997}] (with $a$ and $b$  slightly modified to fit multiples of the lattice constant). By definition, the number of propagating modes (or open channels), is of the order of $2w\sqrt{2m^*E}/\pi\hbar$, with $w$ the transverse width. It is clear the relevance that the finite transverse dimension $w$ has in the splitting of the propagating modes, which number is limited also by the energy $E$. The calculation  shows not only the origin of the shoulder in the negative resistance domain of a biased DB observed in the experimental reports but also the ability of the multichannel transfer matrix method to solve faithfully and to explain features that previous theoretical attempts failed. It is evident also from figure \ref{MultichannelAiryDB} a) that the propagating modes with higher longitudinal energy transmit at higher bias, but with lower transmission probability, a well-known effect that led to envision new devices where the symmetry of the DB, broken by the bias field, could be restored ``not by applying an electric field, but by injecting minority carriers or ballistic electrons"\cite{CapassoJAP1985}.

\begin{figure}
\begin{center}
\includegraphics[width=8cm]{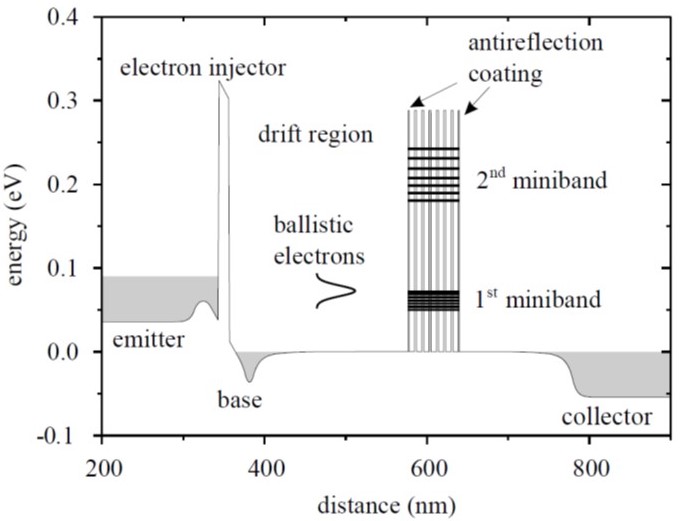}
\caption{Schematic band structure of the
three-terminal resonant tunneling transistor with superlattice in the base with anti-reflection coating
barriers at the ends. Horizontal lines correspond to energy levels intra-miniband. Figure reproduced with permission.\textsuperscript{[\onlinecite{Pacher2002}]} 2002, Physica E.}\label{ResonantTransistor}
\end{center}
\end{figure}
\subsection{Ballistic electrons through heterostructures and SLs}

In the 80s of the last century, when the size of semiconductor devices reached  dimensions of the  mean free path of charge carriers, it became natural to think that  the carriers' current is not only less diffusive, but ballistic or  quasi ballistic. \cite{Shur1979, Heiblum1985} Thus, new structures and devices were proposed, in which ballistic electrons pass resonantly through the quantized energy levels of symmetric double barriers or the minibands of SLs located in the base \cite{Levi1985,Capasso1986} or in the emitter region.\cite{S-YCheng1997} The main objectives were to increase the transmission probability and to obtain high peak-to-valley ratios, by restoring the symmetry of DBs or avoiding the loss of periodicity and the ensuing phase coherence in SLs. Among the numerous resonant tunneling transistors with SLs, we will review the three-terminal devices, with anti-reflecting coating, studied experimentally and theoretically by Pacher et al.\cite{Pacher2002}

\begin{figure}
\begin{center}
\includegraphics[width=14cm]{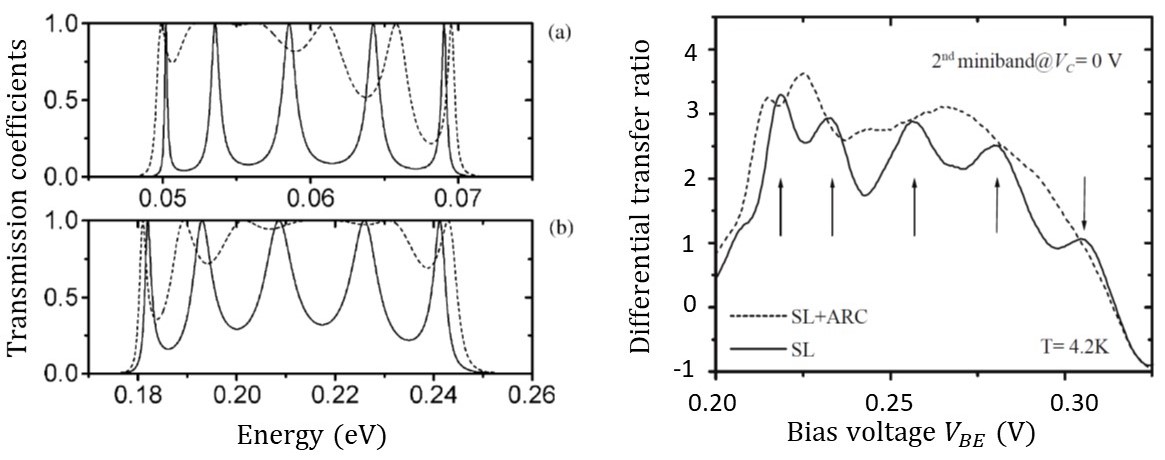}
\caption{Transmission coefficients. In  (a) and (b), the left side panel, the theoretical transmission coefficients for SL without (full curves) and with coating layers (dotted curves). In the right hand side panel the measures transmission coefficients. Figure reproduced with permission.\textsuperscript{[\onlinecite{Pacher2002}]} 2002, Physica E.}\label{PacherCoefficients}
\end{center}
\end{figure}
Figure \ref{ResonantTransistor} shows the schematic band structure of the
three-terminal device with the superlattice studied by Pacher et al. The width of a $GaAs$ valley is $a=6.5$nm, the width of the $Al_{0.3}Ga_{0.7}As$ barrier is $b=2.5$nm, and the number of unit cells is $n=6$. The width $b'$ of the first and last barriers, which act as anti-reflection coating, is $b'=b/2$. Pacher et al calculated and measured the transmission coefficients shown in figure \ref{PacherCoefficients}. For the calculation of the transmission coefficient of the uncoated and coated superlattice, in the 1D one-propagating mode limit, one has first to determine the transfer matrix given by
\begin{equation}
M_{c}= M_{b'}M_nM _{b'}
\end{equation}
where $M_n$ is the SL transfer matrix
\begin{equation}
M_{n}= \left(
\begin{array}{cc}
 U_{n}-\alpha^*  U_{n-1} & \beta U_{n-1} \\
\beta^* U_{n-1} &  U_{n}-\alpha  U_{n-1}
\end{array}
\right)
\end{equation}
defined in (\ref{TM1Dncells}) and $M_{b'}$ the transfer matrix of a coating layer, defined as
\begin{equation}
M_{b'}= \left(
\begin{array}{cc}
\alpha_{b'} & \beta_{b'}  \\
\beta_{b'}^*&  \alpha_{b'}^*
\end{array}
\right).
\end{equation}
In these matrices, we have
\begin{eqnarray}
\begin{array}{ll}\displaystyle{
\alpha= e^{ika}\left( \cosh
qb + i\frac{k^{2} - q^{2}}{2qk}\sinh qb   \right)} & \hspace{0.2in}
\displaystyle{ \beta=-i\frac{k^{2} + q^{2}}{2qk}\sinh qb}\cr\cr
\displaystyle{\alpha_{b'}=e^{ika}\left( \cosh
qb' + i\frac{k^{2} - q^{2}}{2qk}\sinh qb'   \right)}& \hspace{0.2in}
\displaystyle{ \beta_{b'}=-i\frac{k^{2} + q^{2}}{2qk}\sinh qb'},
\end{array}
\end{eqnarray}
with $k = \sqrt{2m^*_w E/\hbar^{2}}$ and  $q = \sqrt{2m^*_b(V_{o} - E)/\hbar^{2}}$ the wave numbers in the wells and barriers, and $m^*_w$ and $m^*_b$ the effective masses and $V_o=0.23$eV. As mentioned before, the Chebyshev polynomials $U_n$ are evaluated at the real part of $\alpha$. After multiplying and simplifying, one can obtain the relevant matrix element of $M_c$ to determine the transmission coefficient, i.e. the element $\alpha_c=M_{c11}$ that can be written in compact form as
\begin{equation}\label{AlphaARC}
 \alpha_c=(\alpha_b'^2-\beta_b'^2)U_n-(\alpha_b'^2\alpha^*-\beta_b'^2\alpha+2\alpha_b'\beta_b'\beta)U_{n-1} =1+h_{b'}U_n+f_{b'}U_{n-1}.
\end{equation}
Pacher et al.  assumed that $f_{b'}U_{n-1}=0$,  and reported a simplified transmission coefficient, denoted as $T_{SL}^{ARC}$ in the equation (1) of Ref. [\onlinecite{Pacher2002}]. They  plotted the simplified transmission coefficient $T_{SL}^{ARC}$ and compared with the transmission coefficient for a SL without coating layers (see graphs reproduced in the left panel of figure \ref{PacherCoefficients}, with dotted and full curves respectively). It is worth noticing that the assumption $f_{b'}U_{n-1}=0$ is correct only for $b'\simeq b/2$. For values of the coating layer width $b'$ smaller than $b/2$, the simplified transmission coefficient overestimates the effect of increased transmission. In the other limit, i.e. $b'\gg b$, the behavior of the superlattice with coating layer is similar to that of  quasi-bounded superlattice discussed in section \ref{Quasi-Bound}.
By applying a DC voltage between the emitter
and the base, hot electrons were injected and transmitted through the superlattice. From the ratio
$I_C/I_E$ of the measured currents at $4.2$K,
the transmission results shown in the right panel of figure \ref{PacherCoefficients} were obtained.

\begin{figure}
  \centering
  \includegraphics[width=12cm]{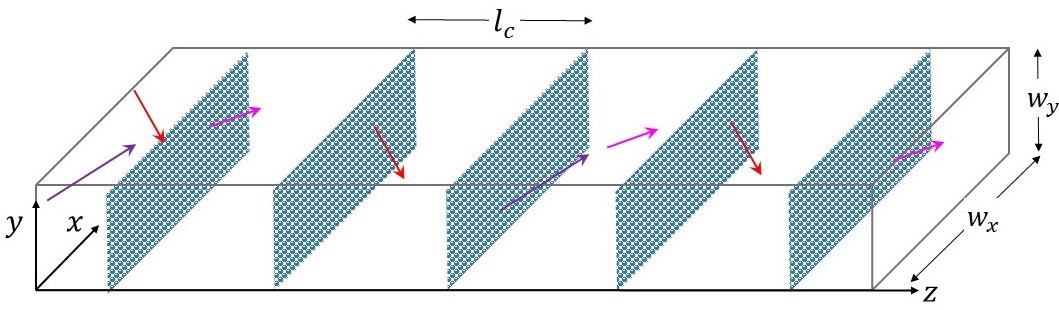}
  \caption{Schematic representation of a 3D SL with 2D arrays of $\delta$ scaterer centers. }\label{3ChannelDeltas}
\end{figure}

\subsection{Multichannel conductance through a 3D SL}

Multichannel conductance calculations through 1DSls with coupled channels were published, among others, by Ulloa et al. [\onlinecite{Ulloa}], based on self-consistent numerical calculations, and using the transfer matrix method in Refs. [\onlinecite{Pereyra2002,AnzaldoPereyra2007,PereyraMendoza,AnzaldoPereyra2005}]. In most of the latter works a transverse electric field couples the propagating modes leading to interference and flux transition from one propagating mode to another, clearly observed in the transmission coefficients $T_{ij}$ and conductance. In Ref. [\onlinecite{Pereyra2002}], 3D superlattices with 2D arrays of $\delta$-scatterer centers, separated by homogeneous layers of width $l_c$, as shown in figure \ref{3ChannelDeltas}, were considered, and the transmission coefficients and conductance $G_{N,n}=(e^2/h)\Tr tt^{\dagger}$, defined in section \ref{ScatAmpConductance}, calculated for different values of the number of propagating modes $N$ and different number of unit cells $n$. Here we will review some of the specific formulas and present new results for $N=3$ and $n=2,3,6$.

For a periodic potential
\begin{equation}
V_{P}(x,y,z)=\gamma \delta \left( z-\eta l_{c}\right) \sum_{\nu =1}^{N_{\nu
}}\sum_{\mu =1}^{N_{\mu }}\delta \left( x-x_{\nu }\right) \delta \left(
y-y_{\mu }\right) \qquad \kappa =1,...,n
\end{equation}
with confining 2D hard walls, infinite outside $\{0\leq x\leq
w_{x},\,\,0\leq y\leq w_{y}\}$,  longitudinal lattice parameter $l_{c}$ and interaction strength $\gamma$, the solutions
\begin{equation}
\phi _{i}\left( x,y\right) =\frac{2}{\sqrt{w_{x}w_{y}}}\sum_{%
\{{i}^{2}=n_{x}^{2}+n_{y}^{2}\}}\sin \frac{n_{x}\pi x}{w_{x}}\sin \frac{%
n_{y}\pi y}{w_{y}},
\end{equation}
and energies
\begin{equation}
E_{i}=\frac{\hbar ^{2}\pi ^{2}}{2m^{\ast }}\left(\frac{n_{x}^{2}}{w_{x}^{2}}+
\frac{n_{y}^{2}}{w_{y}^{2}}\right)\leq E
\end{equation}
with $E$ the incoming particles energy, define the transverse propagating modes labeled by the channel-index $i$, which is determined by the pairs of quantum numbers $n_{x},n_{y}$ $=1,2,3,...$ The functions $\phi _{i}\left( x,y\right)$ constitute a realization of the set of orthonormal functions, mentioned in  section \ref{mMultichannelApproach}, which allow us to obtain the channel coupling matrix
\begin{equation}
K_{ij}=\frac{8\pi ^{2}m\gamma }{h^{2}}\delta \left( z-\eta l_{c}\right)
\sum_{\nu =1}^{N_{\nu }}\sum_{\mu =1}^{N_{\mu }}\phi _{i}^{\ast }\left(
x_{\nu },y_{\mu }\right) \phi _{j}\left( x_{\nu },y_{\mu }\right) =\delta
\left( z-\eta l_{c}\right) \Gamma _{ij},
\end{equation}
Replacing and integrating the Schr\"odinger equation upon the variable $z$ and following the usual procedure with  $\delta $-potentials, to impose the continuity conditions, one can straightforwardly determine the transfer matrix
\begin{equation}
M_{\delta }=\left(
\begin{array}{ll}
\alpha _{\delta } & \beta _{\delta } \\
\beta _{\delta }^{\ast } & \alpha _{\delta }^{\ast }
\end{array}
\right) .
\end{equation}
with
\begin{equation}
\alpha _{\delta }=I_{N}+\beta _{\delta },\text{\qquad }\beta _{\delta }=%
\frac{1}{2i}\left(
\begin{array}{ccc}
\begin{array}{ll}
\frac{\Gamma _{11}}{k_{1}} & \frac{\Gamma _{12}}{k_{1}} \\
\frac{\Gamma _{21}}{k_{2}} & \frac{\Gamma _{22}}{k_{2}}
\end{array}
&  &  \\
& . &  \\
&  & .
\end{array}
\right) ,\quad \text{and\quad }\frac{\Gamma _{ij}}{\Gamma _{ji}}=\frac{k_{i}%
}{k_{j}}
\end{equation}

To use the polynomials and invariant functions mentioned in previous sections, it is
necessary to determine the eigenvalues of the $2N\times 2N$ transfer matrix $
M$. A specific procedure to evaluate the matrix-polynomials, for this type of transfer matrix, was shown in Ref. [\onlinecite{Pereyra2002}], and we refer the reader to this reference.

\begin{figure}[tbp]
\includegraphics[width=14cm]{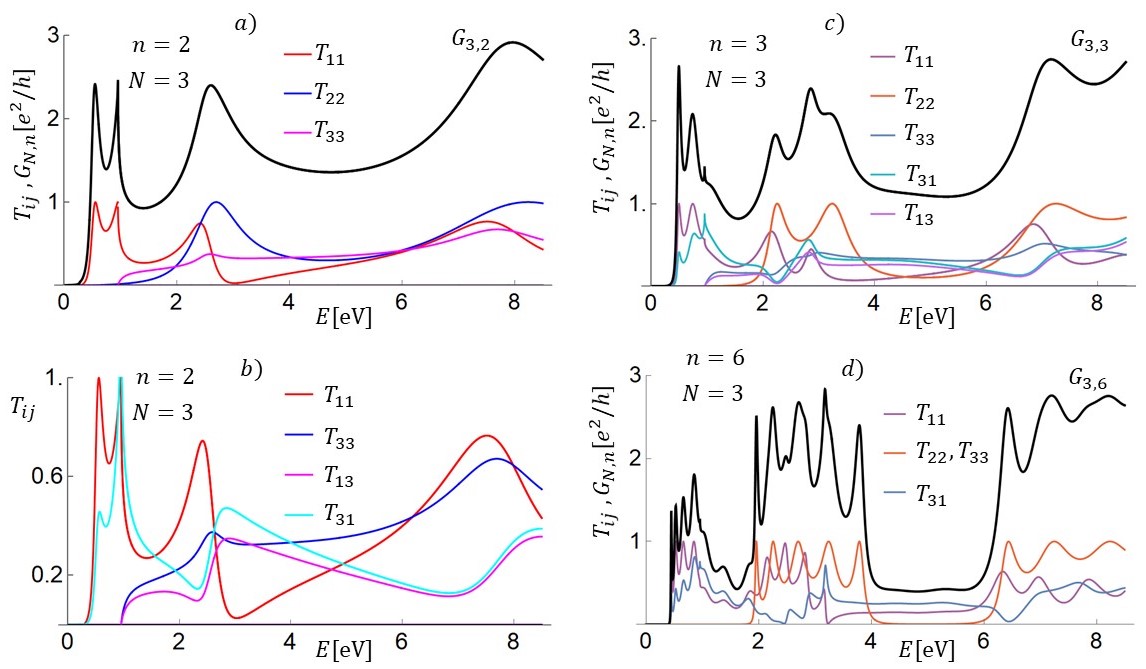}
\caption{Transmission coefficients and conductance for 3 propagating modes. The coupling effects on the transmission
coefficients $T_{ij}$ between the channels with indices $i=1$ and $i=3$, is evident from figures in panels a) and b). Notice that the transition probability $T_{ij}$, from channel $j$ to channel $i$, rises as soon as the energy $E$ allows to open the incoming channel $j$. In figures of panels c) and d), transmission coefficients and conductance for $n$=3 and for $n=$6, respectively.}\label{Conductance3Chann}
\end{figure}

Given the transfer matrix and applying the TFPS, we can evaluate the transmission coefficients $T_{i,j}$ and the conductance $G=(2e^2/h)\Tr tt^{\dagger}$ for any set of parameters. In figure \ref{Conductance3Chann} we show transmission coefficients and conductance for the 3 propagating modes, the channels with the lowest energy, and for different number of unit cells. For this figure we consider $l_{c}=20$\AA , $w_{x}=100$\AA,  $w_{y}=50$\AA , $N_u=30$, $N_{\nu }=15$,  and $\gamma =-200eV$. For these parameters, the channel thresholds are: $E_{T1}$=0.281eV; $E_{T2}$=0.449eV and $E_{T3}$=0.954eV. In figures \ref{Conductance3Chann}a) and b), the number of unit cells is $n=2$. In figures \ref{Conductance3Chann}a) we plot the transmission coefficient $T_{ii}$ and the conductance $G$, while in figure \ref{Conductance3Chann}b) we plot the transmission coefficients $T_{ij}$ for channels indices $i,j=1$ and 3, with the strongest channel coupling. It is clear from this figures that the transition probability $T_{ij}$, from channel $j$ to channel $i$, starts to rise once the energy $E$ is enough to open the incoming channel $j$. In figures \ref{Conductance3Chann}c) and d) we plot the transmission coefficients when the number of units cells are $n=3$ and $n=6$, respectively.

\section{Superluminal and optical antimatter effects. Injection, filter and spin inversion.}

The theory of finite periodic systems has been applied also to study the space-time evolution of wave packets through superlattice structures. Among the various examples we shall review first the transmission of electromagnetic waves through normal media and through metamaterial SLs where normal media alternates with left-handed layers. We will then review some topics of spintronics where the transfer matrix method and the TFPS has bee applied with success. Specifically, to understand the negative resistance behavior in the IV characteristics of spin injection and detection in all-semiconductor contacts, and to show a couple of results of interest in spintronics, which show  the space-time evolution of spin wave packets through homogeneous magnetic superlattices and clear examples of spin filter and spin inversion..

\subsection{Electromagnetics wavepackets through SLs. Evidences of superluminal effects.}

Simanjuntak et al. applied the TFPS to study the space-time evolution of electrons\cite{Simanjuntak2003} and of electromagnetic\cite{Simanjuntak2007} wave packets, and to discuss the presumption of the so-called generalized Hartmann effect.\cite{Simanjuntak2013} We will briefly review here the time evolution
(described by Maxwell equations) of Gaussian wave packets (WPs) with centroids
in the allowed and in the photonic band gaps, which was studied with more detail in Ref.[\onlinecite{Simanjuntak2007}]. The calculation reviewed here shows that the time spent by the wave packet with centroid in the photonic gap is half of the time it will require moving with light velocity, a result compatible with the experimental and theoretical results in Refs. [\onlinecite{Steinberg1993,Spielmann1994,Pereyra2000}], referred to in the last section.

A Gaussian wave
packet of electromagnetic fields, with parallel polarization and normal incidence from the left, was defined as
\begin{eqnarray}\label{PgaussT}
\Psi_E(z,t) &=& \int dk \, e^{-\gamma (k - k_o)^2} e^{ikz_o}
E(z,k) e^{-i\omega t}
\end{eqnarray}
with the packet-peak at $t=0$ located at $-z_o$. To determine $\Psi_E(z,t)$, for any $t$ and at any point $z$, one needs the
$k$-component $E(z,k)$ outside and inside the SL, which was assumed to
contain $n$ cells of length $l_c = l_1 + l_2$, being $l_1$ and
$l_2$ the lengths of dielectric layers with permittivities
$\epsilon_1, \epsilon_2$, refractive indices $n_1, n_2$ and
permeabilities $\mu_1, \mu_2$. At any point, the electric and magnetic fields contain a
right- and a left-moving parts. Hence
\begin{eqnarray}
E(z,k_i) &=&A_r e^{ik_iz} + A_l e^{-ik_iz}=E_r(z,k_i)+E_l(z,k_i), \nonumber\\
H(z,k_i) &=&\frac {n_i}{\mu_i c}
\left(E_r(z,k_i)-E_l(z,k_i)\right)\nonumber ,
\end{eqnarray}
where $k_i = n_i k$ ($i=0,1,2$), $k=\omega /c$, with $c$ the
speed of light in vacuum, and $n_0$ the refractive index outside the SL. To use the transfer matrices, it was convenient to define the field vectors
\begin{eqnarray}
\mathcal{E}(z,k) \ = \ \left(
\begin{array}{c}
E_r(z,k) \\
E_l(z,k)
\end{array}
\right), \label{psiAB}
\end{eqnarray}
and the transfer matrix $M(z',z)$
\begin{eqnarray}
\mathcal{E}(z',k) \ = \ M(z',z)\mathcal{E}(z,k).
\end{eqnarray}
Inside the SL, a vector
$\mathcal{E}_{j+1}(z,k)$ at $z=jl_c+z_p$ in cell $j+1$ (with $j =
0, 1, \ldots \, , (n - 1)$ and $0<z_p\leq l_c$) is related to
$\mathcal{E}(0^-,k)$ at $z=0^-$ by
\begin{eqnarray}
\mathcal{E}_{j+1}(z,k) = M_p(z, jl_c) M_j(jl_c,0^+)M(0^{+},0^{-})
\mathcal{E}(0^-,k). \nonumber \label{EWFz}
\end{eqnarray}
$M_p(z, jl_c)$, as in section \ref{EigenvalEigenvec}, is the transfer matrix for part of a unit cell, and $M_j(jl_c,0^+)$ the
transfer matrix for the first $j$-cells, thus
\begin{equation}
M_j(jl_c,0^+)=[M(l_c,0^+)]^j.
\end{equation}
Here $M(l_c,0^+)$ is the single cell transfer matrix
\begin{eqnarray}
M(l_c,0^+) \equiv \left(
\begin{array}{cc}
\alpha &
\beta \\
\beta^* &\alpha^*
\end{array}
\right)= \left(
\begin{array}{cc}
\alpha_b e^{ik_1l_1} & \beta_b \\ \beta_b^* & \alpha_b^*
e^{-ik_1l_1}
\end{array}
\right).
\end{eqnarray}
Assuming $\mu_1=\mu_2=\mu_0=1$, the unit cell matrix elements reduce to
\begin{eqnarray}
\alpha_b = \cos (k_2l_2) +\frac {i}{ 2} \left(\frac{n_2}{ n_1} +
\frac{n_1}{
n_2}\right) \sin (k_2l_2), \hspace{0.2in}{\rm and} \hspace{0.2in}
\beta_b = \frac{i}{2} \left(\frac{n_2}{n_1} - \frac{n_1}{n_2}\right) \sin (k_2l_2),
\end{eqnarray}
and, for the transfer matrix elements of $j$ cells, we have
\begin{eqnarray} \alpha_j =
U_j(\alpha_R) - \alpha^* U_{j-1}(\alpha_R), \hspace{0.2in}{\rm and} \hspace{0.2in}
\beta_j = \beta U_{j-1}(\alpha_R)  =  \beta_b
U_{j-1}(\alpha_R) .
\end{eqnarray}
Outside the optical superlattice (OSL) the electric field, assuming an incoming wave of unit amplitude at
$z < 0$, is
\begin{eqnarray}\label{EFleft}
E(z,k) = e^{ikz}\left(1 + r_T \right).
\end{eqnarray}

\begin{figure}
\begin{center}
\includegraphics[angle=0,width=15cm]{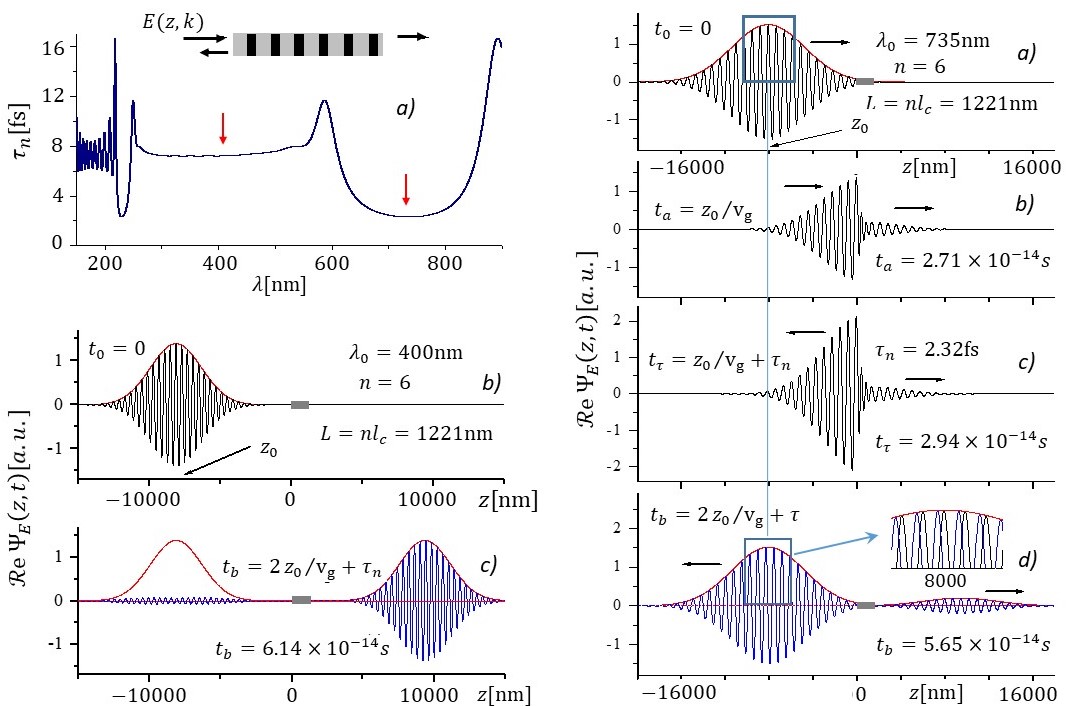}
\caption{Superluminal effect. Phase time for an optical SL with six unit cells with silica and titanium oxide alternating layers, and snapshots of WPs, which centroids are in the allowed and forbidden regions of the OSL. In $a)$, upper left, the phase time for the field component $E(z,k)$ as function of the wavelength $\lambda=k/2\pi$ through the OSL with refractive indices $n_1$=1.41 and $n_2$=2.22 and layer widths $l_1$=124.46nm, $l_2$=79.06nm, respectively. In panels $b)$ and $c)$, left, the WP in the allowed region, almost completely transmitted. At $t=0$, the WP centroid at a distance $z_0$ from the left edge of the OSL, and at $t=2z_0/v_g+\tau_n$, with $v_g$ the light velocity. The centroids of the enveloping red curves are at distances $z_0$ and $z_0 + n l_c v_g$ from the origin. In the panels at the right, a WP in the photonic gap. In $a)$, the WP at $t=0$, centered at $\lambda_0 = 735nm$ and
located at $-z_o=-40l_c$. In $b)$ and $c)$ the WP at
$t_a=z_0/v_g$ and $t_{\tau}=z_0/v_g + \tau_n$, where $\tau_n=2.32fs$ is
the phase time. The WP at $t_b=2 z_0/v_g + \tau_n$ has the same
enveloping curve as the WP at $t=0$, despite the phase shift shown
in the inset, which shows that the actual spent time $\Delta t$ described by
Maxwell equations agrees with the phase time $\tau_n$ and with
superluminal velocities. The enveloping curves are just for
a guide. Figure reproduced with permission.\textsuperscript{[\onlinecite{Simanjuntak2007}]} 2007, Physical Review E.}\label{TimSeref}
\end{center}
\end{figure}
Here $r_T = - {\beta_T^*/\alpha_T^*}$ is the reflection amplitude,
with
\begin{eqnarray}
\alpha_T^* &=& \alpha_{nR} - \frac{i}{2} \left(\frac{n_0}{n_1} +
\frac{n_1}{n_0}\right) \alpha_{nI} + \frac{i}{2} \left(\frac{n_0}{
n_1} - \frac{n_1}{n_0}\right) \beta_{nI}\nonumber\\
\beta_T^* &=& \frac{i}{2} \left(\frac{n_0}{ n_1} -\frac {n_1}{n_0}\right) \alpha_{nI} -\frac {i}{2} \left( \frac{n_1}{n_0} +
\frac{n_0}{n_1}\right) \beta_{nI}.
\end{eqnarray}
and the subindices $R$ and $I$, as used before,  stand for the real and imaginary parts.
The electric field in cell ($j+1$) takes the form

\begin{eqnarray}\label{EFin}
E_{j+1}(z,k) &=& \frac{1}{2} \left[(\alpha_p + \gamma_p) \alpha_j +
(\beta_p +
\delta_p)\beta_j^*)\right] \times \left[\left(1 + \frac{n_0}{ n_1}\right) + \left(1 -
\frac{n_0}{n_1}\right) r_T \right] \nonumber\\
&& + \frac{1}{2}\left[(\alpha_p + \gamma_p)\beta_j + (\beta_p +
\delta_p)\alpha_j^*)\right]\times\left[\left(1 - \frac{n_0}{ n_1}\right) + \left(1 +
\frac{n_0}{n_1}\right)r_T \right]
\end{eqnarray}
where $\alpha_p, \beta_p, \gamma_p, \delta_p$ are the elements of
$M_p(z, jl_c)$. In the transmitted field region, i.e. at $z > nl_c$, the  electric
field becomes
\begin{eqnarray}\label{EFright}
E(z,k) = \frac{1}{\alpha_T^*} e^{ikz}.
\end{eqnarray}

Using Eqs. (\ref{EFleft}), (\ref{EFin}) and (\ref{EFright}) in Eq.
(\ref{PgaussT}) we have the WP described by Maxwell equations in
space and time. It is worth mentioning that in these equations,
the multiple scattering processes are rigorously taken into account.
As time develops the WP evolves in a rather complicated {\it but
describable} way. The actual behavior depends critically upon the
OSL parameters and the WP characteristics. To visualize the space-time evolution, crucial snapshots at some specific values of time will be taken. If the wave packet centroid at $t$=0 is at $z_0$ and the left edge of the SL is at $z=0$ Important snapshots are  at: 1) $t_0=0$,
2) $t_a=z_o/v_g$, 3) $t_{\tau}=t_a+\tau$, when (according to the
phase time prediction) the centroid must be just {\it leaving} the
SL and, 4) at $t_b=2z_o/v_g+\tau$, when the reflected (transmitted)
WP should be located at $-z_0$ (or at $z_0+L$). The phase time was defined by Bohm \cite{Bohm1951} and Wigner \cite{Wigner1955} as the frequency
derivative of the transmission amplitude’s phase $\theta_t$, i.e., as
\begin{equation}
\tau=\frac{\partial \theta_t}{\partial \omega}
\end{equation}
where $\theta_t $ is the phase of the transmission amplitude. This is one of various formulas proposed for the theoretical calculation of tunneling times, among them was the B\"uttiker-Landauer tunneling time.\cite{Buttiker1982,Buttiker1983,Hauge1989} It was shown in Ref. [\onlinecite{Pereyra2000}], reviewed briefly in the next section, that the phase time describes within the error bars of the experimental results of Steinberg et al.\cite{Steinberg1993} and those of Spielmann et al.\cite{Spielmann1994}.

In Ref. [\onlinecite{Simanjuntak2007}] the time series was calculated for a specific superlattice where silica with low ( $n_1$=1.41) and titanium oxide with high ($n_2$=2.22) refractive indices alternate. In figure  \ref{TimSeref} some quantities are plotted for the electromagnetic WP through this low-high index superlattice (LHSL). In the upper left panel $a)$ of figure  \ref{TimSeref}, the pase time $\tau_n$ is plotted as function of the field wavelength, when the silica and titanium oxide layers  widths are $l_1$=124.46nm, $l_2$=79.06nm, respectively. In this phase-time spectrum, two regions of almost constant phase time are indicated with red arrows. One in the allowed band and one in the photonic gap. The wave packets shown in the other panels are defined in these wavelength regions.

In panels $b)$ and $c)$, at the left, the WP centroid is in the allowed band. In these panels the WP is shown at $t_0$ and when it comes back at $t_b$. The Gaussian envelopes (red curves) are plotted BY assuming that the WP centroid moves with light velocity outside the SL and inside the SL spends the phase time $\tau_n\simeq 7.2$fs, instead of $\simeq$4fs that would be required if it were to move with light velocity. The agreement with the assumption that the time spent is $\tau_n$, is perfect. More details are shown in Ref. [\onlinecite{Simanjuntak2007}].

In the time series of the panels in the right side of figure \ref{TimSeref}, we have the snapshots of the WP defined in the photonic band gap, with centroid at $\lambda_0 = 735 nm$. The
width is adjusted such that the main part of the WP lies in the photonic gap and in a
region of almost constant phase time, $\tau \approx 2.32 fs$, which
is close to the experimental\cite{Steinberg1993} tunnelling time
$\tau_{ex}\simeq2.1\pm0.2fs$ and the predicted phase
time\cite{Pereyra2000} ($\tau \simeq 2.3 fs$), for a slightly different
system with $n=5$. As is well-known, these results imply the striking
{\it superluminal velocities}. In \ref{TimSeref} (b) we have the WP
at $t_a$ touching the left-hand edge of the SL. In \ref{TimSeref} (c)
the WP at $t_{\tau}=z_0/v_g+\tau$. Because of the phase shift
due to the interference between the ``arriving" and ``leaving" packets, the coincidence
is not easy to visualize. But it is much easier to compare the wave packets in (a)
and (d). In the inset both
packets are plotted, and even though the packets coincide there is a phase shift. However, the Gaussian enveloping of the reflected WP at $t_b=2z_0/v_g+\tau$ is
also the enveloping of the WP at $t=0$. This coincidence shows that the actual time spent by the WP inside the SL,
$\Delta t=t_b-2t_a$, agrees with the phase time prediction.

\subsection{EM wavepackets through metamaterial SLs. Optical antimatter effects.}

\begin{figure}
\begin{center}
\includegraphics[angle=0,width=9cm]{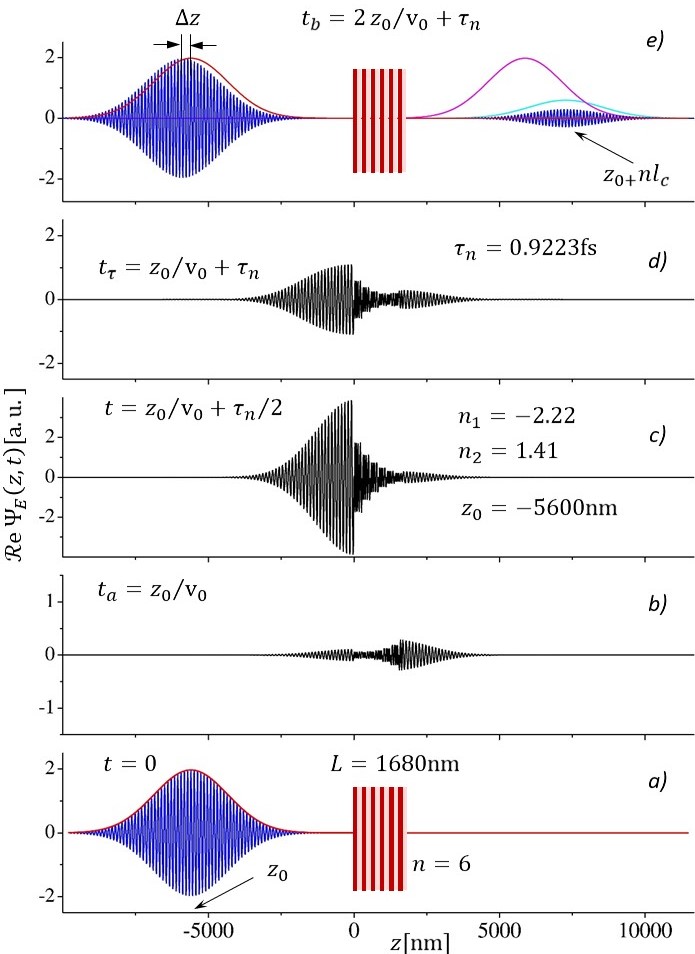}
\caption{Space-time evolution of a WP through an $LRSL$ with indices $n_1$=-2.22 and $n_2$=1.41, $l_1$=$l_2$=140nm, and $n=6$. In $a)$, the wave packet at $t=$0. In panel $b)$ the WP at $t_a=z_0/v_g$, when theoretically the centroid is reaching the left side of the SL. To visualize the WP in this panel, its amplitude was slightly amplified. As will be seen in the next figure, because of canceled layer, due to the antimatter effect, the wave packet gets transmitted slightly before it really reached the SL. In panel c) the WP at $t_a+\tau_n/2$, with the phase time $\tau_n=$0.9223fs much smaller than the time $L/c=$5.6nm, it would require if it moves with light velocity. In panel c) the WP should be leaving the SL. It certainly seems that the halves of the WP are already outside the SL. Notice that in this panel the reflected WP amplitude is larger than the incoming one. In d), at $t_b=2 z_0/v_g + \tau_n$ the WP should be back, i.e. under the
enveloping curve of the WP at $t=0$, however, the WP centroid is a distance $\Delta z$ farther because of the optical antimatter effect better visualized in the next figure.Figure reproduced with permission.\textsuperscript{[\onlinecite{Simanjuntak2007}]} 2007, Physical Review E.}\label{TotRefMetMat}
\end{center}
\end{figure}
The possibility of having negative refractive index, in left-handed materials (LHM), becomes a controversial and active research field, where striking and new effects are expected. Veselago,\cite{Veselago} already in 1968, predicted unique properties of EM wave propagation in LHM: a) waves appear to propagate towards the source and not away from it, b) negative group velocity and, c) converging and diverging lenses exchange their roles because waves incident on one side of the normal to right/left-hand interfaces are refracted to the same side. For the energy transferred from the source to the load to be positive and to avoid causality violations, Veselago proposes the constraints:
\begin{eqnarray}
\frac{\partial \epsilon \omega}{\partial \omega} > 0 \,\,\,\,\,\,\,\,\,\,\,\,\,\,\,\,\,\,\frac{\partial \mu \omega}{\partial \omega} > 0 \label{veselago}
\end{eqnarray}
Here $\epsilon$ and $\mu$ are the material's permittivity and permeability, respectively. Smith and Kroll,\cite{Smith} maintain that while a reversed $k$ resembles a time-reversed  propagation towards the source, the work done is nevertheless positive.

Besides the blazing presumption of perfect lenses,\cite{Pendry} and problems like the sign selection and directions of motion of the energy and the electromagnetic field, an overwhelming research activity, from basic to applied science, has developed, with significant advances in  manufacturing artificial left-handed material and structures, as well as in specific calculations for transmission of electromagnetic waves through LH media.\cite{Gupta,Zhang2005,Perrin2005,grzegorczyk2006,vinogradov2008,Moser2011}  Robledo et al. \cite{RobledoMorales2008,Pereyra2011} studied  the transmission of electromagnetic waves through LH and RH structures by using the transfer matrix method and, by applying the TFPS for left and right handed superlattices (LRSL). It has been explicitly shown that the transmission amplitude
\begin{eqnarray}
t_L=\left[\cos {(k_1d_1\cos \theta_1)}+\frac{i}{2|n_1|n_2}\left(n_2^2\frac{\cos {\theta_1}}{\cos {\theta_2}}+n_1^2\frac{\cos {\theta_2}}{\cos {\theta_1}}\right)\sin {(|k_1|d_1\cos \theta_1)}\right]^{-1}
\end{eqnarray}
of a single left-handed slab is just the complex conjugate of the transmission amplitude \begin{eqnarray}
t_L=\left[\cos {(k_1d_1\cos \theta_1)}-\frac{i}{2n_1n_2}\left(n_2^2\frac{\cos {\theta_1}}{\cos {\theta_2}}+n_1^2\frac{\cos {\theta_2}}{\cos {\theta_1}}\right)\sin {(k_1d_1\cos \theta_1)}\right]^{-1}
\end{eqnarray}
of a similar but right-handed slab. In these equations, $\theta_1$ is the incidence angle and $d_1$ is the slab width. An important consequence of this property is that the phase time $\tau$ of a single slab, becomes the negative of the corresponding phase time of a right-handed slab. This implies, in general, negative transmission times, and warnings on possible causality violation.

\begin{figure}
\includegraphics[width=15cm]{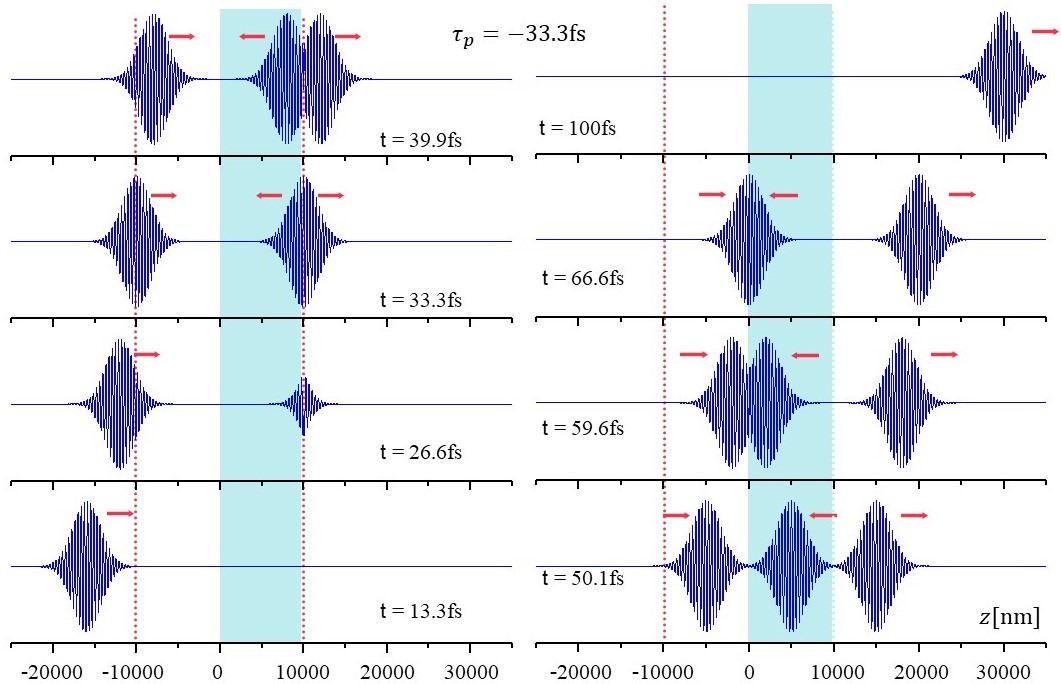}
\caption{Optical antimatter effect. Snapshots of the Gaussian wave packet at different times. At $t=0$, left bottom, the WP moving towards a negative-indices slab (with thickness $d=10\mu$m) and negative phase time $\tau_p=-33.3$fs. As soon as the WP tail reaches the point at $z=-d$, the WP tail starts to appear in the far edge of the slab, as transmitted WP. At time $t=33$fs, two Gaussian packets are seen, one at the far edge of the slab, and one  touching the edge of the ``canceled layer", between the dotted lines. It is also noticeable that the gaussian WP at the far edge contains two half packets, one moving towards the left and the other, the transmitted packet, moving towards the right. The transmission occurs as if the negative-indices slab and the canceled layer (between the dotted lines) would not exist. Figure reproduced with permission.\textsuperscript{[\onlinecite{Pereyra2011}]} 20011, Europhysics Letters. }\label{timeseries}
\end{figure}
To follow the space-time evolution of a gaussian electromagnetic wave packet (GEWP), we plot in figure \ref{TotRefMetMat} the function ${\Re} {\rm e} \Psi_E(z,t)$ defined in (\ref{PgaussT}) for different values of $t$. The electric fields $E (z,k)$, outside and inside the negative indices medium, are solutions to the corresponding Maxwell equations. In the time series of figure \ref{TotRefMetMat}, the WP is defined with centroid in the middle of a forbidden region of the transmission coefficient of its field components $E(z,k)$ through a left/right superlattice $LRSL$ with indices $n_1$=-2.22 and $n_2$=1.41, $l_1$=$l_2$=140nm, and $n=6$. In $a)$, the wave packet at $t=$0 and an enveloping Gaussian for reference in panel d). In panel $b)$ the WP at $t_a=z_0/v_g$, when theoretically the centroid is reaching the left side of the SL. To visualize the WP it was slightly amplified. As will be seen  in figure \ref{timeseries}, the cancellation of a normal medium layer, due to the antimatter effect, causes the wave packet to transmit and reflect a little before it actually reaches the SL. In panel c) the WP at $t_a+\tau_n/2$, with $\tau_n=$0.9223fs the phase time, much smaller than the time $L/c=$5.6nm IT  would require if it were to move with light velocity. In panel c) the WP should be leaving the SL. It seems certainly that halves of the WP are outside the SL. Notice that the reflected WP amplitude is larger. In d), at $t_b=2 z_0/v_g + \tau_n$ the WP should be back, under the enveloping curve of the WP at $t=0$. The WP centroid is distance $\Delta z$ farther which can be explained by the optical antimatter effect.

In Ref. [\onlinecite{Pereyra2011}] the optical antimatter effect was analysed. To understand the results shown in figure \ref{timeseries} where a GEWP evolves near and through a negative refraction index slab, it is worth a brief comment on the implications of the phase time predictions for this system. Outside the slab, where the  relative
electric permittivity and the relative magnetic permeability,
$\epsilon_{r1}=1$ and $\mu_{r1}=1$, the wave packet moves with the velocity of light $c$. This means that the packet centroid, which at $t=0$ is located at $z_o=-2d=-20\mu$nm, will reach the left-hand side of the slab at  $t_a=66.6$fs. The phase time predicts that the WP centroid will spend a negative time $\tau_p=-33.3$fs to pass through a slab with relative
electric permittivity and relative magnetic permeability,
$\epsilon_{r2}=-1$ and $\mu_{r2}=1$. This prediction implies that the wave-packet centroid should be leaving  the right edge of the slab at $t_{lp}=t_a+\tau_p=66.6-33.3$fs $=33.3$fs. This result is counterintuitive because it implies that,  at $t=33.3$fs,  the wave packet peak would be half-way
to the planar slab (i.e. at  $z=-10\mu$nm), but also departing from the right hand side of the slab (at  $z=10\mu$nm). In other words, we will see the WP  at two different points at the same time!

In the left and right columns of figure \ref{timeseries} we present a set of
snapshots of the first $100$fs. In the bottom left panel, at $t=13.3$fs, the WP is on its way to the negative indices slab. At $t=26.6$fs, when the Gaussian peak is not yet half-way to the slab, one can observe the formation, in the neighborhood of the right hand side of the slab, of a kind of image of the Gaussian wave packet. If we observe the snapshot at and after $t=39.9$fs, it is clear that the transmitted WP started to be formed before the centroid of the incoming WP reaches the left handed medium. It is interesting to observe the GEWP at $t=t_{lp}=33.3$fs. At this time we have two Gaussian wave packets. The Gaussian packet in the left, with peak at $z=-10\mu$m, is the original GEWP half-way to the slab. The other Gaussian packet contains two halves separated by the dotted red line at the interface between the slab and the semi-infinite medium. The half packet inside the slab moves to the left, and the half packet outside the slab (inside the semi-infinite medium) moves to the right. The former half corresponds to the image of the GEWP, and the other half corresponds, as can be seen in the other panels (for $t>33$fs), is the transmitted Gaussian wave packet. These graphs are compatible with the phase time prediction and show a way in which the incoming wave packet, its image and the transmitted one can be seen, at the same time, at different points of the space. Mathematically, this is a natural consequence of the continuity conditions imposed at the interfaces. Through these conditions each part of the system becomes aware of the physical properties of the other parts.

\begin{figure}[t]
\includegraphics[width=14cm]{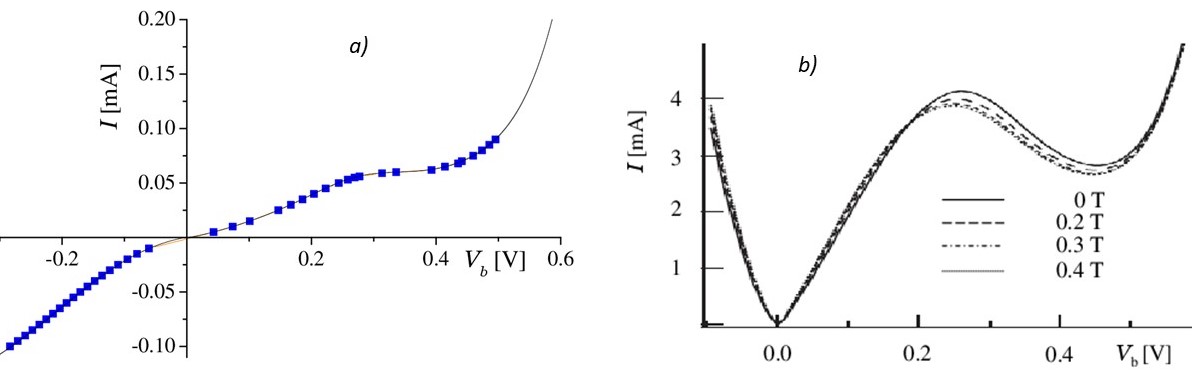}
\caption{I-V characteristics in Ga(Mn)As/GaAs spin injectors. In a), the I-V Characteristics measured in Ref. [\onlinecite{Shiogai}] for the spin injector $p^+Ga_{0.95}Mn_{0.05}As/n^+$-$GaAs$ with stop layer. In b), magnetic field effects on the  I-V characteristic of the  Ga(Mn)As/GaAs structure measured in Ref. [\onlinecite{Holmberg}]. Figure reproduced with permission.\textsuperscript{[\onlinecite{Shiogai}]} 2014, Physical Review B.}\label{figShiogai}
\end{figure}

\subsection{Exchange energy and spin injection through Esaki barriers}

The ability to inject, manipulate, and detect spin-polarized
carriers has been the main goals of
semiconductor spintronics.\cite{Datta1990} Different proposals by countless research groups have been published\cite{Matos2007}. Spin injection and detections represent an important part of the problem, and different in nature to the spin manipulation problem.  Despite the big progress, the problem of producing an efficient all-semiconductor all-electrical injection and detection device, remained rather elusive. This problem has slowed down the development of actual devices to control and manipulate the spin-filter and spin-inversion mechanisms.

The theoretical suggestion that larger contact resistance $R_c$  may increase the ferromagnetic/nonmagnetic (F/N)  spin injectors efficiency,\cite{Rashba2000,Fert1, Smith2001,Takahashi} promoted the research on  all-semiconductor  Mn-based  ferromagnetic semiconductors,\cite{VanEsch1997,Ohno1998, Ohno1999} and opened up the possibility to inject spin-polarized electrons using the Esaki diode $p^+(Ga,Mn)As/n^+GaAs$.

In  figures \ref{figShiogai} $a)$ and $b)$, low-bias IV characteristics obtained by Shiogai et al.\cite{Shiogai} and by Holmberg et al.\cite{Holmberg}, respectively, are shown. A feature of interest in these graphs is the structure below and around 0.4eV. This and the negative resistance behavior as a function of bias, lead us to study the injection and detection as a spin tunneling problem through bias and spin-dependent Esaki-barriers. To understand the IV characteristics for spin-up and spin-down through a dynamical barrier where, besides the bias potential and build-up interface potential, one has to consider the strong exchange interaction energy and the impurity states in the gap, we undertake a transfer matrix calculation based on the transfer matrix method. It was shown in Ref.[\onlinecite{Pereyra2014}] that the turning points of the negative magnetoresistance  are closely related  with the exchange energy.  The spin-splitting on the ferromagnetic side leads, naturally, to vanishing of tunneling probabilities at bias threshold points, where the propagating modes become evanescent. In  forward bias, the spin up transmission coefficient, $T_{E\,\uparrow}$, vanish when the Fermi level, of the $n$ side, aligns with the spin up valence band edge.

\begin{figure}[ht]
\includegraphics[width=16cm]{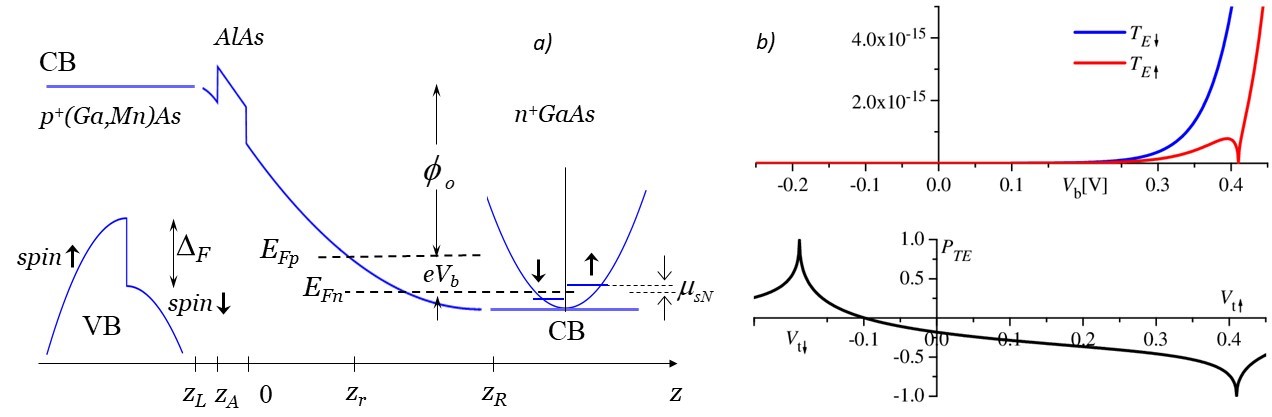}
\caption{ In $a)$, schematic band structure of a biased Esaki potential barrier at the interface of the ferromagnetic/stop layer/nonmagnetic structure, with spin-split offset $\Delta_F$ in the valence band of the $p$ side and the non-equilibrium spin accumulation electrochemical potential  $\mu_{sN}$ in the $n$ side. In $b)$, the pin-up and spin-down detection transmission coefficients, when $\Delta_F$= 0.6eV and $E_{Fp}$  is 0.112eV below $E_{p}^0$. The vanishing of $T_{E\,\sigma}$ at the threshold potentials $V_{t\uparrow}$ and $V_{t\downarrow}$, gives rise to  the transmission  spin polarization in the lower graph. Figure reproduced with permission.\textsuperscript{[\onlinecite{Pereyra2014}]} 2014, Physical Review B.}\label{structureF/I/N}
\end{figure}
In figure \ref{structureF/I/N} $a)$, the band structure at the interface   $p^+Ga_{0.95}Mn_{0.05}As/Al_{0.36}Ga{0.64}As/n^+GaAs$ is shown. At this interface, an Esaki barrier is formed with classical return points at $z_L$ and  $z_r$. The quantum nature of this contact has been systematically ignored by semi-classical approaches, which generally restrict themselves to recognizing it as a resistor with some numerical value. The physics in this contact, however, is a truly quantum phenomenon.
For the accurate calculation of   transmission coefficients, the transfer matrix method (TMM) in the  WKB approximation was used (details can be seen in [\onlinecite{PereyraWeiss}]). In reverse  (injection) and forward (extraction) configuration, the corresponding $4\times 4$ transfer matrices fulfill the relations
\begin{eqnarray}
\Phi(z_r)=M_I(z_r,z_L)\Phi(z_L)=\left(\begin{array}{cc}\alpha_{I}
&\beta_{I}\cr \gamma_{I} & \delta_{I}\end{array}\right) \Phi(z_L)
\end{eqnarray}
and
\begin{eqnarray}
\Phi(z_L)=M_E(z_L,z_r)\Phi(z_L)=\left(\begin{array}{cc}\alpha_{E}
&\beta_{E}\cr \gamma_{E} & \delta_{E}\end{array}\right) \Phi(z_L).
\end{eqnarray}
The state vectors
\begin{eqnarray}
\Phi(z)=\left( \varphi_{\uparrow}^+, \varphi_{\downarrow}^+,\varphi_{\uparrow}^-, \varphi_{\downarrow}^- \right)^T,
\end{eqnarray}
in the propagating modes representation, are written in terms of the propagating wave functions ($\sigma$ stands for spin up $\uparrow$ or spin down $\downarrow$)
\begin{eqnarray}
 \varphi_{\sigma}^+(z)=\frac{a_{\sigma}}{\sqrt{k_{\sigma}}}e^{i k_{\sigma}z}\chi_{\sigma} \hspace{0.1in} {\rm and }\hspace{0.1in} \varphi_{\sigma}^-=\frac{b_{\sigma}}{\sqrt{k_{\sigma}}}e^{-i k_{\sigma}z}\chi_{\sigma} ,
\end{eqnarray}
and the evanescent functions
\begin{eqnarray}
 \varphi_{\sigma}^+(z)=\frac{a_{\sigma}}{\sqrt{q_{\sigma}}}e^{q_{\sigma}z}\chi_{\sigma} \hspace{0.1in} {\rm and }\hspace{0.1in} \varphi_{\sigma}^-=\frac{b_{\sigma}}{\sqrt{q_{\sigma}}}e^{-q_{\sigma}z}\chi_{\sigma} .
\end{eqnarray}
Here
$
k_{\uparrow,\downarrow}\!=\!\sqrt{\frac{2m^*}{\hbar^2}\bigl(E_{Fi}-U_{\uparrow,\downarrow}(x,V_b)\bigr)}$ and $
q_{\uparrow,\downarrow}\!=\!\sqrt{\frac{2m^*}{\hbar^2}\bigl(U_{\uparrow,\downarrow} (x,V_b)-E_{Fi}\bigr)}$, the corresponding wave numbers with $E_{Fi}=E_{Fp}$ and $E_{Fn}$ the quasi-Fermi energies. Given the transfer matrices and the relations with the scattering amplitudes,
$t_I=( \delta_{I}^{\dagger})^{-1}$ and $ t_E=( \delta_{E}^{\dagger})^{-1}$,  the transmission coefficients
\begin{eqnarray}
T_{\sigma,\sigma'}=|t_{\sigma,\sigma'}|^2=|( \delta^{\dagger})^{-1}_{\sigma,\sigma'}|^2.
\end{eqnarray}
and the transmission spin polarization for spin extraction, defined as
\begin{eqnarray}
P_{TE}=\frac{
T_{E\uparrow}-
T_{E\downarrow}}{T_{E\uparrow}
+T_{\downarrow}},
\end{eqnarray}
were calculated. These quantities are plotted in the upper and lower panel of figure \ref{structureF/I/N}, respectively. The current spin polarization, spin injection efficiency and the spin accumulation defined as
\begin{eqnarray}\label{Spin-injectionAccum}
P_{j}\!\simeq\!\frac{j_{\uparrow}\!-\!
j_{\downarrow}}{j_{\uparrow}
\!+\!j_{\downarrow}} , \hspace{0.3in}   P_{jQ}\!\simeq\!\frac{j_{Q\uparrow}\!-\!
j_{Q\downarrow}}{j_{Q\uparrow}
\!+\!j_{Q\downarrow}}   \hspace{0.3in}{\text {and}}\hspace{0.15in} \delta \mu_s\propto-\!P_j,
\end{eqnarray}
where $j_{\sigma}$=$j_{I\sigma}$+$j_{E\sigma}$, and $Q=I,E$ (for injection and extraction), have similar behavior and, they are also very sensitive to exchange energy and the threshold potentials $V_{t\uparrow}$ and $V_{t\downarrow}$.

\begin{figure}[ht]
\includegraphics[width=16cm]{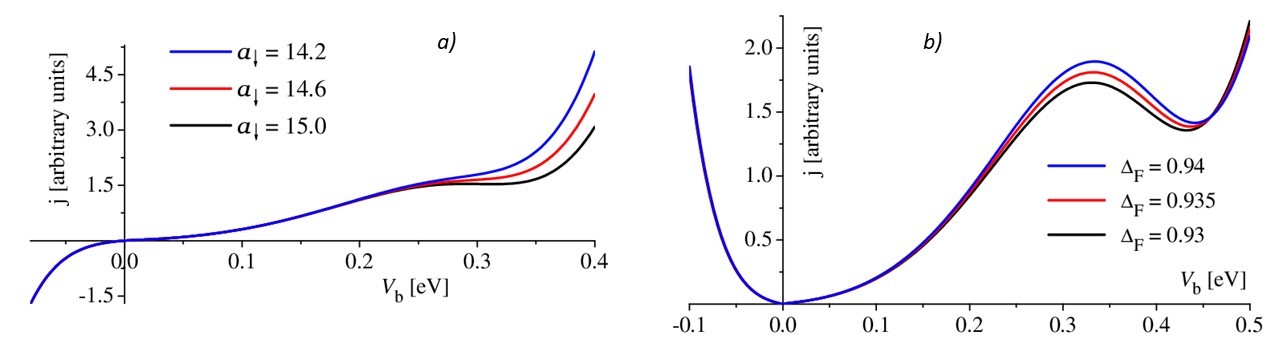}
\caption{a) The effect of the parameter $p$ on $\mathfrak{D}_{F\sigma}$,  in the neighborhood of the valence band edges. b) Total current through the Esaki barrier for the same parameters of figure \ref{structureF/I/N}, and the densities of states explained in the text. Figure reproduced with permission.\textsuperscript{[\onlinecite{Pereyra2014}]} 2014, Physical Review B.} \label{CurrentsDifp}
\end{figure}
The currents were evaluated, as usual, from
 \begin{eqnarray}
 j_{\sigma}^Q\!=\!\mathcal{G} \!\int \! \mathfrak{D}_{N\sigma}(E)\mathfrak{D}_{F\sigma}(E\!-\!eV_b) T_{\sigma\sigma}^Q(E,\!V_b)f_F(E)g_N(E)dE \nonumber
 \end{eqnarray}
where $\mathcal{G}$ is a geometrical factor, $\mathfrak{D}_{N\sigma}$ and $\mathfrak{D}_{F\sigma}$ the densities of states in the non-magnetic and ferromagnetic sides, and $f(E)$ and $g(E)=$1-$f(E)$ the occupation probabilities. Since the experiments temperature is $\sim 4K$, the occupation probabilities factor becomes $\delta (E-E_F)$. The currents shown in figure \ref{CurrentsDifp} were calculated assuming that in the ferromagnetic side the DOS is
$
\mathfrak{D}_{F\sigma}\propto e^{-a_{\sigma}\epsilon_{\sigma}}(a_{\sigma}\epsilon_{\sigma})^{p/2}
$
for $\epsilon_{\sigma}\!=\!E\!-\!E_{Fp}\!-\!\sigma \Delta_F/2>0$, and
$
\mathfrak{D}_{F\sigma}\propto (-\epsilon_{\sigma})^{q/2}
$
with $q\!=\!1$ for $\epsilon_{\sigma}\!<\!-\Delta_F$, $q\!=\!p$ for $\!-\Delta_F\!<\!\epsilon_{\sigma}\!<\!0$, being $a_{\sigma}$ and $p$ parameters chosen to fit the DOS's obtained numerically by Turek et al.\cite{Turek}, and $\mathfrak{D}_{N\sigma}= \mathfrak{D}_{K\sigma}\propto (E-E_{\rm o\sigma})^{p/2}$ (for $p=1$). Here $\mathfrak{D}_{K\sigma}$ is the Kane's $y$ function\cite{KaneEO1963}
\begin{eqnarray}
y(E)=\int_{-\infty}^E (E-E_{0\sigma}-\xi)^{p/2}\mathfrak{F}(\xi)d\xi.
\end{eqnarray}
As can be seen in the figures \ref{figShiogai}, currents account for the experimental behavior. These results show also that in the presence of exchange, the spin splitting and the DOS are strongly correlated quantities, and define the shape and  low-bias I-V characteristics of the F/N structures.

\subsection{The homogeneous magnetic superlattices}

\begin{figure}
\begin{center}
\includegraphics[width=11cm]{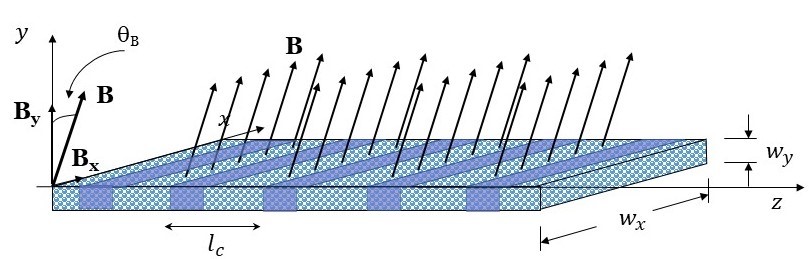}
\caption{A homogeneous magnetic superlattices where a tilted external magnetic field $\textbf{B}$, blocked in alternate (darker) stripes, acts also only on quasi-twodimensional stripes of thickness $d$ and transverse widths $w_x$ and $w_y$. Figure reproduced with permission.\textsuperscript{[\onlinecite{Cardoso2008}]} 2008, Microelectronics Journal. \label{HMSL}}
\end{center}
\end{figure}
The transfer matrix method and TFPS approach to spintronics is a different approach than mainstream, mostly semi-classical, ones. We will briefly review the TFPS applied to study the transport and spin control in magnetic superlattices.

The aim to control spin carriers transport through layered
magnetic structures, has been of much interest.\cite{Seikin,Young,Oestreich}. With this aim, in the presence of the phase coherence of a periodic magnetic field, Cardoso et al. carried out an research project that addressed different aspects of the interaction of spin $1/2$ carriers with magnetic-field superlattices, including issues such as Shubnikov-de Haas oscillations, the Land\'e $g$ factor\cite{Cardoso2005},  Fano resonances and the magnetic-field-tilting angle as a
significant mechanism for the spin transitions in magnetic superlattices. Taking advantage of these
spin-transport phenomena, simple and efficient spin-filter and spin-inverter devices were proposed.\cite{Cardoso2008}

For practical reasons, a
simple system was conceived, where only the external field changes periodically, that is, the same semiconductor under an external magnetic field whose amplitude varies periodically. \cite{Cardoso2001} We refer to this superlatice  as the homogeneous magnetic superlattice (HMSL), and represent it as $(L_FL_H)^n$, where $L_F$ stands for a stripe free of field and $L_H$ for a stripe under an external magnetic field.  As shown in figure \ref{HMSL}, the HMSL is basically a quasi two-dimensional semiconductor wave guide subject to an external magnetic field, which is blocked on stripes that alternate along the growing direction $z$. As is well-known,
the spin 1/2 electrons, moving in those regions subject to an external magnetic field $\bf B$, are described by
\begin{eqnarray}\label{pauli2}
\left[\frac{1}{2m^*}\left(\textbf{p} -\frac{e}{c}\textbf{A}\right)^2+\left(V(y,z)-E_{f}\right)\right]\Psi(\bf r) +
\left(\frac{1}{2}g^*\mu_{B} {\mbox{\boldmath{ $\sigma$}}}\cdot\textbf{B}  + V_{R} +V_{D} \right)\Psi(\bf r)=0
\end{eqnarray}
where $V(x,y)$ is a lateral confining potential, $V_{R}$ and $V_{D}$ are the Rashba and Dresselhaus spin-orbit
interaction. If we have a tilted magnetic field, say $ {\bf{B}}= B \left(  \sin \theta_{B}, \cos\theta_{B},0 \right) = B \hat{\textbf{B}}$, the spin precession and spin transitions induced by the Rashba and Dresselhaus interactions are negligible compared with those induced by the Zeeman interaction. Thus, for this review we drop those terms. It was shown in Ref. [\onlinecite{Cardoso2001}] that after some
analytical calculation, one is left with the equation
\begin{eqnarray}\label{ecx2}
\left(\frac{d^2}{dz^2}-\frac{z^{2}\cos\theta_{B}}{l_{B}^4}+ k_{z}^2-\frac{g}{2l_{B}^{2}}\!{\mbox{\boldmath{
$\sigma$}}}\!\cdot\!\hat{\textbf{B}}\!\right)Z(z)=0\mbox{,}
\end{eqnarray}
which solutions are the hypergeometric matrix functions\cite{Pereyra2002}
\begin{eqnarray}
{\cal{A}}_{\sigma}\left(z\right)=\mbox{}_{1}\textmd{F}_{1}\left(-\frac{{\textbf{b}}_{\sigma}}{2};\frac{1}{2};\frac{\cos\theta_{B}}
{l_{B}^2}z^2\right)\displaystyle{ e^{-z^{2}\!\cos\!\theta_{\!\small B}/2l_{B}^2}},
\end{eqnarray}
and
\begin{eqnarray}
{\cal{B}}_{\sigma}\left(z\right)=z\;\mbox{}_{1}\textmd{F}_{1}\left(\frac{{\textbf{1}}2-{\textbf{b}}_{\sigma}}{2};\frac{3}{2};\frac{\cos\theta_{B}
}{l_{B}^2}z^2\right)\displaystyle{ e^{-z^{2}\!\cos\!\theta_{\!\small B}/2l_{B}^2}},
\end{eqnarray}
In these functions, the matrix function ${\textbf{b}}_{\sigma}$ is
\begin{equation}\label{matriz-hyper}
{\textbf{b}}_{\sigma}=\frac{l_{B}^2}{2\cos\theta_{B}}\left(k_{z}^{2}-\frac{g}{2l_{B}^{2}}\!{\mbox{\boldmath{
$\sigma$}}}\!\cdot\!\hat{\textbf{B}}-\textbf{1}\right)\mbox{.}
\end{equation}
with $l_B$ the magnetic length and $k_z$ the longitudinal wave number. We will show here examples of distinct spin-projection transmission coefficients. To better visualize the effect of the spin-dependent transmission coefficients, we will plot the space-time evolution of Gaussian pulses of spin 1/2 electrons. To this purpose we need to calculate the wave packet
\begin{equation}\label{paquetegauss}
\psi_s\left(z,t\right)\!=\!\int_{-\infty}^{\infty}\! dke^{\!-\gamma\left(k-k_{0}\right)^2}e^{ik z_{0}}
{\phi}_s\left(z,k \right)e^{-iwt}
\end{equation}
which  centroid at $t_o=0$ is located in the position $-z_o$. The subindex $s$ refers to the spin projection.
For the evaluation of this integral, one requires to know analytically the wave functions ${\phi}_s\left(z,k \right)$, inside and outside the SL, based on the solutions  ${\cal{A}}_{\sigma}\left(z\right)$ and ${\cal{B}}_{\sigma}\left(z\right)$, and the transfer matrix method and the TFPS, to determine the transfer matrix $M(z,z_0)$ that connects the state vector at $z_0$ with  the state vector at any point $z$ inside and outside the superlattice, one can evaluate the scattering amplitudes
\begin{eqnarray}
\textbf{t}_{n}=\left[
               \begin{array}{lcr}
               t_{\uparrow\uparrow}&t_{\uparrow\downarrow}\\
               t_{\downarrow\uparrow}&t_{\downarrow\downarrow}
               \end{array}
               \right] \,\,\,\,\, \mbox{and}\,\,\,\,\,
\textbf{r}_{n}=\left[
               \begin{array}{lcr}
               r_{\uparrow\uparrow}&r_{\uparrow\downarrow}\\
               r_{\downarrow\uparrow}&r_{\downarrow\downarrow}
               \end{array}
               \right]
\end{eqnarray}
and the wave functions $\phi_{\uparrow}\left(z,k \right)$,
$\phi_{\downarrow}\left(z,k \right)$ inside and outside the superlattice.

In this approach, schematically reviewed here, one can obtain different type of effects by varying the SL parameters ad the magnetic field orientation. The behavior of the spin-up and spin-down transmission coefficients, $T_{uu}$ and $T_{dd}$, respectively, is an important guide to visualize in advance the outcome of desired effects for the WP through the HMSL. We will illustrate the HMSL acting as a spin filter, the tilting angle effect and the HMSL acting as spin inverter.

\subsubsection{The HMSL as spin filter}

\begin{figure}[ht]
\begin{center}
\includegraphics[width=17cm]{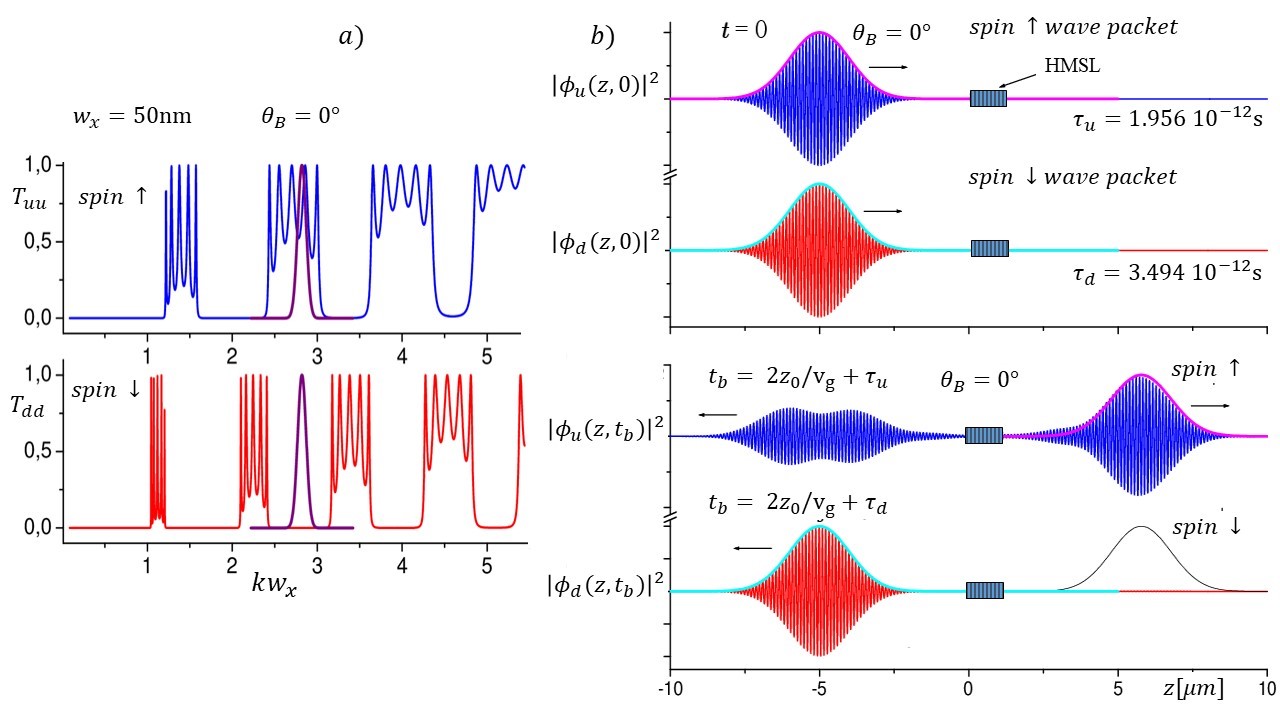}
\caption{Spin-up and spin-down transmission coefficients and Gaussian wave packets through a HMSL. In $a)$ the spin-up and spin-down transmission coefficients for $\theta_{\rm B}=0$ on a HMSL with $w_x=50$nm and $w_y=3$nm, $l_c=130$nm, $n$=6, and $B=$0.18T. The Gaussian curves in the gap of $T_{dd}$ and the allowed region of $T_{uu}$ show the position and shape of the unpolarized spin wave packet that will move through the HMSL. In the upper frame of $b)$, the spin$\uparrow$ and spin$\downarrow$ wave packets at $t=0$ and $-z_0=-40l_c$, moving towards the
HMSL. In the lower frames the spin$\uparrow$ and spin$\downarrow$ packets at $t=2t_a+\tau_u$ and $t=2t_a+\tau_d$,
respectively, when presumably, both packets should be
back at $-z_o$ (a distance $z_o$ from the HMSL). It is clear from these graphs {\it the filter effect} manifest by the absence of transmitted spin-down wave packet components.  The Gaussian curves centered at $-z_o$ and at $L+z_o$ are
plotted as reference. Here $L=nl_c=0.78\mu$m is the HMSL length. Figure reproduced with permission.\textsuperscript{[\onlinecite{Cardoso2008}]} 2008, Microelectronics Journal.}\label{spinfilter}
\end{center}
\end{figure}

In figure \ref{spinfilter} we have the transmission coefficients $T_{uu}$ and $T_{dd}$, in the left, and the spin wave packets in the right. The HMSL is a quasi-2D wave guide with widths $w_x=50$nm and $w_y=3$nm, under an external magnetic field with $B=0.18$T acting on stripes of width $L_H=10$nm, unit-cell width $l_c=130$nm, and $n=6$ unit cells. We consider first that $\theta_B=0$. In this case, no spin$\uparrow$ $\rightarrow$ spin$\downarrow$ transitions occur, i.e. $T_{ud}=T_{du}=0$. This implies well defined band structures for the spin$\uparrow$ and the spin$\downarrow$ transmission coefficients, as shown in figure
\ref{spinfilter}$a)$. In this example, an unpolarized electrons WP is prepared with centroid and wave-packet components entirely in the gap of $T_{d}$ and in the resonant region of $T_{u}$. We expect
that the HMSL will work as a spin filter for the spin$\downarrow$ electrons. Indeed, the snapshots in  figure \ref{spinfilter}$b)$, at $t=0$ and at $t=2z_0/v_g+\tau_{\sigma}$, show the filter effect.
At $t=0$, the spin$\uparrow$ and spin$\downarrow$ Gaussian wave packets with centroids located at $z_0=-40l_c$,
moving towards the HMSL. The centroids of these packets reach the left edge of the HMSL at $t_a=x_o/v_g$, with a group velocity $v_g$. In the lower frame we plot the spin packets at $t=2t_a+\tau_u$ and at
$t=2t_a+\tau_d$, when presumably they should be back at the starting point, having spent a time $\tau_u$ or $\tau_d$, respectively, in the HMSL. As expected, from the transmission coefficient behavior, the spin$\uparrow$ wave-packet components are partially reflected and partially transmitted, while those
of the spin$\downarrow$ packet are completely reflected.

It is interesting to notice that, even though a shape distortion is seen for the spin-up wave packet, the transmitted gaussian wave-packets
are in the position that one would expect them, if the phase-time prediction were correct and the same for most of the
spin$\uparrow$ Gaussian components. The tails of the Gaussian in figure \ref{spinfilter}$a)$ transmit less and have also different tunneling times, while the spin$\downarrow$ components, all have the same transmission coefficient and the same tunneling time (see lower panel in figure \ref{spinfilter}$a)$). For the centroid of the spin$\uparrow$ electrons,
the phase time predicts a tunneling time $\tau_u=1.995\,10^{-12}$s, while for spin$\downarrow$ electrons, it
predicts $\tau_d=3.483\,10^{-13}$s. These are quite different times and they are also rather different from the time
$\tau_f\simeq 1.625\, 10^{-12}$s that they would spent if they were to move (the same distance) in a region free of magnetic
fields. This example shows clearly
that the system behaves as a spin$\downarrow$ filter and the phase time predictions are correct.

\begin{figure}[ht]
\begin{center}
\includegraphics[width=16cm]{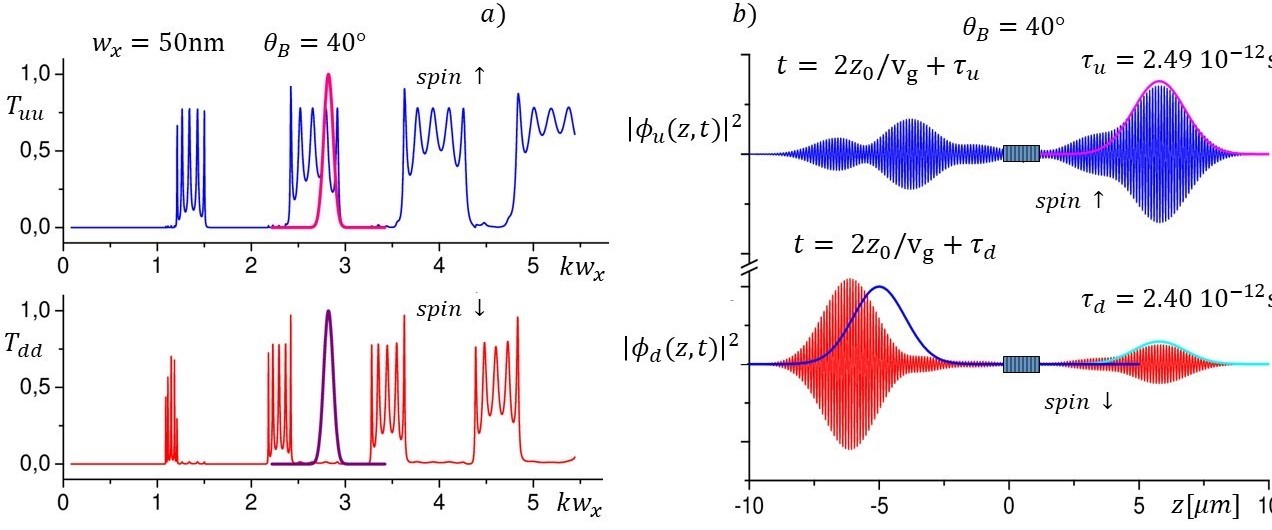}
\caption{Transmitted and reflected spin$\uparrow$ and spin$\downarrow$ wave packets when the magnetic field
tilting angle is $\theta_B=40$. The incoming packets are the same as in figure \ref{f.3}. The spin$\uparrow$ and
spin$\downarrow$ wave packets are plotted for $t=2t_a+\tau_u$ and $t=2t_a+\tau_d$, respectively. Although the
centroid phase time predictions, describes well the transmitted spin-packet tunneling time, it does not represent
the reflected spin-packet tunneling time because of the high wave-packet distortion. The Gaussian curves centered
at $-x_o$ and at $l_s+x_o$ are plotted for reference. Figure reproduced with permission.\textsuperscript{[\onlinecite{Cardoso2008}]} 2008, Microelectronics Journal.}\label{WPHMSL40}
\end{center}
\end{figure}

\subsubsection{Spin mixing and space-time evolution when $\theta_B=40^{o}$}

We consider again the unpolarized  wave packet of the previous example  but now with the magnetic field in the HMSL tilted with $\theta_B=40^{o}$. In \ref{WPHMSL40} $a)$ the transmission coefficients $T_{uu}$ and $T_{dd}$. Now, the reflection coefficient $R_{dd}$ (which was $\simeq 1$) reduces to $\simeq 0.7$ and the transmission
coefficients are smaller than 1, as shown in \ref{WPHMSL40} $a)$. An important amount of flux passes from the spin$\uparrow$ to the
spin$\downarrow$ packet, and vice versa. The spin$\uparrow$ $\rightarrow$ spin$\downarrow$ transitions are
induced by the in-plane field component $\bf{B}_y$. In addition, the band structures experience small red an blue shifts,
respectively. The snapshots in figure \ref{WPHMSL40}$b)$ show
the wave-packets at $t=2t_a+\tau_u$ and at $t=2t_a+\tau_d$, when their centroids are back at $-z_o$ or transmitted at $nl_c+z_o$. It is worth noticing the wave packet distortion is more evident in this example than in the previous
one. This is compatible with the phase time behavior. In the lower panel of figure \ref{WPHMSL40}$b)$, we observe a kind of
spin$\downarrow$ wave-packet hastening, as the reflected and transmitted packets appear beyond the position
expected if they were to move in the HMSL with the centroid's velocity. This can be explained because the phase
time of an important fraction of wave-packet components is smaller that the centroid's phase time. Those components leave the HMSL a bit earlier.
Because of the spin transitions, there are now reflected and transmitted wave packets for spin$\uparrow$ and for spin$\downarrow$ electrons.

\subsubsection{Spin inversion}

\begin{figure}
\begin{center}
\includegraphics[width=18cm]{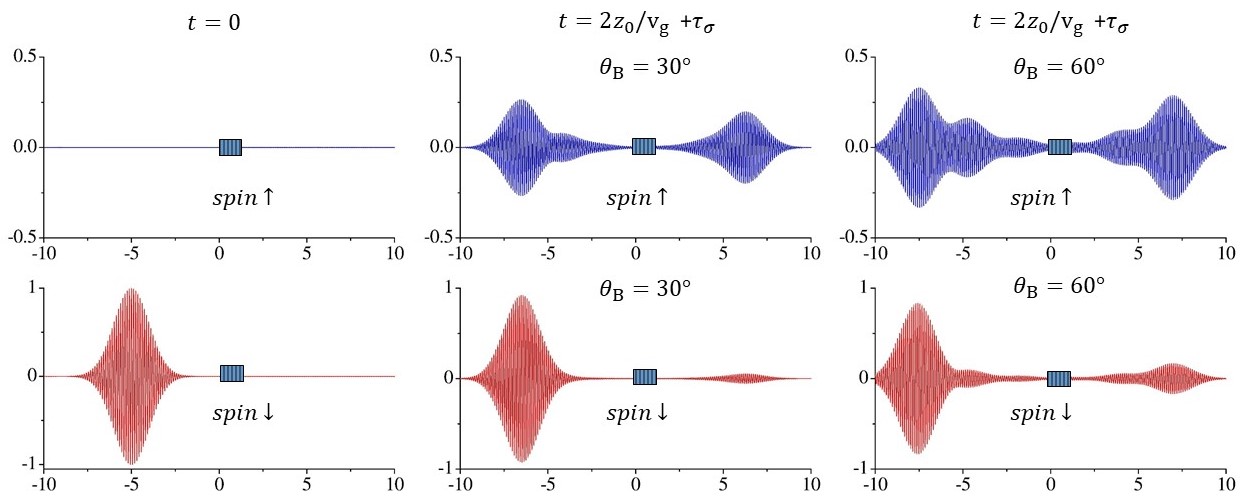}
\caption{Tilted magnetic field effect. In these graphs a completely polarized WP is sent at $t=0$ towards a HMSL. The reflected and transmitted WP remains polarized when $\theta_B=0^{o}$,  but a strong spin mixing is observed when  $\theta_B=60^{o}$.}\label{TiltingHMSL}
\end{center}
\end{figure}

In the previous examples we had at $t=0$ an unpolarized Gaussian wave packet defined in a spin$\downarrow$ gap. The spin mixing made possible to have a small fraction of spin$\downarrow$ electrons transmitted. We will see now that a spin inversion is also possible. In fact, if at $t=0$ we have a polarized Gaussian wave packet, say a spin spin$\downarrow$ WP in lower graph of \ref{WPHMSL40}$a)$. Tilting the field, as we have just seen, and as shown  in figure \ref{TiltingHMSL} an important amount of electrons with spin$\uparrow$ transmit the HMSL. This fraction grows with the tilting angle $\theta_{B}$. Eventually the population of transmitted spin$\uparrow$ electrons is equal and even larger than the population of spin$\downarrow$ electrons. All one needs to complete the spin inversion is a second step of spin filter.

\section{Highly accurate descriptions of high-resolution experiments: tunneling and blue emitting lasers}\label{TunnelingAndBlueLaser}

We will outline here two examples where the theoretical predictions not only compare extremely well with high-resolution experimental results, they provide also the correct theoretical description. The first example is on the tunneling time through the photonic-band-gaps of $(silica/titanium$-$oxide)^n$ superlattices. This example implies calculations of scattering amplitudes and transmission coefficients, and  was shown in Ref. [\onlinecite{Pereyra2000}], that using the results of the TFPS, the experimental results of Steinberg et al. \cite{Steinberg1993} and those of Spielmann et al. \cite{Spielmann1994}, are reproduced within the error bar of $10^{-16}$s. The second example is on the optical response of $(In_xGa_{1-x}N/In_yGa_{1-y}N)^n$ superlattices. This example implies the calculation of superlattice energy eigenvalues and eigenfunctions. In the 90's, S. Nakamura\cite{Nakamura3},  using monochromators with a resolution of $0.016$nm, reported the optical response with new features in the spectra that could not be explained with the standard approach. It was shown recently\cite{Pereyra2018}, that using the TFPS it was possible to explain the experimental spectrum features within the accuracy of the experimental results.

\subsection{Tunneling through the photonic-band gaps}

In the 90's, enlightening and accurate measurements of single-photon and optical-pulse delay times in the photonic band gap of multilayer media were performed by Steinberg et al. \cite{Steinberg1993} and by Spielmann et al. \cite{Spielmann1994}, and stimulated again the interest on the concept of tunneling time, that was highly elusive and controversial\cite{Wigner1955,Bohm1951,Buttiker1982,Hauge1989,Ranfagny1991,Enders1992,
Landauer1994}. The multilayer used to measure the tunneling time were based on quarter wave superlattices $(HL)^n$  where $L$ represents a quarter-wave layer of silica, with refractive index $n_L=1.41$, and $H$ a quarter-wave  titanium oxide layer with refractive index $n_H=2.22$. Since the issue of tunneling time through a potential barrier was first approached in 1932 by MacColl,\cite{MacColl1932} the stationary phase time and other definitions
for the tunneling time were proposed. In the absence of experimental results, the
concern about violations of the special theory of relativity and causality  prevented theorists from accepting the phase time as the tunneling time. The direct measurement  of tunneling times through optical superlattices  and the possibility of calculating accurately the superlattice scattering amplitude by using the TFPS raised the possibility to check whether the phase time describes or not the tunneling time.

\begin{figure}
\begin{center}
\includegraphics[width=16cm]{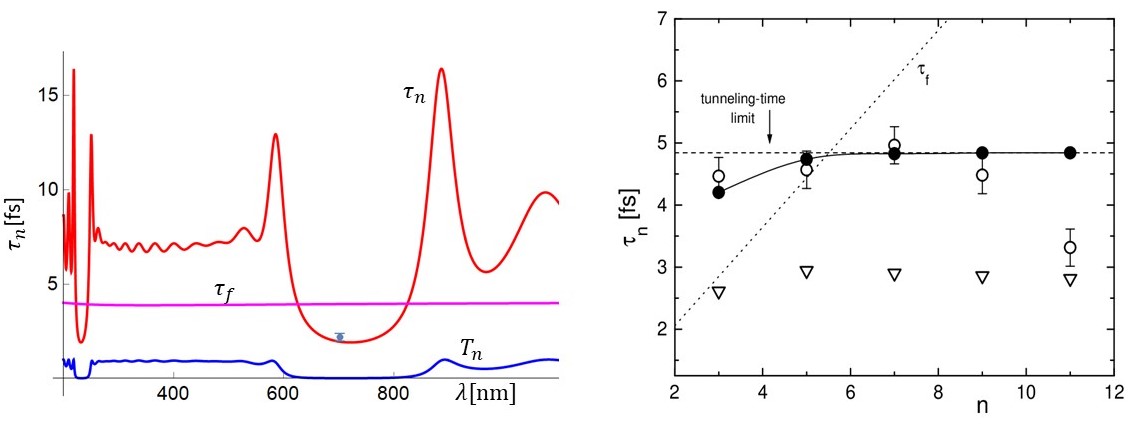}
\caption{Tunneling times through $(HL)^5H$ and $(LH)^nL$ for wavelengths in the photonic band gaps. In a) the tunneling times reported by Steinberg et al. through the SL $(HL)^nH(substrateL)$ for $n=5$, the calculated tunneling time $\tau_n$, and the transmission coefficient $T_n$. $\tau_f$ is the time that would be spent by moving with velocity $c$. In b) the tunneling times (open circles) reported by Spielmann et al. through $(LH)^nL$ SLs made of titanium oxide (H) and silica (L) for different values of $n$, accuracy of 10$^{-16}$s. The black circles are the predictions of the TFPS.  Figure reproduced with permission.\textsuperscript{[\onlinecite{Pereyra2000}]} 2000, American Physical Society.}\label{FigTTime}
\end{center}
\end{figure}
Given the scattering amplitude $t_n$ ($=t_{nr}+i t_{ni}$) in equation (\ref{1DScattAmplit}), and the definition of tunneling time
\begin{equation}
\tau_n =\left| t_n\right| ^{-2}\left( t_{nr}\partial t_{ni}/\partial \omega
-t_{ni}\partial t_{nr}/\partial \omega \right)
\end{equation}
the general and closed formula for the evaluation of this TIME IS given by
\begin{equation}
\tau _n=\frac \hbar {(U_n-\alpha_rU_{n-1})^2+(\alpha_iU_{n-1})^2}\left( A_r%
\frac{d\alpha_r}{dE}+A_i\frac{d\alpha_i}{dE}\right)
\end{equation}
with
\begin{equation}
A_r=\frac{\alpha_i}{1-\alpha_r^2}\left( \left( \alpha_rU_{n-1}+nU_{n-2}\right) U_n-(n+\alpha_r^2)U_{n-1}^2\right)
\end{equation}
and
\begin{equation}
A_i=(U_n-\alpha_rU_{n-1})U_{n-1}
\end{equation}
was reported in Ref. [\onlinecite{Pereyra2000}]. For the specific superlattice where the tunneling time was measured, the unit-cell transfer matrix elements are
\begin{eqnarray}
\alpha &=&\frac{e^{ik_Hd_H}}{4n_Hn_L}%
(e^{ik_Ld_L}(n_H+n_L)^2-e^{-ik_Ld_L}(n_H-n_L)^2)  \nonumber \\
\beta &=&i\frac{n_H^2-n_L^2}{2n_Hn_L}\sin k_Ld_L=i\beta _i
\end{eqnarray}
and adjusting the transfer matrix to include the extra layer $H$ or $L$ and the substrate of length $l_s$, it was possible to calculate the tunneling times shown in Figure \ref{FigTTime}. In Figure \ref{FigTTime}a) we show the tunneling time reported by Steinberg et al. through the SL $air(HL)^nH(substrateL)air$, for $n=5$,  together with the calculated tunneling time $\tau_n$, and the transmission coefficient $T_n$. $\tau_f$ is the time that would be spent by moving with velocity $c$. In Figure \ref{FigTTime}b) the tunneling times (open circles) reported by Spielmann et al. through $air(substrateL)(HL)^n air$ SLs made of titanium oxide (H) and silica (L) for different values of $n$, with error bar of 10$^{-16}$s. The black circles are the predictions of the TFPS. The open triangles are the calculated phase times by Spielmann et al.

\subsection{Optical response  of the blue emitting $(In_{0.05}Ga_{0.95}N/In_{0.2}Ga_{0.8}N)^10$ SLs. }

\begin{figure}
\begin{center}
\includegraphics[width=16cm]{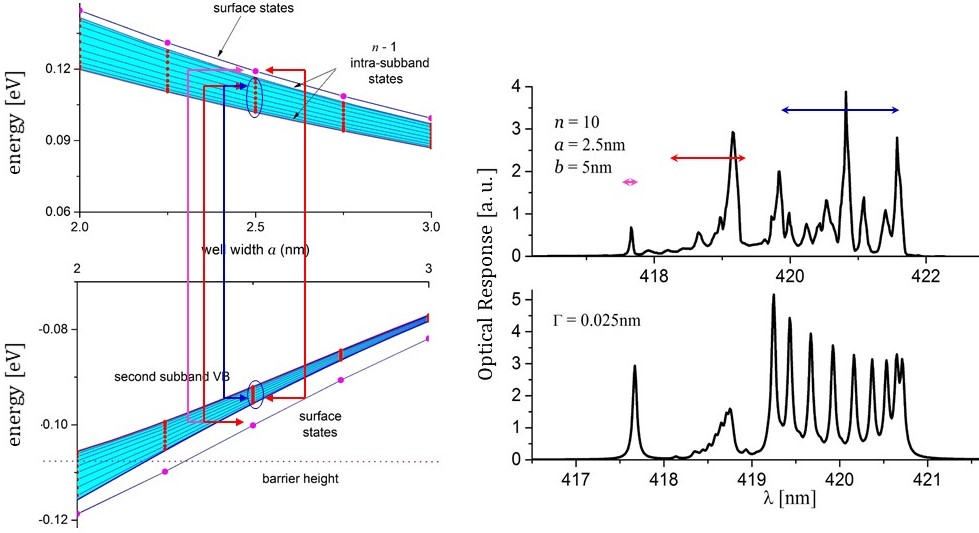}
\caption{SL spectra and optical response. In a) the energy levels in subbands of the conduction and valence bands. The continuous subbands (light blue) are from the Kronig-Penney model.\cite{Kronig1931} The discrete lines are the predictions of the TFPS. The set of transitions responsible of the group structure in the optical spectrum are also shown.  In b), upper panel IS the optical response  of the blue emitting $(In_{0.05}Ga_{0.95}N/In_{0.2}Ga_{0.8}N)^{10}$  SLs, and in the lower panel are the predictions of the TFPS. The experimental resolution was of 0.016nm.  Figure reproduced with permission.\textsuperscript{[\onlinecite{Pereyra2018}]} 2018, Annals of Physics.}\label{SLSpectAndOptTrans}
\end{center}
\end{figure}
An important development of the last years of the last century has been the production of blue light-emitting diodes, based on $GaN/InGaN$ superlattices. Following the discovery of the blue-emitting diodes, an overwhelming research activity began on the so-called wide-gap nitrides. At first, the attempts to produce the announced devices became a real challenge, afterwards, the explanation of the characteristics of the optical response spectra also became a challenge for the theorists. From a theoretical point of view, the continuous subbands predicted by the standard approaches were unable to explain and describe the main characteristics of the high-resolution optical spectra reported by Nakamura et al.\cite{Nakamura3,Nakamura2,Nakamura}, characterized by multiple and close resonances (see the upper right panel in the Figure \ref{SLSpectAndOptTrans}). Nakamura et al. suggested that the multiple resonances could be identified as longitudinal modes, but it was also argued that they could be fluctuations in the cavity field. Recently, using the theory of finite periodic systems and the ability to evaluate the eigenvalues and eigenfunctions of superlattices, it has been possible to account for the optical response of SLs in the active zone of light-emitting devices,  by evaluating the optical susceptibility in the approximation of the golden rule given by
\begin{eqnarray}\label{susceptPL}
\chi^r_{\small PL} = \sum_{\nu,\nu'\!,\mu,\mu'}f_{eh}\frac{\displaystyle \Bigl{|}\int dz
[{\it \Psi}^{r,v}_{\mu',\nu'}(z)]^*\frac{\partial}{\partial
z}{\it \Psi}^{r,c}_{\mu,\nu}(z)\Bigr{|}^{2}}{(\hbar \omega-E_{\mu,\nu}^{c}+E_{\mu',\nu'}^{v}+E_B)^{2}+\Gamma^{2}}\hspace{0.2in}
\end{eqnarray}

Here the eigenvalues and eigenfunctions from the conduction and valence bands are obtained from the explicit expressions  outlined in the last sections. As can be seen in the lower panel of figure \ref{SLSpectAndOptTrans}, the optical susceptibility is faithfully reproduced\cite{Pereyra2005,Pereyra2018}. An important characteristic of the subband spectra to understand the structure of the optical response is the detachment of the surface energy levels. The surface states
in the subband of the conduction band, with energies $E_{1,10} =$ 0.119074 eV and $E_{1,11}=$ 0.119082 eV, measured from
the band edge, are detached 3.2 meV from the subband energy levels, grouped in an energy interval of 13.9 meV. The transitions to the
first subband of the valence band are forbidden by the new selection rules. The transitions responsible of the spectrum around 420nm, are those to the second subband
of the valence band. In this subband, the surface energy levels around $-0.1001244$eV are detached from the other levels that are grouped between $-0.092$eV and $-0.0954$eV. The detachment of the surface states is an important feature that allows ONE to identify the transitions responsible for the isolated high energy peaks (as transition from surface to surface energy levels) and the groups of peaks in the optical spectra, see the sketch in the left panel of Figure \ref{SLSpectAndOptTrans}. This calculations show that the multiple resonances correspond to optical transitions between truly quantum states within subbands of the conduction and valence bands. Lately, we have also shown that the 1nm difference between the observed and predicted spectrum width, is due to the piezoelectric effect with a local field of $~0.0055$eV/nm.

\section{Conclusion}
We review the theory of finite periodic systems from basic properties to general results for quasi-1D periodic systems. We introduce the transfer matrix method as a tool for dealing with dynamic equations of systems with an arbitrary number $N$ of propagation modes, where we define the transfer matrices $M_N$ and $ W_N $ of dimension $2N \times 2N$. We illustrate matrices of dimension $2 \times 2$ for specific examples of one propagating mode, and we introduce the transfer matrix in the WKB approximation. Then we review the general relations of  transfer matrices $M_N$ and $W_N$ with the scattering amplitudes, the representations of the transfer matrix in the orthogonal, unit, and symplectic universality classes, and their group structures. Based on the transfer matrix properties, particularly on the transfer-matrix combination rule, we outlined the derivation of the matrix recurrence relation, which solutions make it possible to determine analytically the transfer matrix $M_{Nn}$ of  a periodic system with $n$-unit cells and $N$ propagating modes, provided that the unit-cell transfer matrix $M_{N}$ is known. Since  most of the applications of this theory deal with the one propagating mode limit, we only present the final results for the matrix polynomials $p_{Nn}$, and show that in the one propagating mode limit, one has the recurrence relation  of  the Chebyshev polynomials of the second kind, which were first found by Jones and Abelès. The transfer matrix $M_{Nn}$ and the matrix polynomials $p_{Nn}$, which carry all the information of the periodic system, the complex processes and the multiple reflections that particles or wave functions experience along of the periodic structure, are rigorously determined for any number of propagation modes $N$, arbitrary potential profile and any number of unit cells $n$. We then show that given the polynomials $ p_{Nn} $ and $p_{1n} = U_n$, the matrix elements and all physical quantities, such as the scattering amplitudes, the resonant energies, and the Landauer conductance, are expressed in a simple way in terms of these polynomials.  In the second part of this review we presented the more recent results related to the calculation of the energy eigenvalues and eigenfunctions for confined superlattices in the 1D limit. These quantities resolve the fine structure in the bands and at the same time give the possibility of determining the true discrete dispersion relation, as opposed to the continuous dispersion relation obtained when the transfer matrix method is combined with the Floquet theorem. We dedicate a brief section to comment on this approach. We end up the TFPS review with a summary of the analysis of parity symmetries of eigenfunctions and resonant functions.

The TFPS was applied to a broad variety of physical systems,  and we have presented here the application of this theory to a number of physical systems. We addressed first the calculation of the conductance of a biased double barrier, and show that an unexplained and frequently observed feature in negative resistance region can be explained using the multichannel transfer matrix method. We then discuss the transmission in ballistic transistors and the conductance in a channel-coupling superlattice with layers of $\delta$-scatter potentials. We devote a section to review the application of the TFPS to study the transmission and space-time evolution of electromagnetic and electron wave packets through optical, left-handed and magnetic superlattices. We have shown clear evidences of superluminal, optical antimatter, spin filter and spin inversion effects. To conclude, we choose two examples were high-resolution experiments have been performed, and the theoretical predictions are extremely good. The tunneling time through superlattices, with error bars of $\sim 10^{-16}$s, and the optical response of blue-emitting $InGaN$ superlattices, where new features in the optical spectra, with a resolution of $\sim 0.038$meV, and a discrete structure of the subband energy levels are revealed. The aim of this review has been to show that the solid state physics continues evolving towards a truly quantum approach, which was somehow obscured by infinite systems, where the energy spectrum becomes continuous. The discrete spectrum, surface states, and other new properties that emerge in the TFPS are important advances, relevant to theoretical physics, experimental physics, and applications.

\medskip

\acknowledgments

I am grateful to Herbert P. Simanjuntak for the careful reading of the manuscript and the valuable suggestions.
\medskip

\medskip

%

\end{document}